\documentclass[12pt]{JHEP3}

\usepackage{amsmath,epsfig}
\usepackage{amssymb,amsfonts}
\usepackage{latexsym}
\usepackage[latin1]{inputenc}
\usepackage{subeqnarray}
\usepackage{multirow}
\usepackage{xcolor}
\usepackage{epsfig}

\usepackage{graphicx}
\usepackage{longtable}

\allowdisplaybreaks[3]

\relax
\renewcommand{\theequation}{\arabic{section}.\arabic{equation}}
\def\be{\begin{equation}}
\def\ee{\end{equation}}
\def\bea{\begin{eqnarray}}
\def\eea{\end{eqnarray}}

\newcommand\fverb{\setbox\pippobox=\hbox\bgroup\verb}
\newcommand\fverbdo{\egroup\medskip\noindent%
                        \fbox{\unhbox\pippobox}\ }
\newcommand\fverbit{\egroup\item[\fbox{\unhbox\pippobox}]}

\newcommand{\bear}{\begin{eqnarray}}

\newcommand{\eear}{\end{eqnarray}}

\newcommand{\bsea}{\begin{subeqnarray}}
\newcommand{\esea}{\end{subeqnarray}}
\newbox\pippobox

\def\6{\partial}

\def\a{\alpha}
\def\g{\gamma}
\def\nn{\nonumber}

\def\le{\left}
\def\ri{\right}

\def\pa{\partial}

\def\m{\mu}
\def\n{\nu}
\def\rr{\rho}
\def\s{\sigma}

\def\sp{\;\;\;,\;\;\;}

\def\sq
\def\a{\alpha}
\def\b{\beta}
\def\l{\lambda}

\def\hri#1#2{\href{http://arxiv.org/abs/#1}{[ArXiv:#1]#2}}
\def\hre#1#2{\href{http://arxiv.org/abs/#1/#2}{[ArXiv:#1/#2]}}

\def\d{\delta}

\newcommand{\al}{\alpha}
\newcommand{\ga}{\gamma}
\newcommand{\Ga}{\Gamma}
\newcommand{\ba}{\beta}
\newcommand{\da}{\delta}
\newcommand{\ka}{\kappa}
\newcommand{\ud}{\mathrm{d}}
\newcommand{\la}{\lambda}
\newcommand{\La}{\Lambda}

\newcommand{\sig}{\sigma}
\renewcommand{\l}[1]{\left#1}
\renewcommand{\r}[1]{\right#1}

\newcommand{\half}{\frac{1}{2}}
\newcommand{\e}{\mathrm{e}}
\def\nn{\nonumber}

\def\s{\sigma}
\def\e{\epsilon}
\def\thema{c}
\title{Generalized Holographic Quantum Criticality at Finite Density}
\author{\large B. Gout\'eraux$^a$ and E. Kiritsis$^{a,b}$\\
~\\
~\\
$^a$Univ Paris Diderot, Sorbonne Paris Cit\'e, \href{http://www.apc.univ-paris7.fr/APC_CS/}{APC},
UMR 7164 CNRS, F-75205 Paris, France.\\
~
\\
$^b$ \href{http://hep.physics.uoc.gr}{Crete Center for Theoretical Physics},
Department of Physics, University of Crete, 71003 Heraklion, Greece.
\\\\
E-mail: \email{blaise.gouteraux@apc.univ-paris7.fr}, \href{http://hep.physics.uoc.gr/~kiritsis/}{http://hep.physics.uoc.gr/~kiritsis/}
}

\preprint{CCTP-2011-20}

\abstract{We show that the near-extremal solutions of Einstein-Maxwell-Dilaton theories, studied in \cite{cgkkm},  provide  IR quantum critical geometries, by embedding classes of them in higher-dimensional AdS and Lifshitz solutions. This explains the scaling of their thermodynamic functions and their IR transport coefficients, the nature of their spectra, the Gubser bound, and regulates their singularities. We propose that these are the most general quantum critical IR asymptotics at finite density of EMD theories.}


\begin{document}

\section{Introduction}
\label{intro}

Strong interactions of realistic finite-density systems have provided an arena for a wealth of techniques, geared
to assess in most cases the qualitative physics. The remaining lot of unsolved problems especially in the realm of strange metals (which are a subset of condensed matter
systems at the border with magnetism) indicates that new techniques may be welcome.
One such technique is the gauge-gravity correspondence, abstracted from the correspondence between non-abelian gauge theories and string theories.
So far it has been explored in several directions, and seems to provide a novel perspective both in the modeling as well as solution of some strongly coupled QFTs.
The hope behind potential applications to condensed matter physics is that IR strong interactions of the Kondo type in materials where spins can interact with electrons,
may provide bound states that behave in a range of energies as non-abelian gauge degrees of freedom
 that may also be coupled to other fields. The gauge interactions are characterized by a number of charges $N_c$
that are conventionally called ``colors". Their actual number depends on the problem at hand but it is typically small.

If this is the case, then in terms of the electrons and spins the YM fields are composite. However, in the regime where the effective YM interaction
is strong, the physical degrees of freedom are expected to be colorless bound states. Their residual interactions, analogous to nuclear forces in high-energy physics, are still strong. On the other hand,
in the limit of a large number of colors  $N_c\to \infty$, although the original interaction of colored sources is strong, the effective interaction between
colorless bound states can be made arbitrarily weak, as it is controlled by $1/N_c\to 0$.
In this limit, the theory simplifies, and may be calculable.
Of course, typically, the original problem has a finite and sometimes small number of effective colors. The question then is: how reliable are the large $N_c$
estimates for the real physics of the system?
The answer to this varies, and we know many examples in both classes of answers. A good example on one side is the fundamental theory of strong interactions,
Quantum Chromodynamics, based on a the gauge group SU(3), indicating $N_c=3$ colors.
It is by now well-known that for many aspects of this theory, $3\simeq \infty$,  the accuracy varying between $3-10\%$.
It is also known that the analogous theory with two colors, SU(2), has some significant differences from its $N_c\geq 3$ counterparts.
There are other theories where the behavior at finite $N_c$ is separated from the $1/N_c$ expansion by phase transitions, making large-$N_c$ techniques essentially useless.

 Of course, large $N_c$ techniques have been applied to strongly coupled systems for decades, so what is new here?
 In adjoint theories in more than two dimensions, it is well-known that until recently, even the leading order in $1/N_c$ could not be computed.
 Although some qualitative statements could be made in this limit, the amount of quantitative results was scarce, to put it mildly.
 On the other hand, 't Hooft observed that the leading order in $1/N_c$ is captured by the classical limit of a quantum string theory, \cite{hooft}.
 Finding and solving this classical string theory was therefore equivalent to calculating the leading order result in $1/N_c$ in the gauge theory.
 Unfortunately, such string theories, dual to gauge theories, remained elusive until 1997, when Maldacena \cite{maldacena} made a rather radical proposal: (a) This string theory
 lives in more dimensions than the gauge theory\footnote{This unexpected (see however \cite{polyakov}) fact can be intuitively understood in analogy with simpler adjoint theories in 0 or 1 dimensions. There it turns out that the eigenvalues of the adjoint matrix in the relevant saddle point become continuous in the large-$N_c$ limit, and appear as an extra dimension.
 In general how many new dimensions are going to emerge in a given QFT in the large-$N_c$ limit is not a straightforward question to answer, although exceptions exist.};
  (b) At strong coupling, it can be approximated by supergravity, a tractable problem. The concrete example proposed
 contained on one hand a very symmetric, scale invariant, four-dimensional gauge theory (N=4 super Yang-Mills), and on the other a ten-dimensional IIB string theory
 compactified on the highly symmetric constant curvature space AdS$_5\times \mathbf S^5$.
  Therefore this correspondence became known as the AdS/CFT, or holographic, correspondence.

Although this claim is a conjecture, it has amassed sufficient evidence to spark a large amount of theoretical
 work exploring the ramifications of the correspondence, on one hand for the dynamics of strongly
 coupled gauge theories and on the other hand of strongly curved string theories.

In the context of holographically dual string theories, many issues are still not fully understood. First and foremost is that the classical string theories, dual to gauge theories, cannot yet be solved.
The only approximation that makes them tractable is the (bulk\footnote{We refer to as the ``bulk", the spacetime in which strings propagate. This is always a spacetime with a single boundary. The boundary is isomorphic to the space on which the dual quantum field theory (gauge theory) lives.}) derivative expansion. This reflects the effect of the string oscillations on the dynamics of the low-lying string modes.

It is known in many cases and it is widely expected that such an expansion is controlled by the strength of the QFT interactions.
In the limit of infinite strength,  the string becomes stiff and the effects of string modes can be completely neglected. The theory then collapses to a gravitational theory coupled to a finite set of fields.
Since we are working to leading order in $1/N_c$, the treatment of this theory is purely classical. Observables (typically boundary observables corresponding to correlators of the dual CFT)  are computed by solving second-order non-linear differential equations.

The effects of finite but large coupling are then captured by adding in the gravitational action higher-derivative interactions. Note that this derivative expansion is not directly related to the IR expansion of the dual QFT.

The bulk theory, as mentioned earlier,  has at least one, and usually more than one dimensions that are extra compared to those of the dual QFT.
One of them is however special: it is known as the ``holographic" or ``radial" dimension, and controls the approach to the boundary of the bulk spacetime. Moreover, it can be interpreted as an ``energy" or renormalization scale in the dual QFT.

The second order equations of motion of the bulk gravitational theory, viewed  as evolution equations in the radial direction, can be thought of as Wilsonian RG evolution equations, \cite{deboer}. The boundary of the bulk spacetime corresponds to the UV limit of the QFT. Although the equations are second order, they need only one boundary condition in order to be solved, as the second condition is supplied by the ``regularity" requirement of the solution at the interior of spacetime.
Here gravitational physics comes to the rescue: a gravitational evolution equation with arbitrary boundary data leads to a singularity. Demanding regular solutions gives a unique, or a small number of options. The notion of ``regularity" can however vary, and may include runaway behavior as in the case of holographic open string tachyon condensation relevant for chiral symmetry breaking, \cite{chi}.

An important evolution of the holographic correspondence is the advent of the concept of Effective Holographic Theories (EHTs), \cite{cgkkm}, in analogy with the analogous concept of Effective Field Theories (EFTs) in the context of QFT\footnote{There are several works that contain a version or elements of the idea of the EHT, \cite{rg}, although they vary in the focus or philosophy.}.
The rules more or less follow those of EFTs with some obvious changes, and most importantly, with much less intuition.

In standard EFTs, there are several issues that are relevant:
(a) Derivation of the low energy EFT from a higher energy theory;
(b) Parametrization of the interactions of an EFT, and their ordering in terms of IR relevance;
(c) Physical Constraints that an EFT must satisfy.
Although we know many things about EFTs thanks to the Wilsonian approach, there are still general questions which can not be answered with our tools, like whether a given EFT can arise as the IR limit of a UV complete QFT.

 What are the ingredients of an EHT?  As in EFTs, we need to select a collection of fields, and an action principle that will determine  the equations of motion. Of course a string theory has an infinite number of fields, corresponding to all the (single trace) gauge invariant operators of the QFT.

In a strongly coupled theory, an infinite number of operators are expected to have non-zero vevs in the vacuum (except if they are forbidden by symmetries as in CFTs). All the dual fields would have non-trivial solutions profiles in the gravitational theory. However, at strong coupling, this (gauge-invariant) spectrum can be truncated, and only a few important fields (dual to important, low-dimension QFT operators) may be kept. In UV or IR perturbation theory near a fixed point, this selection is clear, and forms the basis of the truncation of the Wilsonian flows to a small number of relevant couplings.

Once the appropriate QFT operators are selected, a dual gravity two-derivative
diffeomorphism\footnote{The diffeomorphism invariance of the gravitational/string theory is an avatar of the translational invariance of the dual CFT. It is simple to show that, coupling the stress tensor of a $d$-dimensional QFT$_d$ to a spin-two source (metric), the effective functional of the source is diffeomorphism invariant. However the source (metric) is $d$-dimensional. The holographic gravitational theory contains a higher-dimensional metric, and is invariant under a higher-dimensional diffeomorphism invariance. How this emerges seems still a mystery.}
invariant action can be written down. Its interactions capture the dynamics of the theory at infinite coupling. Finite coupling corrections can be subsequently added as higher-derivative corrections to the gravitational action.

After this is done, we may revisit the questions posed earlier for the EFTs.
The first one was to  derive the low energy EFT from a higher-energy theory.
The analogous question here would be to determine low-energy correlation functions, from the knowledge of the UV theory, which on the gravitational side is captured by the boundary conditions (at the AdS boundary) of the bulk equations of motion. The process of integrating out degrees of freedom \`a la Wilson, here amounts to running the RG-like equations of motion from the AdS boundary to the deep IR. There are many subtleties with this interpretation as RG flow, and they have been discussed by several authors, \cite{deboer,rg}. Despite this, the main conceptual point made here remains valid.

There is a related issue however that needs to be examined. Consider an RG evolution from the ultimate UV, to an intermediate scale $\Lambda$ that is the defining cutoff scale for an EHT. This is done by solving the second order equations, supplied by two boundary conditions, in order to obtain the values of the fields and their derivatives at the intermediate scale $\Lambda$.
This is the analogue of the calculation of the effective couplings in terms of the UV ones with a caveat: the effective couplings at $E=\Lambda$ now depend not only on the UV values of the couplings but also on their derivatives\footnote{The derivatives can be translated equivalently via the holographic dictionary to the expectation values of the appropriate operators.}.

These values can be the UV values of the EHT, that can be used to solve further the equations of motion in the interior of the space (corresponding to energies smaller than $\Lambda$).
However, it is only in this last evolution that a regularity condition may be imposed, reducing the boundary conditions from two to one. This can be translated back into the relation of the fields at $E=\Lambda$ and those at the boundary, fixing eventually one of the two boundary conditions, and restoring the Wilsonian intuition.

However, according to the Wilsonian intuition, the RG flow of a fixed set of couplings in a QFT is an irreversible process, as short distance information is integrated out. It can therefore run only in one sense: from the UV to the IR.
In the holographic context however, using local operators, it seems that this evolution is reversible as it relies in the solution of standard second order differential equations.
This impression is in fact misleading as it does answer the proper question.
In QFT, the reason that the RG flow is irreversible is the following:
Starting with a theory and a finite set of couplings at scale $E_1$, and integrating out to obtain the theory at scale $E_2<E_1$, we will obtain an infinite number of couplings for all possible operators that are generated in the process. So far there is nothing irreversible. However, we now keep a finite set of the couplings at $E_2$ neglecting the rest of the (usually irrelevant) couplings. It is this act that amounts to coarse-graining, and removes information. Once we write the RG evolution for a finite set  of coupling, the evolution is irreversible. To put it precisely. Given a finite set of couplings at scale $E_2$, there is an infinity of theories at scale $E_1$ which under an RG transformation match the finite set of couplings at $E_2$
 Therefore in a holographic context we should ask a similar question: given the values of a finite set of bulk fields and their derivatives at $r=\Lambda$, how many holographic theories are there that provide these values at $r=\Lambda$, starting from $r=\e<<\Lambda$. The answer is again obviously infinite, and therefore, holographic RG evolution is as irreversible as the QFT one is.

On the other hand, as the dynamical condition of regularity emerges in the IR, the complete IR dynamics can be discussed in terms of IR quantities and actions alone.
Consider for example an action (the EHT action) involving a finite (or infinite) collection of fields, and valid up to $r=1/\Lambda$ in the holographic coordinate.
This EHT action is enough to describe all the relevant IR physics, in terms of a set of couplings (the boundary conditions at the cutoff boundary). The regularity conditions in the IR guarantee that all vevs have been fixed.
In particular the physics is described as a function of the IR parameters (usually called "physical" in QFT parlance). What is missing is the connection of the physical to the bare (UV) parameters.
However, for the study of the non-trivial IR behaviors at strong coupling, this approach is quite powerful, as it can determine most of the interesting IR physics.

An important issue involves the number of relevant bulk fields.
In any QFT the number of operators is infinite, corresponding to the infinite number of the holographically dual string modes. Although all except a finite number have no sources in the UV, they are expected to develop vevs (such vevs can only be redefined away in the presence of exact conformal symmetry.).
This implies that a holographic saddle point solution will in general involve an infinite number of bulk fields. In the UV scaling regime  this phenomenon is known as operator mixing. However, from standard Wilsonian arguments we know that at least in the UV scaling region such flows can be successfully truncated to a small number of fields, namely the ones that (a) are sourced or (b) have small dimensions. This notion is perturbative and can eventually fail in the IR if the theory is strongly coupled.

In the reverse perspective, an EHT is determined by a finite set of bulk fields and an action.
The set of fields is decided as the ones that are instrumental in the determination of the IR saddle point (and therefore participate non-trivially in the saddle-point solution), or give rise to observables via their correlation functions. For example, at zero density, an associated gauge field is trivial, but its fluctuations determine the conductivity at zero density.

The truncation of the infinite spectrum of bulk fields to a finite number that may contribute to the saddle-point solution,  has often important implications  for the solution: a naked singularity may appear in the IR. Singularities are in general not allowed in saddle-point solutions, and typically they should be cloaked by regular horizons. The presence of an horizon reflects an effective coarse-graining in the underlying theory, and typically represents a state with thermal character.
However, naked singularities may be acceptable if their presence is due to the truncation of the relevant fields. To put it differently, the presence of Kaluza-Klein (KK) or stringy states may resolve the singularity.
Examples of singularity resolution by KK states abound in string theory (see \cite{gibbons} for an early reference). In the context of holography, many non-trivial flows become singular upon dimensional reduction. This has led to the development of the Gubser criterion\footnote{Which states that in the case of a relevant neutral scalar operator (such as in EMD theories, \eqref{1}), the scalar potential evaluated on-shell should be bounded from below (note the opposite sign convention we use in \eqref{LiouvillePot} compared to \cite{gubser}). Equivalently, the extremal background must be the limit of a finite temperature system, that is the naked singularity should be coverable by an event horizon and made into a black hole.} for the acceptability of a naked singularity, \cite{gubser}. The resolution of singularities by stringy states is also possible (see for example \cite{kk,dabholkar}),
but is a harder effect to control quantitatively.
In both cases the Gubser criterion is a necessary condition, but it is not known whether it is also sufficient.

A related question to the presence of a naked singularity at zero temperature, is to what extent calculations in the singular saddle point solutions, like correlators, are reliable and not dependent on the eventual singularity resolution. The criterion of a ``repulsive singularity" was developed systematically in \cite{gkn,gkmn}. A repulsive singularity is determined by the fact that the correlators in the singular saddle-point  are well-defined and do not need additional boundary conditions at the singularity. A similar criterion was defined for the evaluation of Wilson loops.

We now come to the second important issue in the EHT, namely the parametrization of the interactions of an EFT, and their ordering in terms of IR relevance.
In usual QFT this is easily achieved by the standard derivative expansion, as well as the notion of an IR anomalous dimension of fields.

The situation in EHTs is more complicated. The reason is that bulk diffeomorphism  invariance mixes radial and transverse derivatives.
In particular, the linearized equations that compute correlators in a given saddle point produce a non-linear function of the space-time derivatives.
Indeed, it is qualitatively understood that what controls derivatives  in the bulk theory is the QFT coupling constants. At strong coupling, higher-derivative interactions are suppressed. Therefore, in an EHT the IR relevant couplings are controllable in the limit of strong coupling.

However, a related issue appears in terms with the same number of bulk derivatives. Consider the scalar potential: in this case, and depending on the flow of scalar fields in the saddle-point solution, different parts of the potential will dominate the IR behavior.
For instance, the potential\footnote{Our conventions on its sign are compatible with (\ref{1}).}
\be
V(\phi)={1\over 2}(e^{2\phi}+e^{-2\phi})
\label{a10}\ee
has an AdS extremum at $\phi=0$. Near this AdS extremum that we can take as the UV CFT, it can be approximated as $V\simeq 1+2\phi^2$, which suffices to construct the UV asymptotics of the holographic RG (hRG) flow of $\phi$.
If the boundary conditions demand that $\phi>0$ along the flow, then in the IR, $\phi\to\infty$ and the potential can be approximated as $V_{IR}\simeq {1\over 2}e^{2\phi}$. For sufficiently low IR cutoff scale, the EHT may use the exponential asymptotics of the potential.

Compactified string theories generate potentials that are typically sums of exponentials, generated by fluxes of various massless forms or the metric. As we will show in this paper, all saddle-point solutions emerging from such potentials are (secretely) conformal (scaling) in the IR.
 Strong IR effects not leading to exponential potentials may also appear in non-trivial compactifications. They will spoil the IR scaling properties
 only in a few concrete but exceptional cases as we will discuss in this paper.

In \cite{cgkkm}, the philosophy of the EHT was applied to the next-to-simplest non-trivial theories, which contain a conserved charge.
In such theories one should always include a bulk graviton and a gauge field dual to the stress tensor and the $U(1)$ conserved current respectively.
The leading relevant IR scalar operator, dual to a neutral\footnote{Most of the original works on AdS/CMT considered charged matter, such as a complex scalar field, in a charged AdS black hole. The latter is unstable to the condensation of the scalar field, which after condensation sources the electric flux on the boundary, \cite{Hartnoll:2011fn}. In this context, the chemical potential drives the charged matter on the boundary. In our setup \eqref{1}, the role of the scalar is not to form a condensate, but to drive the IR asymptotics away from a conformal IR fixed point (such as AdS$_2\times\mathbf R_{(p-1)}$) while retaining a non-zero electric flux to the boundary.} bulk scalar $\phi$ was also included, and a two-derivative action was advocated at strong coupling, which after various field redefinitions takes the form
\be
S=M^{p-1}\int d^{p+1}x\sqrt{-g}\left[R-{1\over 2}(\partial\phi)^2+V(\phi)-{Z(\phi)\over 4}F_{\m\n}F^{\m\n}\right].
\label{1}\ee
This is an Einstein-Maxwell-Dilaton (EMD) theory in $p+1$ dimensions. The general functions include the potential $V(\phi)$ and the bulk inverse gauge coupling $Z(\phi)$.
The IR behavior of such a system may be controlled by an extremum of $V$, leading to an AdS$_{p+1}$ saddle-point solution at zero density.
We will not study such standard IR asymptotics but we will focus on the runaway behavior of the scalar, $\phi\to\pm \infty$ in the IR. To study the IR behavior in such a case it suffices to pick the most important term in the potential that we parametrized as\footnote{With these conventions, AdS fixed points for $\delta=0$ will correspond to \emph{positive} $\La$.}
\be
V\simeq 2\Lambda e^{-\delta\phi}.
\label{LiouvillePot}\ee
For consistency, $\d\phi\to-\infty$ in the IR.
Similarly we parametrize the IR asymptotics of $Z$ as
\be
Z\simeq e^{\g\phi},
\label{LiouvilleGaugeCoupling}\ee
where we absorbed the overall constant in a redefinition of the gauge field.

  It turns out that the exponential parametrization of the IR asymptotics, \eqref{LiouvillePot} and \eqref{LiouvilleGaugeCoupling}, is quite general, unless
  \be
C_p(\g,\d)\equiv 2(p-1)+\ga^2+2(p-2)\ga\da-(2p-3)\da^2=0\,.
\label{4}  \ee
For $C_p(\ga,\da)\neq0$, subleading changes to the potential do not affect the qualitative features of the IR behavior and can be considered as the effects of irrelevant perturbations driving the flow to the IR fixed point.
As we will see later, the leading solution is conformal.

 On the contrary,  for $C_p(\ga,\da)=0$, we expect subleading changes of the potential to affect the qualitative IR physics. The curve (\ref{4}) is the line where the near-extremal black-hole solutions turn from stable to unstable.

In the case of zero density, this phenomenon occurs at
\be
\delta^2=\delta_c^2=\frac2{(p-1)}\,.
\ee
The detailed modifications near these asymptotics, namely the spectra as well as the finite density behavior,  were investigated in \cite{gkn,gkmn}, and  shown to match the behavior of asymptotically free theories, associated to potentials with asymptotics of the form $V_n\sim \phi^n~e^{2\phi\over p-1}$.
Such asymptotics are distinct from exponential asymptotics and lead to non-conformal IR dynamics. In particular, when $0<n<1$, the scale factor vanishes exponentially in the IR as $e^{-{r^{1\over 1-n}}}$ as $r\to \infty$, which is a naked (but holographically acceptable) singularity.
The case $n=1/2$ was argued in \cite{gkn,gkmn} to describe the IR physics of large-N$_c$ YM. When $n>1$, the naked (but holographically acceptable) singularity is at a finite distance $r=r_0$, and the scale factor of the metric vanishes there as $e^{- (r-r_0)^{1\over 1-n}}$.
The IR theory is not conformal, but it is not known whether this case corresponds to any realistic QFT.

Therefore, apart from the theories satisfying (\ref{4}), for all other values of $\gamma,\delta$ the expected qualitative behavior does not change, as argued in \cite{cgkkm} and explicitly checked recently in \cite{cadoni}.

With these reservations, the near T=0 behavior of such EHTs was analyzed based on the generic near-extremal solutions that were found.
In particular, the near-extremal behavior of entropy, energy and conductivity (both AC and DC) were derived in \cite{cgkkm}.
They all implied a power-law behavior characteristic of scale invariant theories, with the difference that hyper-scaling was violated. This a priori looks strange, as the naive interpretation of the saddle-point solutions is as flows by a strongly-relevant scalar operator.

\section{Results and outlook}

In this section, we streamline our results, leaving a more detailed and technical description to subsequent sections and appendices.

In this work, we will explain the near-extremal behavior of charged EMD solutions, by indicating that it is consistent with the fact that such EHTs can be obtained from
higher-dimensional conformally or Lifshitz invariant theories at finite density, \emph{via} 'generalized dimensional reduction'\footnote{A given reduction will be generalized if the number of reduced dimensions can be analytically continued to the real line and give rise to a continuous parameter.}, \cite{Kanitscheider:2009as,skg}. The former will be said to admit \emph{generalized conformal symmetry}, while the latter will have \emph{generalized Lifshitz symmetry}.
So far, it does not seem possible to include the whole $(\gamma,\delta)$ plane (though our results cover a large part of it, see Fig. \ref{Fig:KKUplifts}), and moreover for arbitrary values of $\gamma,\delta$ the higher dimension must be real (in the spirit of the $\e$-expansion), \cite{Kanitscheider:2009as}. Despite this, this observation

\begin{enumerate}

\item Explains the near T=0 scaling behavior.

 \item Explains the qualitative difference between EHTs with $C_p<0$ and $C_p>0$. In the neutral case it explains the crossover value, $\delta_c$.

 \item Provides an alternative view of the Gubser bound.

 \item      Provides one possible resolution of the zero temperature naked singularity of the original solution.

     \item Gives a direct and efficient way to compute the scaling transport coefficients along the lines of \cite{gpr,ske, Kanitscheider:2009as,skg}.

         \item Provides a quick way to embed bottom-up EMD theories into supergravities emerging from string theory. All it takes is to choose the data so that the uplift dimension is integer and in a range (up to 11) that allowed embedding in the known supergravity theories.

\end{enumerate}

In particular, we will analyze several classes of higher-dimensional actions and associated solutions, and show that they dimensionally reduce to our near-extremal solutions. Moreover we will develop a network of theories that share extremal solutions, and we think this is just the tip of an iceberg. In the process we will provide more analytic solutions that are applicable to potentials containing two and three exponentials and have therefore AdS completions.

All higher-dimensional theories considered over the course of this work may be encompassed by the following generic action in $p+q+1$ dimensions:
\be  \label{GenericAction}
	S=\frac{1}{16\pi G_D}\int\ud^{p+q+1}x\,\sqrt{-g}\l[R+2\Lambda-\frac12\partial\Phi^2-\frac{1}{2(n+2)!}G^2_{[n+2]}-\frac{e^{\Gamma\Phi}}{4}F^2_{[2]}\r].
\ee
In Table \ref{Table1}, we give an overview of which (sub)section examines which theory, as well as the type of compactification. This can be diagonal (no KK vectors) or not, and over a curved space or not. For instance, for diagonal reductions, we use either of the two following metric Ans\"atze\footnote{The reader can look up the details of the non-diagonal compactifications in the corresponding sections \ref{sch} and \ref{rot}.}:
\be
	\ud s^2_{(p+q+1)} = e^{-\da\phi}\ud s^2_{(p+1)}+ e^{\frac{\phi}{\da}\l(\frac2{p-1}-\da^2\r)}\ud K^2_{(q)}\,,\quad \da^2={2\over p-1}~\frac{q}{(p+q-1)}\leq \da_c^2\,, \label{KKAnsatz1}
\ee
\be
	\ud s^2_{(p+q+1)} = e^{-\frac{2\phi}{(p-1)\da}}\ud s^2_{(p+1)}+ e^{\frac{\phi}{\da}\l(\da^2-\frac2{p-1}\r)}\ud K^2_{(q)}\,, \quad  \da^2=\frac2{p-1}+\frac2q\geq \da_c^2
\label{KKAnstaz2}
\ee
which cover two complementary ranges of the reduction exponent $\delta$. If the number of reduced dimensions $q$ can be traded for the continuous parameter $\da$ in the lower-dimensional theory, then the reduction is generalized in the sense described in \cite{Kanitscheider:2009as,skg}.

\TABULAR{|c|c|c|c|c|ccccc|}{
\hline
Section&$\Lambda$&$G_{[n+2]}$&$F_{[2]}$&$\Phi$&\multicolumn{5}{|c|}{$KK$}\\
\hline\hline
\ref{section:NeutralAds}&$\checkmark$&\O&\O&\O& $\mathbf D$ & $\mathbf T^q$ & $\mathbf{G}$ & $\mathbf C$ & $\da^2\leq\da_c^2$\\
\hline
\ref{section:NearExtrBoostedAds}&$\checkmark$&\O&\O&\O& $\mathbf{ND}$ & $\mathbf S^1$ & $\mathbf{NG}$ & $\mathbf C$ & $\da^2\leq\da_c^2$\\
\hline\hline
\ref{n0}&\O&$n=0$&\O&\O& $\mathbf D$ & $\mathbf K^q$ & $\mathbf{G}$ & $\mathbf C$ & $\da^2\geq\da_c^2$\\
\hline
\ref{nq}&\O&$n=p-1$&\O&\O& $\mathbf D$ & $\mathbf K^q$ & $\mathbf{G}$ & $\mathbf C$ & $\da^2\geq\da_c^2$\\
\hline\hline
\ref{qb1}&$\checkmark$&$n=q$&\O&\O& $\mathbf D$ & $\mathbf K^q$ & $\mathbf{G}$ & $\mathbf C$ & $\da^2\leq\da_c^2$\\
\hline
\ref{b}&$\checkmark$&$n=p-1$&$\checkmark$&\O& $\mathbf D$ & $\mathbf K^q$ & $\mathbf{G}$ & $\mathbf C$ & $\da^2\leq\da_c^2$\\
\hline\hline
\ref{Lifshitz}&$\checkmark$&\O&$\checkmark$ & $\checkmark$ & $\mathbf D$ & $\mathbf T^q$ & $\mathbf{G}$ & $\mathbf{NC}$ & $\da^2\leq\da_c^2$\\
\hline
\ref{NEdilatonic}&\O&\O&$\checkmark$&$\checkmark$& $\mathbf D$ & $\mathbf K^q$ & $\mathbf{G}$ & $\mathbf{NC}$ & $\da^2\geq\da_c^2$\\
\hline\hline
\ref{rot}&\O&$n=p-1$&\O&\O& $\mathbf{ND}$ & $\mathbf S^q$ & $\mathbf{NG}$ & $\mathbf C$ & $\da^2\leq\da_c^2$\\
\hline
}
{A guide to the various higher-dimensional theories considered in this work. The symbol \O\ means that the field is absent from the higher-dimensional theory,  $\checkmark$ that it is present. From left to right, the second up to the fifth columns indicate respectively the presence of a higher-dimensional cosmological constant, a form $G_{[n+2]}$ (if so, we only indicate the value of its rank $n$), a Maxwell field $F_{[2]}$, a scalar field $\Phi$ (always accompanied by a non-zero gauge coupling $\Gamma$ to $F_{[2]}$). In the sixth column, $KK$, we indicate whether the reduction is diagonal (if it is we denote it by  $\mathbf D$, and therefore no KK vectors are produced), or non-diagonal ($\mathbf{ND}$); whether it is over a torus ($\mathbf T^q$) or a curved space ($\mathbf K^q$); whether it is generalized ($\mathbf G$) or not ($\mathbf{NG}$) as explained in the main text; whether it is consistent ($\mathbf C$) or not ($\mathbf{NC}$); as well as the range of the $\da$ exponent in the reduction Ansatz.
\label{Table1}}

We now turn to the lower-dimensional theories examined in this work, which can all be encompassed by the following $(p+1)$-dimensional action:
\be
	S_{(p+1)}=\int\frac{\ud^{p+1}x\sqrt{-g}}{16\pi G_{(p+1)}}\l[R-\frac12\l(\partial\phi\r)^2-\frac{e^{\g\phi}H^2_{[n+2]}}{2(n+2)!}+V_1 e^{-\da_1\phi}+V_2 e^{-\da_2\phi}+V_3 e^{-\da_3\phi}\r],
\ee
and summarize in table \ref{Table2} which (sub)section examines which theory. The exponential terms in the potential have various origins:
\begin{itemize}
	\item The higher-dimensional cosmological constant;
	\item The curvature of the reduced compact space, in a (non-)diagonal reduction;
	\item A dualised $0$-form originating from the reduction of a $p$-form gauge potential.
\end{itemize}

\TABULAR{|c|c|c|c|c|}{
\hline
Section & $\Lambda_1$ & $\Lambda_2$ & $\Lambda_3$ & $H_{[n+2]}$\\
\hline\hline
\ref{section:NeutralAds} & $\checkmark$ & \O & \O & \O\\
\hline
\ref{section:NearExtrBoostedAds} & $\checkmark$ & \O & \O & $n=0$ $\ga\da_1=\d_c^2$\\
\hline\hline
\ref{KKsm} & $\checkmark$ & \O & \O &  $\ga\da_1=(n+1)\d_c^2$\\
\hline
\ref{n0} & $\checkmark$ & \O & \O &  $n=0$ $\ga\da_1=\d_c^2$\\
\hline
\ref{nq} & $\checkmark$ & $\da_1\da_2=p\da_c^2$ & \O & \O \\
\hline\hline
\ref{qb1} & $\checkmark$ &   $\da_1\da_2=\da_c^2$ & \O &  $n=0$ $\ga=-(p-2)\da_1$\\
\hline
\ref{b} & $\checkmark$ &   $(p-1)\da_2=\da_1 - \ga$ & $\da_3=\frac2{\da_1 - \ga}$ &   $n=0$ \\
\hline\hline
\ref{Lifshitz} & $\checkmark$ & \O & \O & $\checkmark$\\
\hline
\ref{NEdilatonic} & $\checkmark$ & \O & \O & $\checkmark$\\
\hline\hline
\ref{rot} & $\checkmark$ &  $\sim\Lambda_1$ &  $\sim\Lambda_1$ & $\checkmark$\\
\hline
}{A guide to the E(M)D theories considered in this work, classified by (sub)section. We indicate the number of exponential terms in the scalar potential and the presence of charge (accompanied by the rank of the corresponding form). If there are several exponential terms in the potential, we enter the relations between the exponents in the table. In the last line, the reader is invited to refer directly to section \protect\ref{rot}, since the expressions are a little cumbersome to enter in a table.
\label{Table2}}

\begin{enumerate}

\item In section \ref{sch}, we examine the reduction of Einstein-AdS gravity
\be
S=\int d^{p+q+1}x\sqrt{G}\left[R+2\Lambda\right],
\label{i1}
\ee
allowing for KK vectors. We first turn them off in section \ref{section:NeutralAds} and obtain a neutral dilatonic theory with a single exponential, which admits a black hole solution with a planar horizon. The uplift of this solution is a higher-dimensional AdS-Schwarzschild  black hole. The scaling symmetry is explained at extremality as it is embedded in the conformal symmetry of the higher-dimensional theory. Note that $\d<\d_c$ in this construction. Therefore the continuous spectrum and absence of mass gap is correlated with the conformal symmetry and the flatness of internal space.

In section \ref{section:NearExtrBoostedAds}, allowing for charge while requiring a single scalar is only consistent for $q=1$.
In that case the higher-dimensional uplift of the near-extremal
 dilatonic charged solutions is a moving (infinite boost limit) AdS-Schwarzschild $(p+2)$-dimensional black hole.

\item In section \ref{section:StaticFlatBranes}, we examine the diagonal reduction over a compact space $\mathbf K^q$ of Einstein gravity plus an $[n+2]$-form field strength
\be  \label{i3}
	S=\frac{1}{16\pi G_D}\int\ud^{p+q+1}x\,\sqrt{-g}\l[R-\frac{1}{2(n+2)!}G^2_{[n+2]}\r],
\ee
leaving the form untouched.
This class of actions contains among others the three main branes of string/M-theory.
The single exponential potential is provided by the curvature of the $\mathbf S^{q}$.
There are two subcases of interest described in sections \ref{n0} and \ref{nq}.

\begin{enumerate}
\item $n=0$:  In this case, the charged near-extremal solution of the lower-dimensional theory lifts to a charged near-extremal RN black-hole with a horizon of topology  $\mathbf S^{q}\times \mathbf T^{p-1}$, in the higher dimensions.

\item $n=p-1$: In this case, a second exponential potential is generated by dualising the $0$-form field strength in the action as in (\ref{EDToroidal2Potentials}). There is no AdS minimum. A new solution is presented, with emergent AdS$_{p+1}$ symmetry in the IR.
        It lifts in the higher dimension to a uniform $(p-1)$-brane wrapped on a torus  with a transverse $\mathbf S^{q}$.

     When the charge of the $(n+1)$-form is zero, this is an uncharged solution of the single exponential theory, but with $\d>\d_c$.
     In this case the lower-dimensional theory has a mass gap and discrete spectrum, \cite{gkn}. The higher-dimensional solution is AdS$_{p+1}\times \mathbf S^{q}$ reduced on $\mathbf S^{q}$, that explains both the scale invariance and the presence of the discrete and gapped spectrum.

     In this case the Gubser bound becomes $q>1$, which is equivalent to the internal sphere having a positive curvature.

\end{enumerate}

\item In section \ref{qb}, we study AdS Einstein gravity with forms:
\begin{enumerate}
\item In section \ref{qb1}, we describe the diagonal reduction over a compact space $\mathbf K^q$ of AdS Einstein gravity plus a $[q+2]$-form field strength:
\be
	S=\frac{1}{16\pi G_D}\int\ud^{p+q+1}x\,\sqrt{-g}\l[R-\frac{1}{2(q+2)!}G^2_{[q+2]}+2\La\r].
	\label{i5}
\ee
In the higher-dimensional theory, the solution is that of a near-extremal  $q$-brane, wrapped on a $q$-sphere.
The extremal metric is AdS$_{q+2}\times\mathbf K^{p-1}$, with the charge of the $q$-brane sourcing the curvature of spacetime. The dimensional reduction happens along the $q$ curved directions of AdS$_{q+2}$ leading to a warped AdS$_2$ geometry.

In lower dimensions, the scalar potential originates both in the higher-dimensional cosmological constant and in the internal curvature of the reduced space.. The theory \eqref{1} with potential \eqref{LiouvillePot} and near-extremal solution \eqref{2}-\eqref{7b} is recovered by setting to zero the curvature of the $\mathbf K^q$ and $\mathbf K^{p-1}$.

\item In section \ref{b}, we describe the diagonal reduction over a compact space $\mathbf K^q$ of AdS Einstein gravity plus a Maxwell field and a $[p+1]$-form field strength:
\be
	S=\frac{1}{16\pi G_D}\int\ud^{p+q+1}x\,\sqrt{-g}\l[R-\frac14F^2_{[2]}-\frac{1}{2(q+2)!}G^2_{[p+1]}+2\La\r],
	\label{i5.1}
\ee
The situation is quite similar to the previous case, except that in the higher-dimensional geometry, the $[p]$-form potential is supported by the transverse compact space $\mathbf K^{p-1}$ together with the time direction. It can be better visualised as a magnetic flux supported solely by the $\mathbf K^{p-1}$ around which the worlvolume of the brane is wrapped. The Maxwell term realises an electric point-field on the the worlvolume of the brane.

The lower-dimensional theory can have up to three exponential potentials, originating from the higher-dimensional cosmological constant, the curvature of the reduced space and the dualisation of the $[p+1]$-form. Again, setting all internal curvatures to zero recovers the action \eqref{1},\eqref{LiouvillePot} and solution \eqref{2}-\eqref{7b}.
\end{enumerate}

Note that the uplift in both case requires imposing a certain relation between the gauge field and the exponential potential so that the higher-dimensional equations of motion hold.

\item In section \ref{generic}, we consider successively two different uplifts of the charged near-extremal solutions to solutions of an EMD theory with higher-dimensional scalar $\Phi$ and gauge coupling $\Gamma$, first with cosmological constant and then without:
\be
S=\frac{1}{16\pi G_D}\int\ud^{p+q+1}x\,\sqrt{-g}\l[R--\half\partial\Phi^2-\frac14e^{\Gamma\Phi}F^2_{[2]}+2\La\r],
	\label{i6.1}
\ee
\begin{enumerate}
	\item In \ref{Lifshitz}, we uplift diagonally over a torus $\mathbf T^q$ the charged near-extremal EMD solutions to asymptotically Lifshitz solutions. The uplift allows to cover part of the $(\ga-\da)(\ga+(p-2)\da)>0$ quadrants.
	\item In \ref{NEdilatonic}, we uplift diagonally over a curved space $\mathbf K^q$ the charged near-extremal EMD solutions to the near-horizon geometry of asymptotically flat dilatonic black $p$-branes with $0$-charge.
This uplift covers part of the complementary quadrants  $(\ga-\da)(\ga+(p-2)\da)<0$.
\end{enumerate}

\item  In section \ref{rot},  we describe connections between three players:

\begin{enumerate}

 \item The first are rotating black brane solutions in Einstein gravity plus a $[p+1]$-field strength:
\be  \label{i7}
	S=\frac{1}{16\pi G_D}\int\ud^{p+q+1}x\,\sqrt{-g}\l[R-\frac{1}{2(p+1)!}G^2_{[p+1]}\r].
\ee
Such solutions include the well-known M$_2$, D$_3$ and M$_5$ brane solutions for $q=7,5,4$ and $p=3,4,6$. Two scaling limits can be taken in these solutions: first, the usual one where the charge parameter is large; second, the one where the rotation parameter is large\footnote{Practically speaking, both mean that the constant factor in a given harmonic function drops out.}. Taking the near-extremal limit where the charge sourcing the brane is large (decoupling limit), such solutions can be specialized in a lower-dimensional reduction to asymptotically AdS non-scaling solutions with non-trivial metric, a single gauge field and a single scalar. Contrarily to the cases examined heretofore, the gauge field now comes from rotation in the higher-dimensional theory, not charge. Generically, the non-diagonal reduction over an $\mathbf S^{q}$ would involve one exponential potential accounting for the curvature of the sphere, plus as many exponential potentials as there were independent rotation planes with a non-trivial gauge field turned on.

\item The second is a lower-dimensional theory of a metric, vector and a single scalar with a potential including three exponential factors:
\be
	S=\frac{1}{16\pi G_D}\int\ud^{p+1}x\,\sqrt{-g}\l[R-\half(\partial\phi)^2-\frac14e^{-(p-2)\da\phi}\l(F_{[2]}\r)^2+V(\phi)\right], \label{i7a}
\ee
\bea V(\phi)&=&\frac{2(p-1)^2(p-2)\da^2V_0e^{-\frac{2\phi}{(p-1)\da}}}{p(1+\frac{(p-1)}2(p-2)\da^2)^2}\l[-\frac{(p-2)}{4(p-1)}\l(1-p\frac{(p-1)}2\da^2\r)+\r.\nn\\
&& \l.+e^{\frac{2+(p-2)(p-1)\da^2}{2(p-1)\da}\phi}+\frac{(p-(p-2)^2\frac{(p-1)}2\da^2)}{2(p-1)^2(p-2)\da^2}e^{\frac{2+(p-1)(p-2)\da^2}{(p-1)\da}\phi}\r]. \label{KKSpPotential2}
\eea
This action has a single parameter $\delta$ entering the potential.

\item The third are the charged, single scalar, near-extremal solutions in section \ref{qb} with $\gamma+(p-2)\d=0$. A single exponential potential is retained in the action \eqref{EMDActionKK1}, stemming from the reduction of some internal $\mathbf{S}^q$.

\end{enumerate}

The connections between these three classes are as follows.

(A) The non-extremal solutions of (5a) in the decoupling limit reduced to $p+1$ dimensions, and specialized to a single gauge field and scalar, are identical to the non-extremal solution of (5b), when $\delta={1\over p-2}\sqrt{2\over p-1}\,$.

The potential in (5b) has an AdS extremum, see Fig. \ref{Fig:AdSPotential}. The non-extremal solutions describe flows from this UV extremum to an IR scaling geometry.

(B) The near-extremal limit of the solutions of (5b) are the same as the near-extremal solution in (5c) when the horizon is \emph{not} flat. In the planar horizon limit, the neutral near-extremal solutions are recovered: this illustrates that the planar and near-horizon limit do not necessarily commute.
Therefore the IR scaling geometry of (5b) is one with an emerging generalized Lifshitz symmetry.
In this sense the theory in (5b)  can be described as a possible UV completion of the EMD solutions. It is interesting that the full flow is known analytically here.

\end{enumerate}

\FIGURE{
\begin{tabular}{cc}
\includegraphics[width=.45\textwidth]{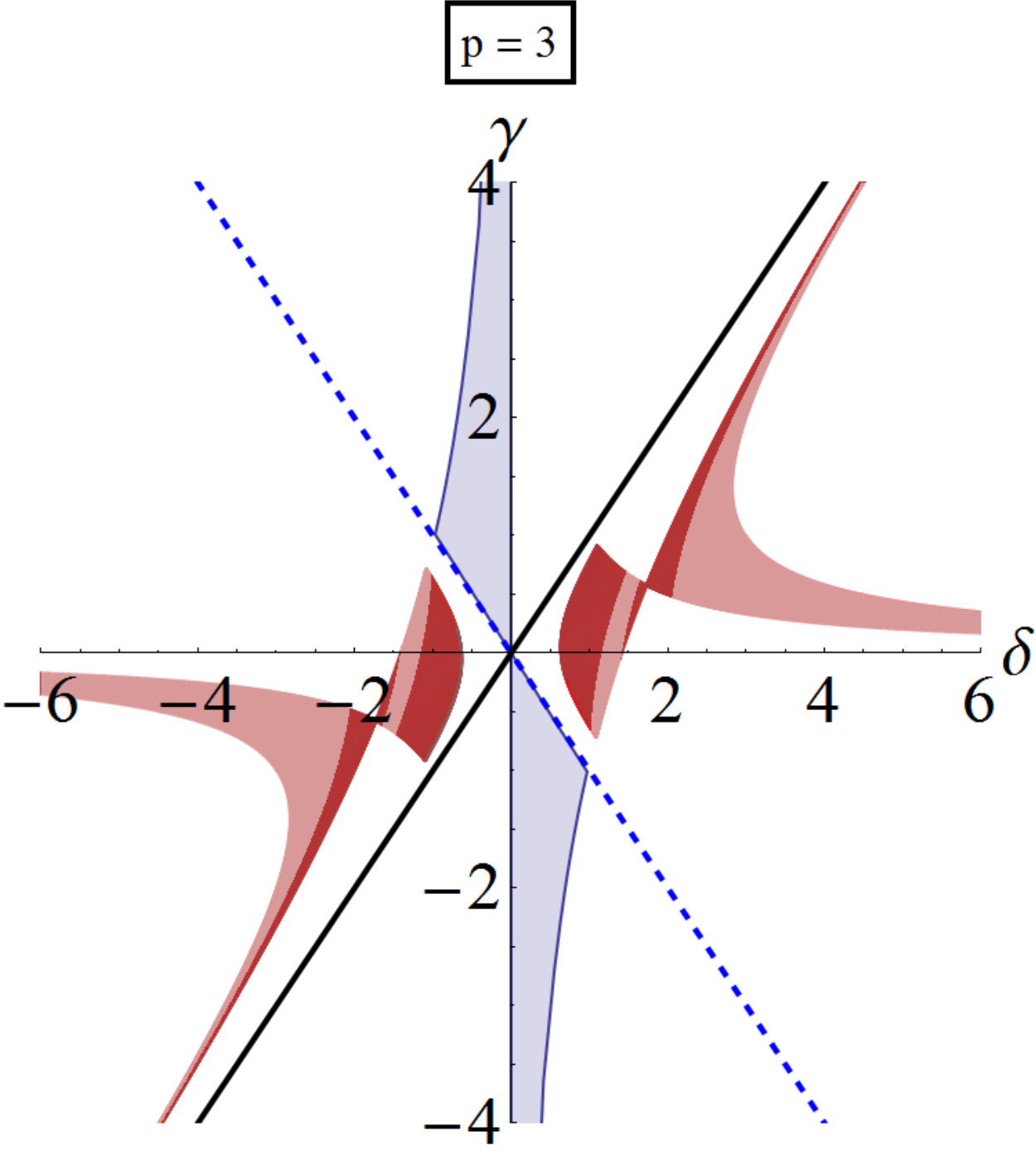}&\includegraphics[width=.45\textwidth]{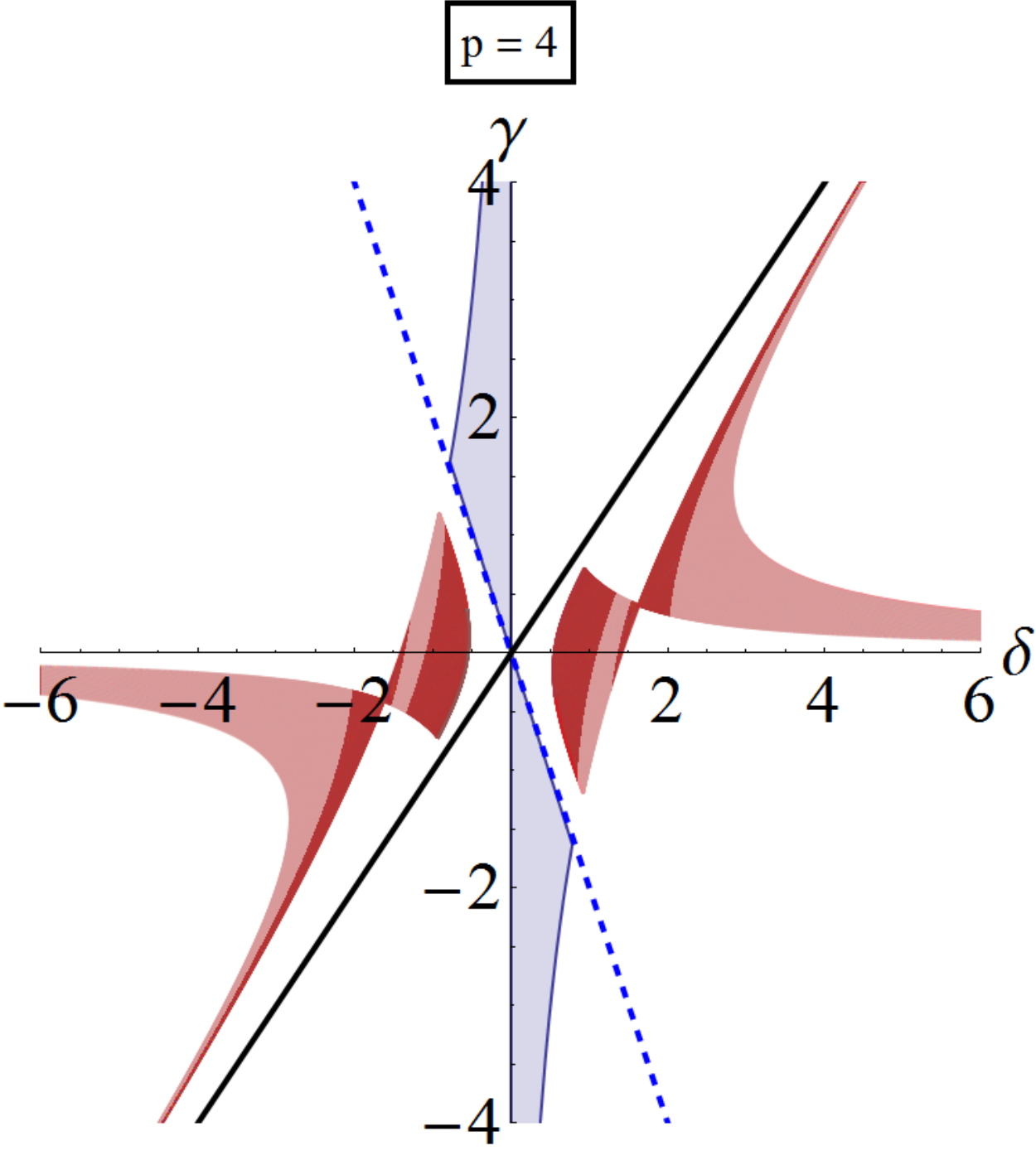}
\end{tabular}
\caption{Parameter spans of the various uplifts of the charged near-extremal solutions \protect\eqref{2}-\protect\eqref{7b} in the $(\gamma,\delta)$ plane for two values of the dimension: on the left, $p=3$; on the right, $p=4$. The solid black line is $\g=\d$, where the uplift is AdS$_2\times\mathbf R^{p+q-1}$. The dashed blue line is $\ga=-(p-2)\da$, where the uplift is AdS$_q\times\mathbf R^p$; on this line, for $\d^2>\d_c^2$, the system is gapped. The blue region is the uplift to Lifshitz solutions with horizon $\mathbf R^{p+q-1}$, while the red (dark+light) one is to dilatonic near-extremal $(p-1)$-branes with horizon $\mathbf R^{p-1}\times\mathbf S^q$. In the light (dark) red region, the system is gapped (gapless). Only the $\g=\d$ and  $\ga=-(p-2)\da$ uplifts are consistent truncations.}
\label{Fig:KKUplifts}
}

The reductions describe above are all \emph{consistent}, except for those in section \ref{generic} and \ref{rot}. By \emph{consistent}, we mean that \emph{every} solution of the lower-dimensional equations of motion are also solutions to the higher-dimensional equations of motion. It is so for the diagonal reductions of sections \ref{section:NeutralAds}, \ref{section:StaticFlatBranes} and \ref{qb}\footnote{In \cite{skg}, the consistency of the curved reduction of Einstein AdS with a Maxwell gauge field was proven. It is straightforward to extend it to the above cases.}, while the non-diagonal $\mathbf S^1$ reduction of section \ref{section:NearExtrBoostedAds} is well-known. In section \ref{generic}, we consider diagonal reductions which are clearly not consistent, since one needs to tune the field strength and the scalar potential in order to truncate one of the two scalars. However, it happens that such a procedure is feasible for the near-extremal charged solutions \eqref{2}-\eqref{7b}, so that the uplift of the \emph{solutions} is consistent even though the uplift of the \emph{theory} is not. The consistency of the non-diagonal reduction of section \ref{rot} has been shown explicitly in \cite{Cvetic:1999xp} in the cases $p=3,4,6$ and $q=7,5,4$ respectively, but not for generic $p$. In section \ref{rot}, we argue that the reduction of the solutions is consistent, although it falls short from proving the full consistency\footnote{Though we do expect that the reduction should be consistent.}.

Although we have not managed to lift  every solution on the $(\g,\d)$ plane (see Fig. \ref{Fig:KKUplifts} for a summary), the pattern above is obvious and leads to propose that {\it all extremal solutions described in section \ref{extremal} correspond to (anisotropic) scale invariant states both at zero and finite density.}
In particular, at zero density they all have conformal AdS symmetry, while at finite density they have conformal Lifshitz symmetry with
\be
z={(\ga-\d)(\g+(2p-3)\d)+2(p-1)\over (\ga-\d)(\ga+(p-2)\d)}\,.
\label{i8}\ee
By varying $\g,\d$ in the range allowed by the Gubser bound
\be
	\g^2-\g\d+2>0\,,\quad -\d^2+\g\d+2>0\,,\quad p\g^2-2\g\d-(p-2)\d^2+2(p-1)>0\,,
\ee
$z$ takes all real values, as can be seen from  Fig. \ref{Fig:z2D}, \ref{Fig:z3D}.

\FIGURE{
\begin{tabular}{cc}
\includegraphics[width=.45\textwidth]{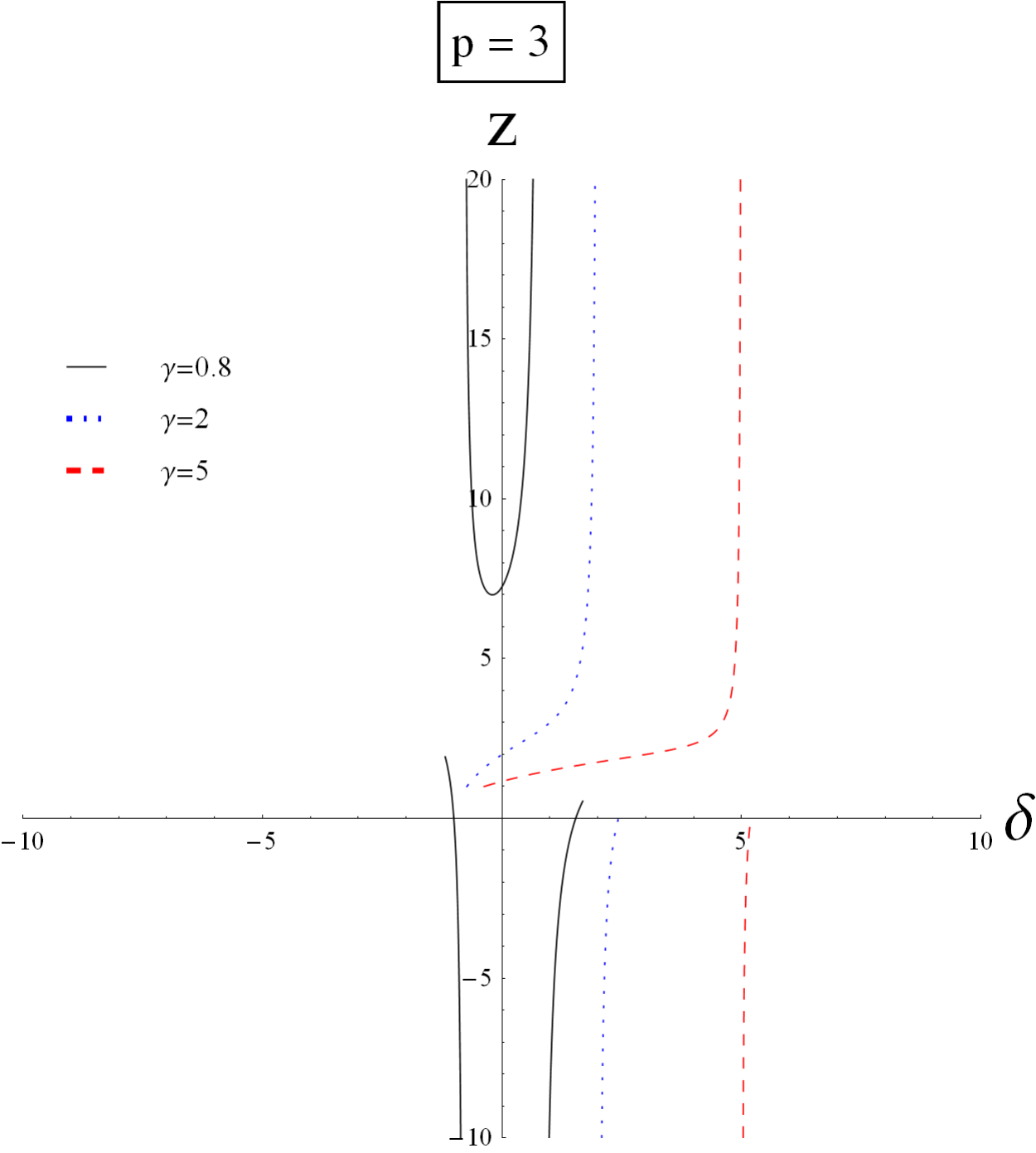}&\includegraphics[width=.45\textwidth]{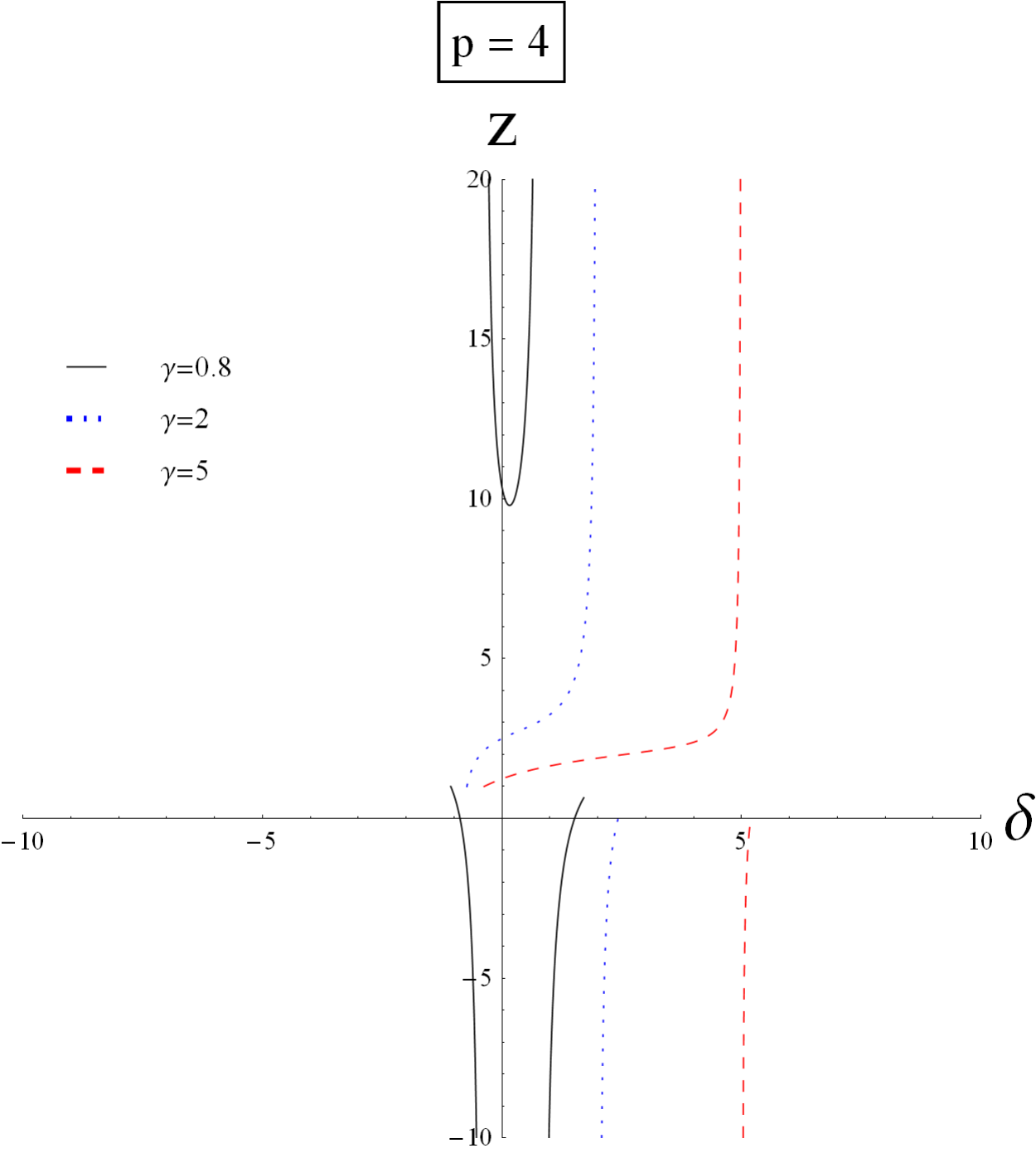}
\end{tabular}
\caption{The dynamical exponent $z$, \protect\eqref{i8}, in terms of $\d$ and for $\g=0.8,2,5$ (solid black, dotted blue and dashed red lines respectively), for $p=3$ (left) and $p=4$ (right), taking into account the Gubser bound.}
\label{Fig:z2D}
}

\FIGURE{
\begin{tabular}{cc}
\includegraphics[width=.45\textwidth]{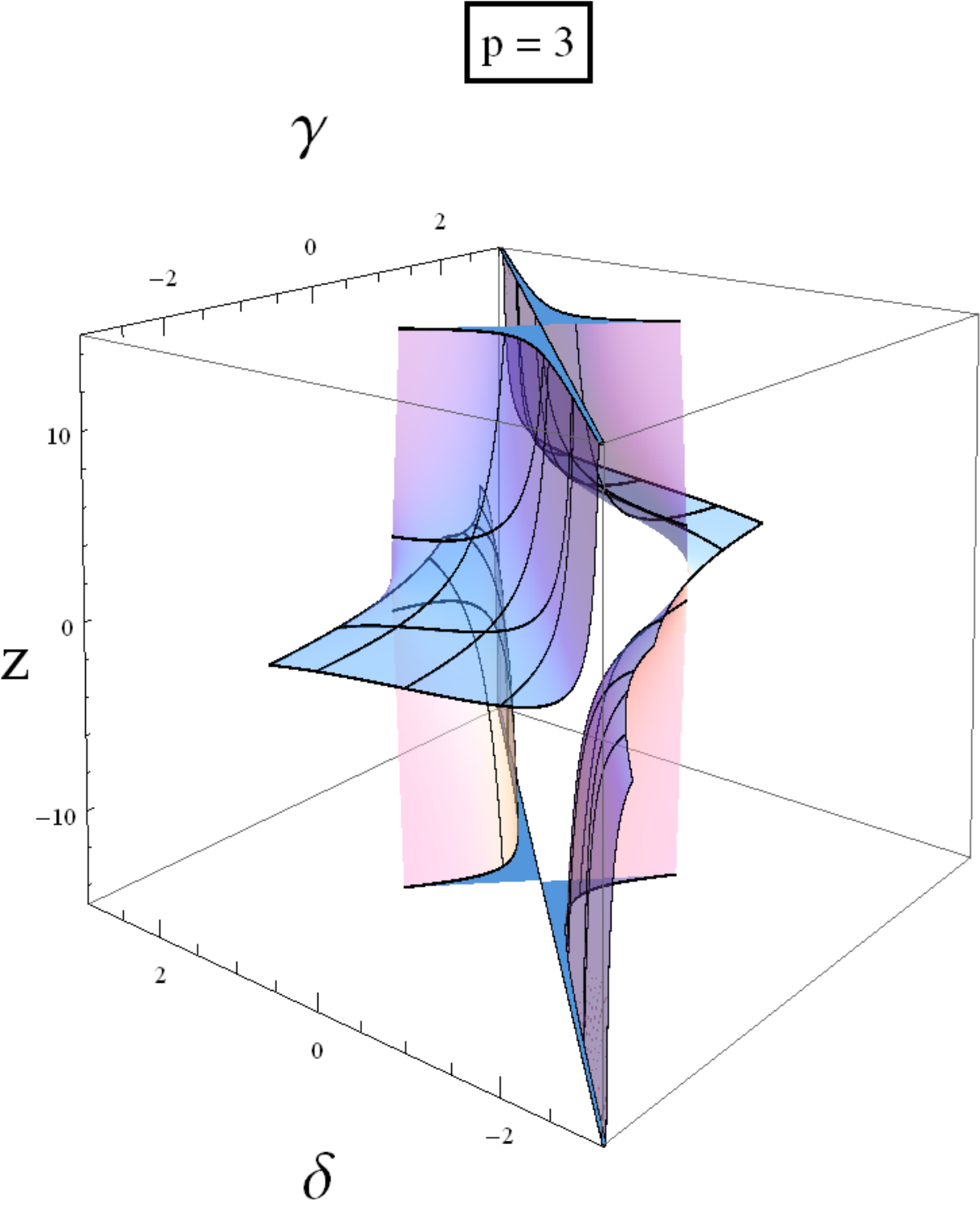}&\includegraphics[width=.45\textwidth]{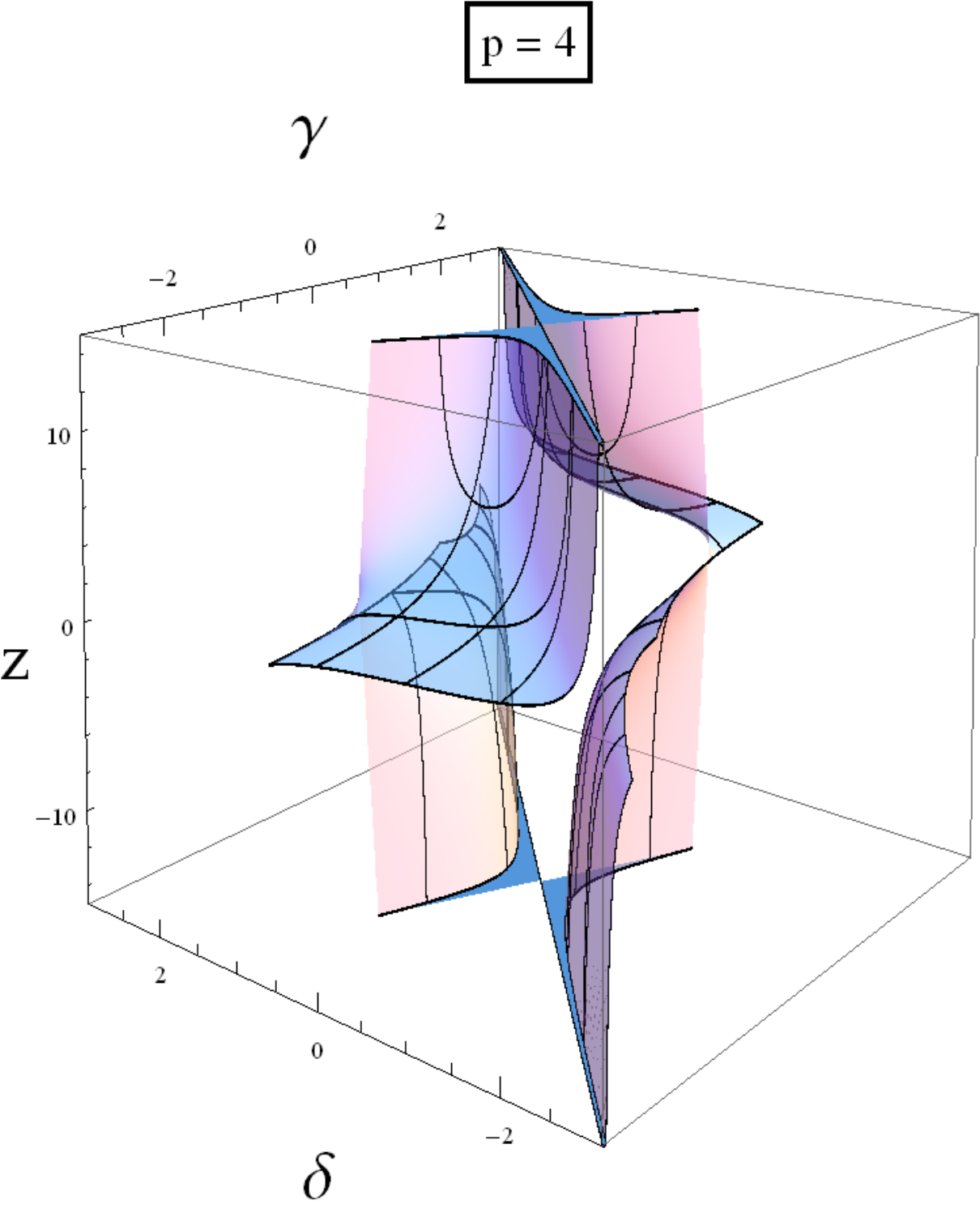}
\end{tabular}
\caption{The dynamical exponent $z$, \protect\eqref{i8}, in terms of $\g$ and $\d$, for $p=3$ (left) and $p=4$ (right), taking into account the Gubser bound.}
\label{Fig:z3D}
}
Some special cases are of note:
\begin{itemize}

\item The case $\delta=0$ corresponds to the Lifshitz solutions with $z=1+{2(p-1)\over \gamma^2}$ found in \cite{taylor}.

\item The case $\gamma=\d$ gives the expected AdS$_2$ near horizon geometry ($z=\infty$), and is the only one with finite entropy at extremality.

\item The case  $\ga+(p-2)\d=0$ also leads to ($z=\infty$) and AdS$_{2}$ symmetry but the extremal entropy vanishes.

\end{itemize}

In the process we have presented several new solutions to actions that go beyond the ones postulated in \eqref{1}, for which it would be interesting to analyze further the holographic physics at finite density.

In all the cases that can be uplifted to higher-dimensional solutions, the latter are regular. Therefore the naked singularity is resolved by the KK states, and the curvature is related to the coefficient of the scalar potential. Nevertheless, this does not preclude a different resolution by stringy states.

Another important issue is the extremal entropy. In the chargeless solutions the extremal entropy is zero. This is also visible in the higher-dimensional incarnation as the uplifted solution is AdS.

At finite density, the extremal entropy is zero except in the case of $\gamma=\delta$ that leads to a charged AdS$_2$ extremal geometry and therefore finite entropy. As the scalar becomes trivial in this limit, so is the uplifting to a higher-dimensional AdS$_2$ extremal geometry, which is known to have finite entropy\footnote{This is the near-horizon limit of charged planar AdS black holes.}. For  the $\g+(p-2)\d=0$ solutions however that also have an AdS symmetry the uplifted geometry is AdS$_{q+2}\times T^{p-1}$ and therefore has zero entropy.
Upon reduction on $T^q$ the size of the torus vanishes at extremality and this preserves the zero-entropy limit in the lower-dimensional solution.

 On the other hand, for the uplifted charged solutions with $\gamma\delta={2\over p-1}$, the entropy is zero at extremality due to the asymmetric scaling symmetry of the higher-dimensional solution.

At finite temperature the dependence of entropy on temperature becomes also transparent in the higher-dimensional setup.
Consider for example the neutral scalar solutions (\ref{28}) with
\be
S\sim T^{2(p-1)\over 2-(p-1)\d^2}.
\label{i9}\ee
Using the relation (\ref{29}) that relates $\d$ to the higher-dimensional theory and substituting in (\ref{i9}) we obtain
\be
S\sim T^{p+q-1},
\label{i10}\ee
which is the natural answer for a $(p+q+1)$-dimensional CFT.
This generalizes to all other cases.

All upliftable cases associated to a discrete gapped spectrum\footnote{Such cases where identified in \cite{cgkkm} as those where the black holes are unstable $C_p(\g,\d)<0$ and the coefficient of the IR Schr\"odinger potential is positive.} uplift to curved internal manifolds $\mathbf K^q$ (though the converse is not true: there are curved uplifts which are gapless), see Fig \ref{Fig:KKUplifts}. In such cases the Gubser bound appears as the limit in which $q\to 1^{+}$ as in this limit the internal curvature that supported the solution vanishes.

The higher-dimensional picture has both conceptual and practical applications.
At the conceptual level in explains many facts that seem mysterious or less motivated in the near-extremal solutions, namely scale invariance, nature of spectra, the Gubser bound, etc.

On the practical side, they can help in simplifying the calculation of transport coefficients. This was already indicated in neutral backgrounds in the case of the calculation of the bulk viscosity in \cite{gpr,ske,Kanitscheider:2009as}.
A simple hydrodynamic reduction related the bulk viscosity of the lower-dimensional solution to the shear viscosity of the higher-dimensional AdS solution.
Such relations generalize in the presence of charge. We will not explore them here in their full generality\footnote{This is done in \cite{skg}.} but in section \ref{transport} we will review the scaling transport coefficients in the IR, and derive the full AC conductivity showing that it is a scaling function of ${\omega\over T}$.

We conclude this exposition by stating that the near-extremal geometries presented in \cite{cgkkm} and analyzed further here provide the most general holographic quantum critical points in such theories, generalizing AdS and Lifshitz geometries. A similar classification of IR holographic quantum criticality is also possible in the superconducting phase. We will report on this is a future publication, \cite{gk} (see also \cite{Liu:2010ka} for related work on the superconducting instability in EMD backgrounds).

The structure of the rest of this  paper is as follows.
In section \ref{extremal}, we review the neutral and charged near-extremal solutions EMD theories described already in \cite{cgkkm}.
In section \ref{sch}, we describe the KK reduction of higher-dimensional Einstein theory giving rise to EMD theories in lower dimensions and present a case with Schr\"odinger symmetry. In section \ref{section:StaticFlatBranes} we describe the dimensional reduction of static, asymptotically-flat black $(p-1)$-branes and their near-horizon limits, giving rise to both charged and neutral solutions of EMD theories.
In section \ref{qb}, we describe successively the dimensional reduction of the near-extremal limit of charged AdS black $(p-1)$-branes in \ref{qb1}, and the same with a magnetic flux on the worldvolume of the brane and a $U(1)$ charge, again providing solutions to EMD theories in lower dimensions.
In section \ref{generic}, we describe the uplift of the charged near-extremal EMD solutions to either Lifshitz solutions or near-extremal dilatonic black $p$-branes.
In section \ref{rot}, we describe the connections between rotating $(p-1)$-branes and AdS dilatonic black holes, via KK reduction to EMD theories.
In section \ref{transport}, we review the scaling properties of thermodynamic functions and transport coefficients of near-extremal EMD solutions, and derive the equation for the AC conductivity exhibiting its scaling in terms of the frequency and temperature.

Appendix  \ref{conv}, contains our notations and conventions.
In appendix \ref{SUGRAAdS}, we describe the effective 4-, 5- and 7-dimensional EMD effective actions leading to AdS dilatonic black holes from the decoupling limit of the M$_2$, D$_3$ and M$_5$ black brane solutions. Finally, in Appendix \ref{section:RotatingBranes},  we give the expression for the rotating black $(p-1)$-brane and its decoupling limit in generic dimension $p+q+1$.

\section{The near-extremal scaling solutions of EMD theories\label{extremal}}

We start with the action
\be
S=M^{p-1}\int d^{p+1}x\sqrt{-g}\left[R-{1\over 2}(\partial\phi)^2+V(\phi)-{Z(\phi)\over 4}F_{\m\n}F^{\m\n}\right]
\label{action}\ee
where
\be
V=2\Lambda e^{-\delta\phi}\sp Z=e^{\gamma\phi}.
\label{2b}\ee
Early studies of the black-hole solutions of such actions may be found in \cite{Chan:1995fr,Cai:1996eg,Cai:1997ii}, while they were later revisited systematically in \cite{Charmousis:2009xr}.
In the domain-wall coordinate system,
\be
	\ud s^2=e^{2A}\left[-f(r)\ud t^2+\ud R^2_{(p-1)}\right]+\frac{\ud r^2}{f(r)}\,,
\ee
the equations of motion derived from \eqref{action} read
\be
(p-1)A'' + {\frac{1}{2}} \phi'^2=0\sp A_t'={Q\over Z e^{(p-2)A}}\,,
  \label{1a}\ee
\be
f'' + pA'f' - \frac{Q^2}{e^{2(p-1)A} Z(\phi)}=0\,,\\
  \label{1b}\ee
\be
(p-1)A'\left({f'\over f}+pA'\right) - {V(\phi)\over f} - {\frac{1}{2}} \phi'^2 + \frac{Q^2}{2e^{2(p-1)A}f Z(\phi)}=0\,.
  \label{1c}\ee

  \bigskip

   \subsection{The neutral solution\label{neutral}}

     \bigskip

  The $(p+1)$-dimensional neutral ($Q=0$) near-extremal solution is obtained using a scaling ansatz, \cite{creal},

  \be
e^A=r^{2\over (p-1)\d^2}\,,\qquad e^{\delta\phi}={\Lambda\d^4\over \left({2p\over p-1}-\d^2\right)}~r^{2}\,,\qquad f=1-\left({r_0\over r}\right)^{{2 p\over (p-1)\d^2}-1}\,.
\label{28}\ee

The metric can be rewritten by a charge of coordinates
\be
w=r^{1-{2\over (p-1)\d^2}}\sp t\to {t\over \Big |1-{2\over (p-1)\d^2}\Big|}\sp x^i\to {x^i\over \Big |1-{2\over (p-1)\d^2}\Big|}\,,
  \label{3b}\ee
  as conformal to an AdS-like black hole,
  \be
  ds^2=e^{2\chi}~{{dw^2\over f}-fdt^2+dx^idx^i\over w^2}
 \label{4b} \ee
with
\be
 e^{2\chi}=2{{2p\over p-1}-\d^2\over \left({2\over p-1}-\d^2\right)^2}{1\over V(\phi)}\sp f=1-\left({w\over w_0}\right)^{2p-(p-1)\d^2 \over 2-(p-1)\d^2}.
\label{5b}\ee
We define these to have \emph{generalized conformal} symmetry, \cite{ske,Kanitscheider:2009as}.

  \bigskip

\subsection{The charged solution\label{charge}}

 \bigskip

  The  $(p+1)$-dimensional charged near-extremal solution is obtained using a scaling ansatz, \cite{Charmousis:2009xr,cgkkm},
  \be
  e^{A} = {r}^{\frac{(\ga-\da)^2}{2(p-1)}}\sp
  e^\phi = e^{\phi_0}{r}^{(\da-\ga)}\sp f(r) =f_0 {r}^{v}h(r)\,,
  \label{2}\ee
  \be
   f_0=\frac{8(p-1)\Lambda~e^{-\da\phi_0}}{uw_p}\label{3}
 \sp
 h(r)= 1 - \le({r_0\over r}\ri)^{\frac{w_p}{2(p-1)}},
\ee
  \be
  A_t = \frac{4(p-1)}{w_p}\sqrt{\frac{\Lambda v}{u}}e^{-\frac{(\ga+\da)}2\phi_0}{r}^{\frac{w_p}{2(p-1)}}~h(r)\,,
  \label{a4}\ee
  \be
  u = \ga^2-\ga\da+2\sp
  v = -\da^2+\ga\da +2\sp u-v\geq 0\,,\label{5}
  \ee
  \be
    w_p = 2(p-1)+p\ga^2-2\ga\da-(p-2)\da^2=p u+(p-2)v-2(p-1)\,,
 \label{555}   \ee
  with
 \be
  F_{rt}={Q\over e^{2(p-2)A}Z}\sp Q=2e^{{(\g-\d)\over 2}\phi_0}\sqrt{v\Lambda\over u}\,.
 \label{7b} \ee
We have set the the IR scale $\ell=1$ in the above solution.

There are constraints on the parameters for the solution to exist and have AdS character
\be
u>0\sp v>0\sp w_p>0\,.
\label{6}\ee
The IR asymptotics correspond to $r\to 0$.
These capture the Gubser criterion for the associated IR naked singularity at extremality.
Moreover, when $C_p$, defined in (\ref{4}), is positive
\be
C_p=u+(2p-3)v-2(p-1)>0\,,
\ee
the system is gapless, \cite{cgkkm}.
We may use $u,v$ to parametrize the solutions instead of $\g,\d$:
\be
\g=\e{u-2\over \sqrt{u-v}}\sp \d=\e{v-2\over \sqrt{u-v}}\sp \e=\pm 1\,.
\ee

We now consider the extremal solution, $r_0=0$, and rewrite the metric as
\be
ds^2={r^{-\da(\ga-\da)}\over f_0}\left[-f_0^2r^{{(\ga-\d)(\ga+(p-2)\da)\over p-1}+v}dt^2+f_0r^{{(\ga-\d)(\ga+(p-2)\da)\over p-1}}dx^idx^i+{dr^2\over r^2}\right].
\label{7}\ee
Let us introduce a new radial coordinate $w$ by
\be
r=w^{\frac{-2(p-1)}{(\ga-\da)(\ga+(p-2)\da)}}\sp {dr\over r}=\frac{-2(p-1)}{(\ga-\da)(\ga+(p-2)\da)}{dw\over w}\,,
\label{8}\ee
and rescale
\be
t\to {2(p-1)t\over f_0\Bigl| (\ga-\da)(\ga+(p-2)\da)\Bigl|}\sp x^i\to {2(p-1)x^i\over \sqrt{f_0}\Bigl| (\ga-\da)(\ga+(p-2)\da)\Bigl|}\,,
\label{9}\ee
to obtain
\be
ds^2=e^{\chi} ~d\hat s^2\sp e^{\chi}={4(p-1)^2w^c\over f_0(\ga-\da)^2(\ga+(p-2)\da)^2}={(p-1)uw_p\over (\ga-\d)^2(\ga+(p-2)\d)^2}{1\over V(\phi)}
\label{11}\ee
\be d\hat s^2={-{dt^2\over w^{2z}}+{dw^2+dx^idx^i\over w^2}}
\label{10}\ee
\be
z={(\ga-\d)(\g+(2p-3)\d)+2(p-1)\over (\ga-\d)(\ga+(p-2)\d)}={u+(2p-3)v-2(p-1)\over u+(p-2)v-2(p-1)}
\label{12}\ee
\be
 c={2(p-1)\d\over \ga+(p-2)\d}={2(p-1)(v-2)\over u+(p-2)v-2(p-1)}\,.
\label{12a}\ee

Finally, the gauge field becomes
\be
A_t={-(p-1)\sqrt{uv\over \Lambda}e^{{(\d-\g)\over 2}\phi_0}\over (\ga-\da)(\ga+(p-2)\da)}r^{w_p\over 2(p-1)}=
{-(p-1)\sqrt{uv\over \Lambda}e^{{(\d-\g)\over 2}\phi_0}\over (\ga-\da)(\ga+(p-2)\da)}~w^{-{{w_p\over (\g-\d)(\g+(p-2)\d)}}}.\label{8b}\ee

Therefore, in the generic case, the metric is conformal to a Lifshitz metric with the conformal factor being the inverse of the scalar potential, \cite{Perlmutter:2010qu}. We define these to have \emph{generalized Lifshitz} symmetry.
The Gubser conditions become,

\be
{(p-1)\thema+2z+2(p-1)\over 2(z-1)-\thema}>0\sp {z-1\over 2(z-1)-\thema}>0\sp {2z-(p-1)\thema+2(p-2)\over 2(z-1)-\thema}>0
\ee
while thermodynamic stability implies that
\be
{z\over  2(z-1)-\thema}>0
\ee
and the blackness factor becomes
\be
h=1-\left({w\over w_0}\right)^{z-{p-1\over 2}\theta+p-1}
\ee

A few special cases are of interest.

\begin{itemize}

\item When $\d=0$, the potential is constant. The metric describes a  Lifshitz spacetime as $\chi=0$, \cite{taylor,Goldstein:2009cv,cgkkm,Goldstein:2010aw}.

\item For $\da\neq 0$, the metric becomes conformal to AdS$_2\times \mathbf R^{p-1}$ when $z\to \infty$.
This happens when:
\begin{itemize}
\item $\ga=\d$: In that case, the space is AdS$_2\times \mathbf R^{p-1}$ with a trivial conformal factor.

\item $\ga+(p-2)\d=0$: In this case the following change of variables in the original metric
\be
r=w^{1\over {(p-1)\d^2\over 2}-1}\sp t\to {t\over f_0\Bigl|{(p-1)\d^2\over 2}-1\Bigl|}\sp x^i\to {x^i\over \sqrt{f_0}\Bigl|{(p-1)\d^2\over 2}-1\Bigl|}
\label{14}\ee
brings it into the form
\be
ds^2=e^{\chi}d\hat s^2={w^{(p-1)\d^2\over {(p-1)\d^2\over 2}-1}\over f_0\Bigl|{(p-1)\d^2\over 2}-1\Bigl|^2}\left[{dw^2-dt^2\over w^2}+dx^idx^i\right]
\label{50}\ee
 with conformal factor
\be
e^{\chi}\sim e^{\d \phi}
\label{15}\ee
\end{itemize}
\item The metric would become conformal to AdS$_{p+1}$ when $z=1$. This happens when $v=-(\g-\d)\d-2=0$. In this case the density is zero and the solution is the uncharged solution which is indeed conformal to AdS$_{p+1}$.

\end{itemize}
Moreover, the near-extremal geometry  (\ref{2})-(\ref{a4}) describes correctly the near-horizon region of the extremal black holes of \cite{Gubser:2009qt, Cadoni:2009xm,cadoni}, even though their potentials comprise more than one exponentials.

\section{Dimensional reduction and emergent Schr\"odinger symmetry\label{sch}}

We will start from an Einstein plus cosmological constant action in $p+q+1$ dimensions,
\be
S=\int d^{p+1}x~d^qy\sqrt{\hat G}\left[\hat R+2\Lambda\right],
\label{21}\ee
and we will make the standard toroidal  KK ansatz, \cite{book},
\be
d\hat s^2=G_{\m\n}dx^{\m}dx^{\n}+G_{\a\b}(dy^{\a}+A^{\a}_{\m}dx^{\mu})(dy^{\b}+A^{\b}_{\n}dx^{\nu}),
\label{22}\ee
\be
S=\int d^{p+1}x\sqrt{G}~e^{2\Phi}\left[R+4(\pa \Phi)^2+{1\over 4}\pa_{\m}G_{\a\b}\pa^{\m}G^{\a\b}-{1\over 4}G_{\a\b}F^{\a}_{\m\n}F^{\b,\m\n}+2\Lambda\right],
\label{23}\ee
with
\be
\Phi={1\over 4}\log\det(G_{\a\b})\sp F^{\a}_{\m\n}=\pa_{\m}A^{\a}_{\n}-\pa_{\n}A^{\a}_{\m}.
\label{24}\ee
We will now perform a series of field redefinitions to bring the $(p+1)$-dimensional action to standard form: first go to the Einstein frame
using
\be
G_{\m\n}=e^{-{4\Phi\over p-1}}g_{\m\n}\sp R(G)=e^{4\Phi\over p-1}\left[R(g)-{4p\over p-1}(\pa\Phi)^2+{\cal O}(\square \Phi)\right],
\label{25}\ee
then define a unimodular scalar matrix by
\be
G_{\a\b}=e^{{4\over q}\Phi}\tilde G_{\a\b}\sp \det \tilde G_{\a\b}=1,
\label{a11}\ee
and finally normalize the internal volume scalar as
\be
\phi=2\sqrt{2}\sqrt{{1\over p-1}+{1\over q}}~\Phi
\label{51}\ee
to obtain
\be
S_{E}=\int d^{p+1}x\sqrt{g}~\left[R-{1\over 2}(\pa \phi)^2+{1\over 4}\pa_{\m}
\tilde G_{\a\b}\pa^{\m}\tilde G^{\a\b}-{e^{\gamma\phi}\over 4}\tilde G_{\a\b}F^{\a}_{\m\n}F^{\b,\m\n}+2\Lambda ~e^{-\d\phi}\right]
\label{27}\ee
with
\be
\d^2={2\over {p-1}}~{q\over p+q-1}\leq {2\over p-1}\sp \gamma={2\over (p-1)\delta}.
\label{29}\ee

The equations of motion stemming from this Lagrangian are
\bea
0&=&R_{\m\n}-{1\over 2}(R+2\Lambda e^{-\d\phi})g_{\m\n}
+{1\over 8}\left[\pa_{\m}
\tilde G_{\a\b}\pa^{\n}\tilde G^{\a\b}+\pa_{\n}
\tilde G_{\a\b}\pa^{\m}\tilde G^{\a\b}-g_{\m\n}\pa_{\m}
\tilde G_{\a\b}\pa^{\m}\tilde G^{\a\b}\right]
-
\label{52}\nn\\
&&\qquad\qquad-{1\over 2}\left[\pa_{\m}\phi\pa_{\nu}\phi-{1\over 2}(\pa \phi)^2g_{\m\n}\right]
-{e^{\g\phi}\over 2}\tilde G_{\a\b}\left[F^{\a}_{\m\rr}F^{\b,\rho}_{\n}-{1\over 4}g_{\m\n}F^{\a}_{\rr\s}F^{\b,\rr\s}\right]
,\\
0&=&\square \phi-{\g\over 4}e^{\g\phi}\tilde G_{\a\b}F^{\a}_{\m\n}F^{\b,\m\n}-2\d \Lambda ~e^{-\d\phi},
\label{53}\\
0&=&{1\over \sqrt{g}}\tilde G^{\a\g}\pa_{\m}\left(\sqrt{g}g^{\m\n}\pa_{\n}G_{\g\d}G^{\d\b}\right)-{e^{\g\phi}\over 2}F^{\a}_{\m\n}F^{\b,\m\n},
\label{54}\\
0&=&{1\over \sqrt{g}}\pa_{\nu}\left(\sqrt{g}g^{\m\rr}g^{\n\s}\tilde G_{\a\b}F^{\a}_{\m\rr}\right).
\label{55}\eea

They are of the general form (\ref{1a}-\ref{1c}), with the addition of a $q\times q$ unimodular matrix of scalars.

\subsection{Neutral solutions\label{section:NeutralAds}}

When $F_{\m\n}^{\a}=0$, from (\ref{54}) we can set $\tilde G$ to a constant and we may therefore forget about $\tilde G$ and $F$ as far as solutions are concerned. The scalar-tensor theory has an exponential potential and the near-extremal solution in (\ref{28}) lifts to the higher-dimensional metric of the form
\be
G_{\m\n}=e^{-\d\phi}~g_{\m\n}\sp G_{\a\beta}=e^{\frac{4\Phi}q}\delta_{\al\ba}=e^{\sqrt{2(p-1)\over q(p+q-1)}\phi}\delta_{\al\ba}.
\label{a12}\ee
Starting from the explicit form of the solution (\ref{5b}) we may compute the higher-dimensional metric $G_{\m\n},G_{\a\beta}$. Not surprisingly, this metric is that of an AdS$_{p+q+1}$ -Schwarzschild  black-hole, \cite{Cai:2004iy,gpr,ske,Kanitscheider:2009as}:
\be
ds^2=G_{\m\n}dx^{\m}dx^{\n}+G_{\a\beta}dy^{\a}dy^{\b}\sim {-fdt^2+{du^2\over f}+dx^idx_i+dy^{\a}dy_{\a}\over u^2}
\label{a13}\ee
with
\be
 f=1-\left({w\over w_0}\right)^{p+q}
\label{a14}\ee
where we used (\ref{29}).

Therefore, this uplift to higher dimensions explains the scaling of the original fields: it is secretly conformally invariant. This conformal invariance is hidden if the ``internal" dimensions disappear.
Moreover, the higher-dimensional geometry is regular, unlike the lower-dimensional geometry that has a naked singularity.

In \cite{gkn}, it was shown that there are two special values of $\d$, the crossover value $\delta_c^2={2\over p-1}$ and the Gubser bound limit $\d_G^2={2p\over p-1}$.
Theories with $0<|\d|<\delta_c$ have a continuous spectrum and no gap.
Theories with $|\d|\geq \d_G$ violate the Gubser bound and are therefore unacceptable. Theories with $\delta_c\leq |\d|\leq \d_G$ have a discrete spectrum and a mass gap.

From (\ref{29}) we observe that the values we obtain for $\delta$ for ``physical" values of the extra dimensions $p+q-1>0$, $q>0$, are below the crossover value $\delta_c$.
The properties of the spectra in this regime  become clear now: the higher-dimensional AdS space indeed has a continuous spectrum and no mass gap.

The values of $\delta>\delta_c$ can be obtained by a different uplift of the same solution described in detail in section \ref{KKsm}(see also \cite{skg}). It involves a compactification on a $q={2(p-1)\over (p-1)\d^2-2}$-dimensional sphere whose radius corresponds to the scalar $\phi$ as in (\ref{a8}). The higher-dimensional solution is regular,
and the Gubser-bound $\delta<\delta_G$ corresponds to the statement that the internal sphere should have dimension bigger than unity.
Indeed when $q\to1^+$, the curvature of the sphere disappears.

\subsection{Finite density solutions \label{section:NearExtrBoostedAds}}

It is clear from (\ref{54}) that in the case $q>1$, the matrix $\tilde G$ is dynamical and non-trivial, and does not allow a consistent truncation to the case we are interested in.
Therefore below we will assume that we have a single extra dimension, $q=1$ (though see \cite{skg} for a 'generalized' version).

In terms of the $(p+1)$-dimensional fields, $A^{\a}_{\m},g_{\m\n}$, in (\ref{22}),  the $(p+2)$-dimensional metric $G_{MN}$ is
\be
G_{\m\n}=e^{-\d\phi}\left[g_{\m\n}+e^{\g\phi}A^{\a}_{\m}A^{\a}_{\n}\right].
\label{31}\ee

Substituting from (\ref{7}), we obtain
\bea
ds^2&=&G_{\m\n}dx^{\m}dx^{\n}\\
&=&{e^{-\d\phi_0}\over f_0}\left[-f_0^2\left(1-{2(p-1)v\over w_p}\right)r^{{(\ga-\d)^2\over p-1}+2v-2}dt^2+
f_0r^{{(\ga-\d)^2\over p-1}+v-2}dx^2+{dr^2\over r^2}\right]
\label{32}\\
&=&{e^{-\d\phi_0}\over f_0}\left[-f_0^2\left(1-{2(p-1)v\over w_p}\right)
{dt^2\over w^{2z}}+
f_0{dx^2\over w^2}+{dw^2\over \left(1-{v\over 2}-{(\ga-\d)^2\over 2(p-1)}\right)^2 ~w^2}\right]\\
&=&e^{-\da\phi_0}\frac{\ud x^i\ud x_i}{w^2}+\frac{p(p+1)\ud w^2}{2\La w^2}\,,
\eea
where we changed coordinates to the Lifshitz coordinate (\ref{8}) and in the last line replaced $\ga$ and $\da$ by their values \eqref{29}.
This is a Lifshitz metric with $z$ given in (\ref{12}).
We also have
\bea
G_{\m\a}&=&e^{(\g-\d)\phi}A^{\a}_{\m}=
\frac{4(p-1)}{w_p}\sqrt{\frac{vu}{\Lambda}}e^{\frac{(\ga-\da)}2\phi_0}
  {r}^{\frac{w_pu}{2(p-1)}-(\g-\d)^2}~\delta_{\mu t}
\label{33}\\
&=&\frac{4(p-1)}{w_p}\sqrt{\frac{uv}{\Lambda }}e^{\frac{(\ga-\da)}2\phi_0}~{\delta_{\mu t}\over w^{z+\hat z}}\\
&=&-\sqrt{\frac{p(p+1)}{2\La}}e^{\frac{(\ga-\da)}2\phi_0}w^{-2}~\delta_{\mu t}\,,\\
 G_{\a\b}&=&e^{(\g-\d)\phi}~\delta_{\a\b}=e^{(\g-\d)\phi_0}~r^{-(\g-\d)^2}~\delta_{\a\b}=e^{(\g-\d)\phi_0}~{\delta_{\a\b}\over w^{2\hat z}}\
\label{35}\\
&=&e^{(\g-\d)\phi_0}w^{p-1}~\delta_{\a\b}\,,\\
\hat z&=&{(p-1)(\g-\d)^2\over (p-1)(v+2)-(\g-\d)^2}\,.
\label{56}\eea
Using (\ref{28}), (\ref{29}), we obtain
\be
z+\hat z=p+q+2{1-p\over q+1}\,.
\label{57}\ee
For $q=1$, $z+\hat z=2$.
{\it Note the coefficient of the $\ud t^2$ in the metric (\ref{32}) is zero when $q=1$.} The metric obtained is indeed the infinite boost limit of the boosted AdS black brane, which has zero $tt$ element but is regular everywhere. This is the so called Kaigorodov metric, which is an Einstein space solution of Einstein plus cosmological constant theory, \cite{Cvetic:1998jf}.

The metric is invariant under the generalized Lifshitz scaling
\be
x^i\to \lambda x^i\sp t\to \la^z t\sp w\to \la w, \sp y^{\a}\to \la^{\hat z}y^{\a}\,.
\label{58}\ee

This higher-dimensional metric will become of Schr\"odinger-like form  if $G_{\a\b}\to 0$.
This happens if $e^{(\g-\d)\phi}\sim r^{-(\g-\d)^2}$, namely near $r\to \infty$.

The non-trivial dilatation operator of a Schr\"odinger symmetry is
\be
D_{\bar z}=\eta^{\m}\partial_{\m}=\bar z~t\pa_t+r\pa_r+x^i\pa_i+(2-\bar z)y^{\a}\pa_{\a}\,.
\label{34}\ee
It acts on the fields as
\be
\delta_{\eta} G_{MN}=G_{MP}\pa_{N}\eta^{P}+G_{NP}\pa_{M}\eta^{P}+\eta^{P}\pa_{P}G_{MN}\,.
\label{36}\ee

The metric $G_{\m\n}$ satisfies (\ref{36}) with $\bar z=z$.
 On the other hand we have
 \be
 \delta_{\eta}G_{\m\a}=(2-\bar z)G_{\m\a}+G_{\rr\a}\pa_{\m}\eta^{\rr}+\eta^{\rr}\pa_{\rr}G_{\m\a}\,,
\label{37} \ee
so that
\be
\delta_{\eta}G_{t\a}=(2+r\pa_r)G_{t\a}=(2+{w_p u\over 2(p-1)}-(\g-\d)^2)G_{t\a}\,.
\label{38}\ee
Finally
\be
\delta_{\eta}G_{\a\b}=\left(2(2-\bar z)+r\pa_r\right)G_{\a\b}=\left(2(2-\bar z)-(\g-\d)^2\right)G_{\a\b}\,.
\label{39}\ee
Both vanish at $r\to\infty$.
This suggests that at $r\to\infty$ these solutions realize $z$-Schr\"odinger symmetry.
The case $p=3$ has been analyzed in \cite{mb}.

\section{Static, asymptotically-flat black branes and their near-horizon limits\label{section:StaticFlatBranes}}

In this section, we will study the static black brane solutions of the action
\be  \label{EFormAction}
	S_{p+q+1}=\frac{1}{16\pi G_D}\int\ud^{p+q+1}x\,\sqrt{-g}\l[R-\frac{1}{2(n+2)!}G^2_{[n+2]}\r].
\ee
The form $G_{[n+2]}=dB_{[n+1]}$ is the field strength of a massless $(n+1)$-form.

Our starting point will be the solution supported by $G_{[n+2]}$ with $n\leq p-1$ field strength, $p-1$ being the number of space dimensions of the brane worldvolume. Such solutions are said to have a smeared charge, \cite{Caldarelli:2010xz}. Performing various diagonal dimensional reductions will allow to make connections with the charged dilatonic theories \eqref{action}.

\subsection{The static, asymptotically flat black brane and its near-horizon limit\label{md}}

The action \eqref{EFormAction} has the following black $(p-1)$-brane solution with a smeared charge, \cite{Caldarelli:2010xz},
\be
	\ud s_{(p+q+1)}^2 = h(r)^{\frac{-2}{n+1}}\l[-f(r)\ud t^2+\ud R^2_{(n)}\r]+h(r)^{\frac{2}{(p+q-n-2)}}\l[\frac{\ud r^2}{f(r)}+r^2\ud K^2_{(q)}+\ud R^2_{(p-n-1)}\r]
	\label{BlackqBraneNCharge},
\ee
\be
	f(r)=k_{(q)}-\l(\frac{r_0}{r}\r)^{q-1},\qquad h(r)=1+\frac{\sinh^2\al}{k_{(q)}}\l(\frac{r_0}{r}\r)^{q-1},\label{a15}
\ee
\be	 B_{[n+1]}=-\sqrt{\frac{2(p+q-1)k_{(q)}}{(p+q-n-2)(n+1)}}\l(1-h(r)^{-1}\r)\coth\al\,\ud t\wedge\ud R_{(n)}\,.\label{PhiB1}
\ee
Our conventions are spelled out in appendix \ref{conv}.
$\al$ is the extremality parameter, while $r_0$ controls the temperature.
The spaces $\mathbf R_{(d)}$ are toroidal of the dimension $d$.
This is an asymptotically black $(p-1)$-brane with a flat worldvolume and the electric charge is smeared over $n$ of the $(p-1)$ worldvolume dimensions. For generic $n$, the rotational invariance of the brane worldvolume is broken by the non-uniform electric charge. It can be restored by setting $n=0$, whereupon one finds back a Maxwell $U(1)$ electric field and a point-charge, while $n=p-1$ corresponds to a black brane where the charge is spread out uniformly on the worldvolume.

Note that in the latter case $n=p-1$, we recover well-known solutions:
\begin{itemize}
 \item $q=5$, $n=p-1=3$: this is the non-spinning D$_3$-brane, which is a solution of $D=10$ supergravity from type IIB string theory, see appendix \ref{SUGRAAdS5S5}.
 \item $q=7$, $n=p-1=2$: this is the non-spinning M$_2$-brane, which is a solution of $D=11$ supergravity from $M$-theory, see appendix \ref{SUGRAAdS4S7}.
  \item $q=4$, $n=p-1=5$: this is the non-spinning M$_5$-brane, which is a solution of $D=11$ supergravity from $M$-theory, see appendix \ref{SUGRAAdS7S4}.
\end{itemize}
It is also a well-known fact that, in the near-horizon, large-charge limit, these solutions asymptote to AdS$_{p+1}\times\mathbf K^{q}$ solutions, and that once the constant curvature space $\mathbf K^{q}$ is reduced, they produce AdS black holes.

 We now take the near-extremal limit of the previous solution \eqref{BlackqBraneNCharge}, $\al\to\infty$, to obtain:
\bea
	\ud s^2_{(p+q+1)} &=&\xi^2\left[-f(\xi)\ud t^2+ \ud R^2_{(n)}\right]+\nn\\
	 &&\qquad\quad+\frac{\xi^{\frac{2(n+1)(p-n-1)}{(q-1)(p+q-n-2)}}}{g^2}\l[\frac{\ud\xi^2}{\xi^2 f(\xi)} + \ud K^2_{(q)}\r]+\frac{\xi^{\frac{-2(n+1)}{p+q-n-2}}}{g^{2}} \ud R^2_{(p-n-1)},\label{AdSQbraneNchargeDecoupling}\\
           f(\xi)&=&\frac{(q-1)^2k_{(q)}}{(n+1)^2}-\l(\frac{\xi_0}{\xi}\r)^{n+1}\,,\label{a16}\\
           B_{[n+1]}&=&\frac{(q-1)}{(n+1)}\sqrt{\frac{2(p+q-1)k_{(q)}}{(n+1)(p+q-n-2)}}\xi^{n+1}\ud t\wedge\ud R_{(n)}\,,\label{a17}
\eea
where we have introduced the inverse radius of the compact space $\mathbf K^{q}$
\be
	g = \left(r_0^{q-1}\frac{\sinh^{2}\al}{k_{(q)}}\right)^{{1\over (n+2-p-q)}},
\label{a141}\ee
and changed coordinates to $r^{q-1}=\xi^{n+1}$. It is obvious that for $p-1=n$, the geometry reduces to AdS$_{p+1}\times \mathbf K^{q}$.
 Notice also that in the case of a Maxwell field $n=p-1=0$, the metric  reduces to the familiar AdS$_{2}\times \mathbf K^{q}$.

\subsection{The Kaluza-Klein reduction of black branes with a smeared charge\label{KKsm}}

Since our aim is to recover solutions of lower-dimensional theories with action \eqref{action}, having in particular a non-trivial  potential, we shall dimensionally reduce along  $\mathbf K^{q}$, thereby generating the potential from the non-zero curvature of the compactified space  $\mathbf K^{q}$. This allows the presence of an $(n+2)$-field strength in the lower-dimensional theory. We postulate a diagonal ansatz for the metric
\be
	\ud s^2_{(p+q+1)} = e^{-\frac{2\phi}{(p-1)\da}}\ud s^2_{(p+1)}+ e^{\frac{\phi}{\da}\l(\da^2-\frac2{p-1}\r)}\ud K^2_{(q)}\,,
\label{a8}\ee
with
\be
 \da^2=\frac2{p-1}+\frac2q\geq \frac2{p-1}\sp \g={\frac{2(n+1)}{(p-1)\da}}\,.
\label{a18}\ee
The reduced action is
\be
	S_{(p+1)}=\frac{1}{16\pi G_{(p+1)}}\int\ud^{p+1}\sqrt{-g}\l[R-\frac12\l(\partial\phi\r)^2-\frac{1}{2(n+2)!}e^{\g\phi}H^2_{[n+2]}+2\La e^{-\da\phi}\r],
\label{EMDnForm}
\ee
where $G_{(p+1)}=G_D/V(\mathbf K_{(q)})$ with $V(\mathbf K_{(q)})$ being the volume of $\mathbf K_{(q)}$.
We can identify the scalar as
\be
	 e^{\phi}=\rho^{(p-1)\da}h(\rho)^{\frac{2(p-1)\da}{2(n+1)+(p-1)(q-n-2)\da^2}},\qquad h(\rho)=1+\frac{\sinh^2\al}{k_{(q)}}\l(\frac{\rho_0}{\rho}\r)^{p-\frac{(p-1)}2\da^2}.
\label{a19}\ee
The metric is now
\bea
	\ud s^2_{(p+1)} &=& \rho^{2}h(\rho)^{\frac{-2(p-1)(p-n-2)\da^2}{(n+1)[2p+(p-1)(p-n-2)\da^2]}}\l[-f(\rho)\ud t^2+\ud R^2_{(n)}\r]+\nn\\
	&&+ \rho^{2}h(\rho)^{\frac{2(p-1)\da^2}{[2(n+1)+(p-1)(p-n-2)\da^2]}}\l[\frac{((p-1)\da^2-2)^2\ud \rho^2}{4\rho^{4-(p-1)\da^2}f(\rho)}+\ud R^2_{(p-n-1)}\r] \label{AnistropicEFormD}
\eea
while
\be
	f(\rho)=k_{(q)}-\l(\frac{\rho_0}{\rho}\r)^{p-\frac{(p-1)}2\da^2},
\label{a20}\ee
	\be B_{[n+1]}=-\sqrt{\frac{2(p-1)^2\da^2}{(n+1)[2(n+1)+(p-1)(p-n-2)\da^2]}}\l(1-h(\rho)^{-1}\r)\coth\al~\ud t\wedge\ud R_{(n)}\,.
\label{a21}\ee
Finally, we have to set the Ricci scalar

\be	 R_{(q)}=q(q-1)k_{(q)}=2(p-1)\frac{2p-(p-1)\da^2}{((p-1)\da^2-2)^2}k_{(q)}=2\La\geq 0
\label{a22}\ee
to ensure that the higher-dimensional Einstein's equations are verified. Thus, we have obtained a dilatonic black $(p-1)$-brane with an $n$-brane charge, which is a generic solution of Einstein-Maxwell-Dilaton theories for $\ga\da=2(n+1)/(p-1)$. At the level of the lower-dimensional theory, the solution above has an anisotropic horizon due to the electric charge, whenever $n\neq0$ and $n\neq p-1$. There are two independent parameters, $\rho_0$ and $\al$, which are related to the temperature and the electric charge.

Let us now examine the two cases that lead to translationally invariant solutions in lower dimensions.

\subsubsection{\large $n=0$\label{n0}}

For $n=0$, the higher-dimensional theory is Einstein-Maxwell gravity, and the solution a RN black hole with horizon topology $\mathbf T^{p-1}\times \mathbf S^{q}$.

After a change of radial coordinates, the solution \eqref{AnistropicEFormD} simply reduces to the $\ga\da=2/(p-1)$ charged dilatonic black brane, which is a solution of action (\ref{action}), first discovered for $p=3$ by \cite{Charmousis:2009xr} and further studied in \cite{cgkkm}.

Moreover, the higher-dimensional theory can be, depending on the dimensional uplift scheme:
\begin{itemize}
	\item Either Einstein with a cosmological constant, and this solution uplifts to a boosted AdS black brane\footnote{The near-extremal, infinite boost limit is presented in section \ref{section:NearExtrBoostedAds} of this paper.} in $p+2$ dimensions \emph{via} a Kaluza-Klein uplift with the lower-dimensional U(1) charge becoming a Kaluza-Klein vector, \cite{skg};
	\item Or Einstein-Maxwell without a cosmological constant, and the solution is a RN black hole with horizon topology $\mathbf T^{p-1}\times \mathbf S^{q}$, \cite{gibbons}. Should one then do a toroidal Kaluza-Klein reduction along the brane directions $\mathbf R_{(p-1)}$, \eqref{BlackqBraneNCharge}, the $2$-form field strength is untouched and no potential is generated, and one recovers the asymptotically flat Einstein-Maxwell-Dilaton black holes of \cite{Gibbons:1987ps}.
\end{itemize}

 We may also take the near-horizon limit $\al\to+\infty$ in \eqref{AnistropicEFormD} to find the lower-dimensional equivalent of \eqref{AdSQbraneNchargeDecoupling}.
 This is the solution of (\ref{action}) with $\g\d={2\over p-1}$, which becomes an extremal RN solution with an anisotropic horizon of topology $\mathbf T^{p-1}\times \mathbf S^{q}$. Here the lower-dimensional dilaton potential in (\ref{action}) is generated by the positive curvature of the internal $\mathbf S^{q}$.

 From (\ref{a18}) and the analysis in \cite{cgkkm}, we find that these solutions are in the regime $\delta>\delta_c$. In this regime, the spin-two fluctuations are gapped while the spin-one fluctuations may or may not be gapped.

\subsubsection{\large $n=p-1$\label{nq}}

Setting $n=p-1$, the expressions simplify considerably, and the $(p+1)$-form becomes dual to a $0$-form, inducing another Liouville potential. More explicitly, one may rewrite the action \eqref{EMDnForm} as
\be
	S_{(p+1)}=\frac{1}{16\pi G_{(p+1)}}\int\ud^{p+1}\sqrt{-g}\l[R-\frac12\l(\partial\phi\r)^2+2\tilde\La e^{-\g\phi}+2\La e^{-\da\phi}\r],
\label{EMDnForm2}
\ee
using the equation of motion of the form field (or equivalently replacing it by its Hodge dual). The solution \eqref{AnistropicEFormD} becomes
\be
	\ud s^2_{(p+1)} = \rho^{2}h(\rho)^{\frac{p^{-1}(p-1)\da^2}{p-\frac{(p-1)}2\da^2}}\l[-f(\rho)\ud t^2+\ud R^2_{(p-1)}\r]+ \rho^{(p-1)\da^2-2}h(\rho)^{\frac{(p-1)\da^2}{p-\frac{(p-1)}2\da^2}}\frac{\ud \rho^2}{f(\rho)}\,, \label{EDToroidal2Potentials}
\ee
\be
f(\rho)=\frac{k_{(q)}}{(\frac{(p-1)}2\da^2-1)^2}-\l(\frac{\rho_0}{\rho}\r)^{p-\frac{p-1}2\da^2},
\ee
\be
 h(\rho)=1+\frac{\sinh^2\al(\frac{(p-1)}2\da^2-1)^2}{k_{(q)}}\l(\frac{\rho_0}{\rho}\r)^{p-\frac{p-1}2\da^2},\label{a5}
\ee
\be
	 e^{\phi}=\rho^{(p-1)\da}h(\rho)^{\frac{2(p-1)\da}{2p-(p-1)\da^2}},
\ee
\be
 2\tilde\La=\frac{(p-1)^2\da^2[2p-(p-1)\da^2]}{2((p-1)\da^2-2)^2k_{(q)}}\cosh^2\al\sinh^2\al~\rho_0^{p-\frac{p-1}2\da^2},\label{a6}
\ee
\be
	2\La=\frac{2(p+1)[2p-(p-1)\da^2]}{((p-1)\da^2-2)^2}k_{(q)}\sp k_{(q)}\Lambda>0\sp k_{(q)}\tilde\Lambda>0\label{a7}.
\ee

This is a neutral solution from the lower-dimensional point of view. The Ricci scalar is well-behaved at infinity and displays a curvature singularity at $\rho=0$ if $\da<\da_G$, while the spacetime is a static black hole for $\La>0$ on top of the previous condition.
It has only one integration constant, namely $\rho_0$, related to the temperature. The Liouville potentials have the same overall sign, as well as different exponents with equal sign: although there are two exponentials, the potential is always a monotonic function of the scalar field and there is no UV AdS fixed point.

 This solution is distinct from the two-exponential, neutral solution first reported in \cite{Chan:1995fr}, whose single exponential, planar version \eqref{28} is shown in section \ref{section:NeutralAds} to uplift to the planar Schwarzschild-AdS black hole for $\da<\da_c$.

However, the solution does asymptote to the neutral, singular background \eqref{28} which is a solution of \eqref{action} with a single Liouville potential, see \cite{cgkkm}. The solution  (\ref{EDToroidal2Potentials})-(\ref{a7}) for $\al=0$ upon a change of the radial coordinate is exactly the neutral solution in  (\ref{28}) but now with $\delta>\delta_c$. This is the range in which the spectrum is gapped and discrete. This is now explained as this solution uplifts to (\ref{EDToroidal2Potentials}-\ref{a7}) with an internal sphere that is responsible for the discrete and gapped spectrum.
 Pursuing the generalisation of \eqref{EDToroidal2Potentials} to a non-planar horizon or adding a Maxwell term such as \eqref{LiouvilleGaugeCoupling} would certainly be interesting, but is not straightforward.

Upon taking the decoupling, near-horizon limit $\al\to\infty$, the solution \eqref{EDToroidal2Potentials} develops an emerging IR AdS$_{p+1}$ geometry, and therefore conformal symmetry is recovered.  Moreover, the scalar field flows to a constant. This is not surprising, given that the higher-dimensional IR geometry is a direct product AdS$_{p+1}\times\mathbf K^{q}$. Neither is it in contradiction with the paragraph above, since now the scalar field is constant everywhere, which is always a trivial classical solution of the action \eqref{EMDnForm2}.

\section{Dimensional uplift to near-extremal charged AdS black branes\label{qb}}

In the previous section, we have studied how solutions to  EMD actions of the kind \eqref{action} might emerge from the KK reduction of solutions to higher-dimensional theories containing Einstein gravity plus electric forms. In such reductions the scalar is associated with the size of the internal dimensions.

It is well-known that another way to generate a non-trivial exponential potential is to include a cosmological constant in the higher-dimensional theory. Combining both effects yields two-exponential potentials, as we show below. In this case the general black $(p-1)$-brane solutions are not known analytically, and one can only find their near-extremal limit, which are Bertotti-Robinson geometries and have been widely used in holography and AdS/CFT. Fortunately, this coincides exactly with our interest in the IR properties of \eqref{action}.

In a second step, we also consider the addition of a Maxwell field in the higher-dimensional theory, which yields the same kind of Bertotti-Robinson geometries as previously, but this time the AdS part of the metric is charged under the $U(1)$.

\subsection{$\ga=-(p-2)\da$: Near-extremal AdS black brane with $q+1$-charge\label{qb1}}

We consider the theory
\be
	S_{(p+q+1)}=\frac{1}{16\pi G_D}\int\ud^{p+q+1}x\,\sqrt{-g}\l[R-\frac{1}{2(q+2)!}G^2_{[q+2]}+2\La\r],
	\label{EinsteinLambdaQform}
\ee
where $G_{[q+2]}=\ud B_{[q+1]}$ is the field strength associated to some $(q+1)$-potential and the $q$-dimensional internal space $\ud K_{(q)}$ is in general curved. We shall take as diagonal reduction ansatz
\be
	\ud s^2_{(p+q+1)} = e^{-\da\phi}\ud s^2_{(p+1)}+ e^{\frac{\phi}{\da}\l(\frac2{p-1}-\da^2\r)}\ud K^2_{(q)}\sp \label{KKAnsatzAdS}
	B_{[q+1]}=A_{[1]}\wedge\ud K_{(q)}\,,
\ee
\be
	\da^2={2\over p-1}~\frac{q}{(p+q-1)}\leq \da_c^2={2\over p-1}\,,
\label{a23}\ee
so that the reduction is carried along $q$ of the legs of the $(q+1)$-form and we are reducing along (warped) worldvolume directions. We find that
\bea
	e^{-\da\phi}R_{(p+q+1)}&=&R_{(p+1)}+\da\square\phi - \half\partial\phi^2+e^{\frac{-2\phi}{(p-1)\da}}R_{(q)}\,,\label{a24}\\
	G_{[q+2]}^2&=&\frac{(q+2)!}2e^{-(p-3)\da\phi}F^2_{[2]}\,,
\label{a25}\eea
where $R_{(d)}$ above stands for the Ricci scalar in the appropriate space.
Discarding a boundary term, the lower-dimensional theory is of the kind \eqref{action} but with a potential containing two exponentials:

\be
	S_{(p+1)}=\int{\ud^{p+1}x\,\sqrt{-g}\over 16\pi G_{(p+1)}}\l[R-\frac12\partial\phi^2-\frac{1}4e^{\ga\phi}F^2_{[2]}+2\La e^{-\da\phi}+2\tilde\La e^{-\frac{2\phi}{(p-1)\da}}\r],\label{EMDActionKK1}
\ee
\be
	\da^2=\frac{2q}{(p-1)(p+q-1)}\,,\quad \ga=-(p-2)\da\,, \quad 2\tilde\La=R_{(q)}=q(q-1)k_{(q)}\,,
\label{a26}\ee
where  $G_{(p+1)}=G_D/Vol(\mathbf K_{(q)})$. Note that one of the exponentials is directly related to the higher-dimensional cosmological constant, while the other is related to the curvature of the compactified space.

The equations of motion derived from \eqref{EMDActionKK1} admit a black hole solution, \cite{Chan:1995fr}, that we reproduce here and generalize to arbitrary horizon topology
\be
	\ud s^2_{(p+1)} =e^{2A}\left[-f(r)\ud t^2+\ud K^2_{(p-1)}\right] +\frac{\ud r^2}{f(r)}\,,\label{EMDNEBH1}\ee
\be f(r)=\frac{\l(2\La+(p-2)^2k_{(p-1)}\r)}{\l(1+(p-1)(p-2)\frac{\da^2}2\r)^2}r^{2-(p-1)\da^2}+\frac{-8\tilde\La(p-1)^{-2}\da^{-2}}{\l(2-p(p-1)\da^2\r)}-\l(\frac{r_0}{r}\r)^{\frac{p}2(p-1)\da^2-1},
\label{a27}
\ee
\be
	e^{2A}=e^{\da\phi}=r^{(p-1)\da^2}\,,
\label{a28}\ee
	\be A_{[1]}=\sqrt{\frac{(2-(p-1)\da^2)\La+(p-1)(p-2)k_{(p-1)}}{2+(p-1)(p-2)\da^2}}\frac{2r^{1+(p-1)(p-2)\frac{\da^2}2}}{1+\frac{(p-1)(p-2)}2\da^2}\ud t\,.
\label{a29}\ee
If we set $k_{(p-1)}=0$ and $\tilde\La=0$ and we recover the generic, single exponential, near-extremal black brane solution \eqref{2} studied in  \cite{Cai:1997ii,Charmousis:2009xr,cgkkm} for specific values $\ga=-(p-2)\da$.

On the other hand, setting $\La=0$ and taking the planar limit $k_{(p-1)}=0$ recovers the neutral solution \eqref{28} after sending $\da\to2/(p-1)\da$. Let us stress that this limit does not recover the charged near-extremal solutions \eqref{2}-\eqref{a4}. In the following section \ref{rot}, we shall return to this statement, and connect it to the non-commutativity of planar and near-horizon limits in some instances.

 Using \eqref{KKAnsatzAdS}
and after changing coordinates with respect to the previous solution \eqref{EMDNEBH1}, the uplifted (aka oxidised) solution, which is a classical solution of the action \eqref{EinsteinLambdaQform}, can be written as
\be
	\ud s^2_{(p+q+1)} =-f(\xi)\ud t^2+\xi^{2} \ud K^2_{(q)}+\frac{\ud\xi^2}{f(\xi)} + \ud K^2_{(p-1)}\,,\label{AdSQbraneWarpedDecoupled}\ee
\be
 f(\xi)=\frac{\l(2\La+(p-2)^2k_{(p-1)}\r)}{(q+1)^2}\xi^2+k_{(q)}-\l(\frac{\xi_0}{\xi}\r)^{q-1},
 \label{a30}\ee
 \be
	B_{[q+1]}=\frac{Q}{(q+1)}\xi^{q+1}\ud t\wedge\ud K_{(q)}\,,
\label{a31}\ee
\be
	Q^2=\frac{4\La}{q+1}+2\frac{(p-2)(p+q-1)}{(q+1)}k_{(p-1)}\,, \label{EinsteinScalarCondition}
\ee
where $\ud K_{(q)}$ is the unit volume of element of the internal space $\mathbf K^q$. The IR geometry is therefore AdS$_{(q+2)}\times \mathbf K^{p-1}$, and has an emerging conformal symmetry.

In order to interpret this solution, let us consider a few special cases:
\begin{itemize}
	\item If $q=0$, the field strength has rank $2$, and so is the usual Maxwell term. What we recover here is the familiar AdS$_2\times \mathbf K^{p-1}$ extremal geometry.
	\item If $\La=0$, $k_{(p-1)}=0$, this is simply the neutral Schwarzschild black hole in $q+2$ dimensions times $p-1$ flat directions.  If $p=2$, this corresponds to the well-known black string. This is indeed consistent with the remark that taking this limit in \eqref{EMDNEBH1}-\eqref{a29} yields the neutral solution \eqref{28}.
	\item If $\tilde\La=0$, $k_{(p-1)}=0$, one recognizes an AdS spacetime times a flat space, with the cosmological constant supported by the electric chage.
	\item If $\La=\tilde\La=0$ (which implies $k_{(q)}=0$), we recognize the near-extremal limit of the generic black $q$-brane \eqref{AdSQbraneNchargeDecoupling} with uniform electric charge and unit radius for the space $\mathbf K^{p-1}$.
\end{itemize}

The general solution \eqref{AdSQbraneWarpedDecoupled} can then be interpreted as the near-extremal  limit of a black $q$-brane in $(p+q+1)$-dimensional Anti de Sitter space, a solution to the equations of motion derived from \eqref{EinsteinLambdaQform}. It is simply a direct product of the $(q+2)$-dimensional AdS black brane with a $(p-1)$-dimensional Einstein space where the internal curvature is supported by the charge (density).

The solution with $\tilde \Lambda=0$, corresponding to a toroidal internal space, gives an uplift of the the $\gamma+(p-2)\delta$ near-extremal solutions
of (\ref{action}). They are in the subcritical $\delta<\delta_c$ regime, because the internal space is flat.
The higher-dimensional solution is a near-extremal black $q$-brane solution
wrapped on $\mathbf T^q$.

The uplift allows us also to understand the fixed value of the charge density for the near-extremal black holes \eqref{2}: it simply reflects the scalar condition on Einstein's equations so that a direct product geometry is admissible. This fact has been known for quite some time and heavily used in supergravity contexts, \cite{Freund:1980xh}.

Moreover, for $k_{(p-1)}=0$, the system will be continuous for $\tilde\La=0$, that is when the solution descends from a planar AdS black $q$-brane via a toroidal reduction; while it will be gapped when $\Lambda=0$, since then it descends from a curved asymptotically flat $q$-brane via a curved reduction. One may also check from \eqref{6} that it always hold in the first case, while it holds in the second one if and only if $q>1$, that is for a compact space of non-zero curvature.

\subsection{$p\ga=\da$: Near-extremal AdS black brane with $p$- and $0$-charge\label{b}}

Consider the theory
\be \label{B1}
	S=M^{p+q-1}\int d^{p+q+1}x\sqrt{-g}\left[R-\frac14 \left(F_{[2]}^2\right)-\frac1{2(p+1)!}\left(G_{[p+1]}\right)^2 + 2\tilde\La\right],
\ee
suppose that both field strengths are only along the $p+1$ external coordinates and reduce along a $\mathbf T^q$
\be
	\ud s^2_{(p+q+1)} = e^{-\tilde\da\phi}\ud s^2_{(p+1)}+ e^{\frac{\phi}{\tilde\da}\l(\frac2{p-1}-\tilde\da^2\r)}\ud R^2_{(q)}\sp \label{B2}
\ee
\be
	\tilde\da^2={2\over p-1}~\frac{q}{(p+q-1)}\leq \tilde\da_c^2={2\over p-1}\,.
\label{B3}\ee
Then, the field strengths reduce trivially, and one obtains the lower-dimensional action
\be
	S=M^{p-1}\int d^{p+1}x\sqrt{-g}\left[R-{1\over 2}(\partial\phi)^2-\frac{e^{\tilde\da\phi}}4\left(F_{[2]}^2\right)-\frac{e^{p\tilde\da\phi}}{2(p+1)!} \left(G_{[p+1]}\right)^2 + 2\tilde\La e^{-\tilde\da\phi}\right].\label{B4}
\ee
Using its equation of motion, the $[p+1]$-field strength can be dualised to a $[0]$-field strength (that is, a scalar potential), by setting
\be \label{B5}
	-\frac{e^{p\tilde\da\phi}}{2(p+1)!} \left(G_{[p+1]}\right)^2=2\La e^{-p\tilde\phi}
\ee
so that \eqref{B4} becomes
\be
	S=M^{p-1}\int d^{p+1}x\sqrt{-g}\left[R-{1\over 2}(\partial\phi)^2-\frac{1}4e^{\tilde\da\phi}\left(F_{[2]}^2\right)+2\La e^{-p\tilde\da\phi} + 2\tilde\La e^{-\tilde\da\phi}\right].\label{B6}
\ee
Generalising slightly to
\be \label{B0}
	S=M^{p-1}\int d^{p+1}x\sqrt{-g}\left[R-{1\over 2}(\partial\phi)^2-\frac14 e^{\ga\phi}\left(F_{[2]}^2\right)+2\La e^{-\da\phi} + 2\tilde\La e^{-\tilde\da\phi}\right],
\ee
a black hole solution can be found: it appears in \cite{Chan:1995fr} for the positive horizon curvature case and we write it for arbitrary curvature.
\be \label{B00}
	(p-1)\tilde \da=\da-\ga\,,
\ee
\be
	\ud s^2_{p+1} =e^{2A}\l[\ud K^2_{(p-1)}-f(r)\ud t^2\r] +\frac{\ud r^2}{f(r)}\,,\label{EMDNEBH3}
\ee
\be
	f(r)=\frac{8\Lambda(p-1)}{w_p}r^{v}h(r)\,,
\ee
\be \label{B01}
	 h(r)=1-\left(\frac{r_0}{r}\right)^{\frac{w_p}{2(p-1)}}+\frac{(p-2)w_puk_{(p-1)}r^{1+\frac{(p-2)}{2(p-1)}(\ga-\da)^2}}{8\Lambda(p-1)\left[\frac{(\ga-\da)^2}{2(p-1)}-1\right]\left[1+\frac{(p-2)(\ga-\da)^2}{2(p-1)}\right]}\,
\ee
\be
	e^{2A}=e^{-\frac{(\ga-\da)}{(p-1)}\phi}=r^{\frac{(\ga-\da)^2}{(p-1)}}\,,
\ee
\be
	A_{[1]}=\frac{4(p-1)}{w_pu}\sqrt{\frac{\La v}{u}}r^{\frac{w_p}{2(p-1)}}\ud t\,,
\ee
\be
	\tilde\La=\frac{(p-1)^2(p-2)}{2(p-1)-(\ga-\da)^2}k_{(p-1)}\,
\ee
\be
  	w_p = 2(p-1)+p\ga^2-2\ga\da-(p-2)\da^2\,, \quad
  u = \ga^2-\ga\da+2\,, \quad
  v = -\da^2+\ga\da +2\,.\label{wdu2}
\ee
This is a generic curvature black hole, which reduces to the black brane \eqref{2} if one takes the planar limit $k_{(p-1)}\sim\tilde\La=0$.

It is interesting to take the limit $\La=0$, and replacing factors of $(\ga-\da)$ by the remaining exponent $\tilde\da$, one recovers a neutral solution complementary to the neutral solution \eqref{28}: in the latter, the potential is comological constant-like, while here it supports the curvature of the horizon. Combining the two effects gives the most general neutral solution with two potentials, the spherical version of which can be seen in \cite{Chan:1995fr}, while the full topological version will appear in \cite{skg}.

This solution is distinct from \eqref{EMDNEBH1}, since upon taking $\ga=-(p-2)\da$, \eqref{EMDNEBH3} reduces to a single exponential potential, whereas \eqref{EMDNEBH1} has two. In some sense, the limit $\ga=-(p-2)\da$ is degenerate in \eqref{EMDNEBH3}: for these values, more freedom is allowed since in \eqref{EMDNEBH1}, one may fix independently $\La$, $\tilde\La$ and $k_{(p-1)}$. The price to pay is fixing all the exponents of the scalar exponentials, whereas two are free in \eqref{EMDNEBH3}.

The relation \eqref{B00} between the exponents is verified by the action \eqref{B6}, and so it is straightforward to uplift the solution \eqref{EMDNEBH3}:
\bea
	\ud s^2_{(p+q+1)}&=&-V(\rho)\ud\tau^2+\frac{\ud\rho^2}{V(\rho)}+\rho^2\ud R^2_{(q)}+\ud K^2_{(p-1)} \label{B8}\,,\\
	 V(\rho)&=&\frac{2\tilde\La\rho^2}{(q+1)(p+q-1)}-\left(\frac{\rho_0}{\rho}\right)^{q-1}+\frac{Q^2}{\rho^{2q-2}}\label{B9}\,,\\
	A_{[1]}&=&-\sqrt{2}Q\rho^{1-q}\ud\tau\label{B10}\,,\\
	B_{[p]}&=&-\frac{\sqrt2 Q}{(q-1)}\rho^{1-q}\ud\tau\ud K_{(p-1)}\label{B11}\,,\\
	Q^2&=&\frac{-2\La}{q-1}\,,\qquad 2\tilde\Lambda=(p-2)(p+q-1)k_{(p-1)}\,,\label{B12}
\eea
after changing to the coordinates
\be \label{B13}
	\tau=\frac{p+q-1}{p-1}t\,,\qquad r=\rho^{\frac{p+q-1}{p-1}}\,.
\ee
These metrics have the topology AdS$_{q+2}\times \mathbf K^{(p-1)}$, with a non-trivial electric field and a $[p]$-form gauge field along the product space. Note that if one dualises the $[p+1]$-field strength, a magnetic flux is obtained on the spatial directions $\mathbf T^q$ of the worldvolume of the brane. They are charged generalisations of the metrics in section \ref{qb1}, which are interpreted as the near-horizon limit of AdS black branes. Having a non-trivial $U(1)$ gauge field turned on naturally allows for particle number conservation in the dual field theory living on the boundary.

They can easily be generalised to a curved horizon $\mathbf K^{(q)}\times\mathbf K^{(p-1)}$ by including a curvature factor in the black hole potential \eqref{B9}:
\be \label{B14}
V(\rho)=\frac{2\tilde\La\rho^2}{(q+1)(p+q-1)}+k_{(q)}-\left(\frac{\rho_0}{\rho}\right)^{q-1}+\frac{Q^2}{\rho^{2q-2}}\,.
\ee
Upon reduction, one would obtain in action \eqref{B0} another exponential potential
\be \label{B15}
	\bar V(\phi) = 2\bar\La e^{-\bar\da\phi}
\ee
with
\be \label{B16}
	2\bar\La = q(q-1)k_{(q)}\,,\qquad \bar\da = \frac{2}{(p-1)\tilde\da}\,.
\ee
and modified \eqref{B01}
\be
	h(r)\to h(r)+\frac{8(p-1)\bar\Lambda r^{\frac{p(\ga-\da)^2}{2(p-1)}-1}}{(\ga-\da)^2\left[2(p-1)-p(\ga-\da)^2\right]}\,.
\ee
The end result is an EMD theory with three exponential potentials, different from that examined in section \ref{rot}. In particular, the lower-dimensional solutions of section \ref{rot} are asymptotically AdS, while those examined in this section are not.

\section{Dimensional uplift  for generic $\ga$ and $\da$\label{generic}}

In this section, we give an interpretation to the result that the charged extremal solutions \eqref{2}-\eqref{555} are generically conformally Lifshitz metrics, except in the cases $\ga=\da$ and $\ga=-(p-2)\da$, by uplifting them to Lifshitz solutions. This will be restricted to the quadrants $(\ga-\da)(\ga+(p-2)\da)>0$. This can be relaxed to the quadrants $(\ga-\da)(\ga+(p-2)\da)<0$ by uplifting the solutions to near-extremal, asymptotically flat \emph{dilatonic} black $(p-1)$-branes with a point-like charge. These metrics are generalisations of those considered in \ref{section:StaticFlatBranes} and include a dilaton, \cite{Caldarelli:2010xz}. We consider each case in turn.

\subsection{$(\ga-\da)(\ga+(p-2)\da)>0$: Lifshitz solutions\label{Lifshitz}}

 We start by considering a higher-dimensional theory of a vector coupled to a scalar in an Einstein-AdS background:
\be \label{ActionLifshitz}
	 S=\int\ud^{p+q+1}x\sqrt{-g}\left[R-\half\partial\Phi^2-\frac14e^{\Gamma\Phi}F^2+2\Lambda\right].
\ee
This action can be identified with the EMD action \eqref{action} with gauge coupling and scalar potential \eqref{2b}, after setting $p\to p+q$, $\delta=0$ and $\gamma\to\Gamma$. It is known to have Lifshitz solutions, \cite{taylor}, which are the $\delta\to0$ limit of the charged extremal solution \eqref{2}-\eqref{555}.

 We now reduce this action over an internal torus of dimension $q$, and denote by  $\varphi$ the scalar that controls its volume factor:
\be \label{LifshitzKKAnsatz}
	\ud s^2 = e^{-\Delta\varphi}\ud s^2_{(p+1)}+e^{\frac{2\varphi}{(p-1)\Delta}\left(1-\frac{p-1}2\Delta^2\right)}\ud R^2_{(q)},\quad \frac{p-1}2\Delta^2=\frac{q}{p+q-1}\,.
\ee
The dilaton $\Phi$ and the volume scalar $\varphi$ equations of motion are
\bea
	\square \Phi&=&\frac{\Gamma}4e^{\Gamma\Phi+\Delta\varphi}F^2 \label{EomKKLifshitz1}\\
	 \square\varphi&=&\frac{\Delta}4e^{\Gamma\Phi+\Delta\varphi}F^2+2\Delta\Lambda e^{-\Delta\varphi}.\label{EomKKLifshitz2}
\eea
They can be derived from the lower-dimensional action with two scalars
\be \label{ActionLifshitz2} S=\int\ud^{p+1}x\sqrt{-g}\left[R-\half\partial\Phi^2-\half\partial\varphi^2-\frac14e^{\Gamma\Phi+\Delta\varphi}F^2+2\Lambda e^{-\Delta\varphi}\right].
\ee
 We now investigate whether we can truncate this theory to a single scalar consistently. This can be done if a combination of the two scalars has a massless Klein-Gordon equation, in which case we can set it to zero. We assume that
\be
	\Phi=\alpha\varphi\,,
\ee
which implies from \eqref{EomKKLifshitz1} and  \eqref{EomKKLifshitz2} that
\be \label{KKLifshitzCondition}
	\left(\Gamma-\alpha\Delta\right)^2 F^2 = 8\alpha\Delta\Lambda e^{-(2\Delta+\alpha\Gamma)\varphi}.
\ee
Defining further
\be \label{7.1.7}
	\phi=\sqrt{1+\alpha^2}\varphi\,,\quad \gamma=\frac{\alpha\Gamma+\Delta}{\sqrt{1+\alpha^2}}\,,\quad \delta=\frac{\Delta}{\sqrt{1+\alpha^2}}\,,
\ee
the action \eqref{ActionLifshitz2} turns into our original EMD action \eqref{action} with \eqref{2b}. We already know its generic charged black hole solutions, \eqref{2}-\eqref{555}. All that is left to do to have a consistent KK reduction is find the proportionality factor $\alpha$ such that \eqref{KKLifshitzCondition} holds. Expressing everything in terms of lower-dimensional parameters $\ga$, $\da$ and using \eqref{2}-\eqref{555} in \eqref{KKLifshitzCondition}, we eventually find that
\be
	\alpha^2 = \frac{2+\ga\da-\da^2}{\da(\da-\ga)} =  \frac{v}{\da(\da-\ga)}
\ee
where $v$ is defined in \eqref{5}. We may reexpress $\gamma$ and $\delta$ in terms of higher-dimensional quantities:
\be \label{7.1.10}
	\delta^2 = \frac{\Gamma^2\Delta^4}{4+\Gamma^2\Delta^2}\,,\quad \gamma=\left(1-\frac2{\Delta^2}\right)\delta
\ee
where $\Delta$ is defined in \eqref{LifshitzKKAnsatz}.

We define the Lifshitz exponent as in \eqref{12}
\be
	 z=1+\frac{(p-1)v}{(\ga-\da)(\ga+(p-2)\da)}=1+\frac{2(p+q-1)}{\Gamma^2}\Longrightarrow z\geq 1
\ee
and then from \eqref{6}
\be
	u=2(p-1)\,,\quad v=\frac{2(p-1)(z-1)}{p+q+z-2}\,,\quad w=\frac{2(p-1)^2(p+q+z-1)}{p+q+z-2}\,,
\ee
where $z\geq1$ allows the Gubser bound to hold.
Changing the radial variable to
\be
	r=\rho^{\frac{-2(p-1)}{(\ga-\da)(\ga+(p-2)\da)}},
\ee
and noticing that
\be
	\Delta\varphi=\delta\phi\,,\qquad e^{2A-\Delta\varphi}=e^{\frac{2\varphi}{(p-1)\Delta}\left(1-\frac{p-1}2\Delta^2\right)},
\ee
the higher-dimensional metric reads
\be
	\ud s^2_{(p+q+1)} =-\frac{V(\rho)\ud\tau^2}{\ell_z^2\rho^{2z}} +\frac1{\rho^2}\left[\frac{\ell_z^2\ud\rho^2}{V(\rho)}+\ud R^2_{(p+q-1)}\right],
\ee
\be
	V(\rho) = 1-\left(\frac{\rho_0}\rho\right)^{z+p+q-1}\,,\quad \ell^{-2}_z=\frac{2\Lambda}{(z+p+q-1)(z+p+q-2)}
\ee
\be
	e^{\Phi}=\rho^{\sqrt{2(p+q-1)(z-1)}}\,,\quad A_{\tau}=\sqrt{\frac{2(z-1)}{(p+q+z-1)\ell^2_z}}\rho^{1-p-q-z}h(\rho)\ud\tau
\ee
after rescaling
\be
	\tau=(z-1)(4+\Ga^2\Delta^2)\frac t4\,.
\ee
This is the expected Lifshitz solution. It has been studied by a variety of authors in the past, \cite{taylor,Bertoldi:2009vn,Bertoldi:2009dt,Goldstein:2009cv,cgkkm,Bertoldi:2010ca,Bertoldi:2011zr,Tarrio:2011de}. Here, we interpret it  as the dimensional uplift  of the generalised EMD solutions\footnote{Note that this procedure could be repeated to generate new EMD solutions, for instance starting from the solutions uncovered in \cite{Tarrio:2011de}, which consider non-planar as well as multiple $U(1)$ charge solutions.}. This is also reminiscent of the work of \cite{Caldarelli:2010xz}, where dilatonic black $p$-branes with horizon $\mathbf T^{p}\times \mathbf S^{n-1}$ are shown to reduce to the dilatonic-Maxwell Gibbons-Maeda black holes with horizon $\mathbf S^{n-1}$, \cite{Gibbons:1987ps}.

Note that the relations \eqref{7.1.7} restrict $\ga$ and $\da$ to part of the two quadrants
\be \label{LifshitzQuadrants}
	(\ga-\da)(\ga+(p-2)\da)>0\,,
\ee
see Fig. \ref{Fig:KKUplifts}.

\subsection{$(\ga-\da)(\ga+(p-2)\da)<0$: Near-extremal dilatonic branes\label{NEdilatonic}}

We now apply the same technique, this time starting from an Einstein-Maxwell-Dilaton higher-dimensional theory
\be
S=M^{p+q-1}\int d^{p+q+1}x\sqrt{-g}\left[R-{1\over 2}(\partial\Phi)^2-\frac14e^{\Gamma\Phi}F_{\m\n}F^{\m\n}\right].
\label{7.2.1}
\ee
This theory admits the following near-horizon black $(p-1)$-brane\footnote{The asymptotically flat Gibbons-Maeda black holes are also solutions to this action, \cite{Gibbons:1987ps}, and are related to the black $(p-1)$-branes by KK uplift along a torus, see Appendix A.1 of \cite{Caldarelli:2010xz}.}, which can be derived from the full geometry in Appendix A.2 of \cite{Caldarelli:2010xz}, for instance:
\bea
	\ud s^2 &=& -\rho^{(q-1)A}f(\rho)\ud\tau^2 + \rho^{-(q-1)B}\left[\frac{\ud\rho^2}{f(\rho)}+\rho^2\ud\mathbf K^2_{(q)}+\ud\mathbf R^2_{(p-1)}\right]\label{7.2.2}\\
	f(\rho)&=&k_{(q)}-\left(\frac{\rho_0}{\rho}\right)^{q-1},\label{7.2.3}\\
	e^{\Phi}&=&r^{\frac{N\Gamma}4},\label{7.2.4}\\
	A_{[1]}&=&-\sqrt{Nk_{(q)}}r^{q-1}\ud\tau\label{7.2.5}
\eea
where we have set the IR scale to unity, and $A$, $B$ and $N$ are given by
\be
	A=\frac{4(p+q-2)}{2(p+q-2)+(p+q-1)\Gamma^2}\,,\quad B=\frac{4}{2(p+q-2)+(p+q-1)\Gamma^2}\,, \quad N=A+B\,.\label{7.2.6}
\ee

We may now reduce the theory \eqref{7.2.1} along a curved Ansatz with $\Delta>\Delta_c$:
\be \label{7.2.7}
	\ud s^2 = e^{-\frac{2\varphi}{(p-1)\Delta}}\ud s^2_{(p+1)}+e^{\frac{2\varphi}{(p-1)\Delta}\left(\frac{p-1}2\Delta^2-1\right)}\ud K^2_{(q)},\quad \frac{p-1}2\Delta^2=\frac{p+q-1}{q}\,,
\ee
which results in the theory
\be \label{7.2.8} S=\int\ud^{p+1}x\sqrt{-g}\left[R-\half\partial\Phi^2-\half\partial\varphi^2-\frac14e^{\Gamma\Phi+\frac{2\varphi}{(p-1)\Delta}}F^2+2\Lambda e^{-\Delta\varphi}\right].
\ee
Again we check that the two scalars can consistently be taken proportional to one another upon imposing a condition analogous to \eqref{KKLifshitzCondition}
\be \label{7.2.9}
	\Phi=\alpha\varphi=\frac{\phi}{\sqrt{1+\alpha^2}}\,,\qquad \al=-\sqrt{\frac{2(p-1)(p+q-1)}{q}}\frac{(q-1)\Gamma}{2(p-1)+(p+q-1)\Gamma^2}.
\ee
Setting further
\be \label{7.2.10} \da=\frac\Delta{\sqrt{1+\alpha^2}}\,,\qquad\ga=\frac1{\sqrt{1+\alpha^2}}\left[\al\Gamma+\frac{2}{(p-1)\Delta}\right],
\ee
after some manipulations, we recover the original EMD theory \eqref{action} with arbitrary $\ga$ and $\da$, and the solution \eqref{7.2.2}-\eqref{7.2.5} now coincides with  \eqref{2}-\eqref{555}. The part of the $(\g-\d)(\g+(p-2)\d)<0$ quadrants this uplift covers is represented as the red region of Fig. \ref{Fig:KKUplifts}.

From \eqref{7.2.9} and \eqref{6}, it can be checked straightforwardly that the Gubser condition is equivalent to $q>1$, that is non-zero internal curvature for the space $\mathbf K^q$. On the other hand, although the reduction is curved, not all the parameter space it covers corresponds to a gapped system, see Fig. \ref{Fig:KKUplifts}.

\subsection{Summary of results in the $(\ga,\da)$ plane}

Let us note that the relations \eqref{7.1.7} and \eqref{7.2.10} do not allow $\ga$ and $\da$ to cover the whole $(\ga,\da)$ plane, but rather only part of it, because the relations \eqref{7.1.10} and \eqref{7.2.10} are not trivial.

To summarise, in the quadrants
\be \label{7.3.1}
	(\ga-\da)(\ga+(p-2)\da)>0
\ee
the charged near-extremal solutions \eqref{2}-\eqref{555} descend from Lifshitz metrics with horizon $\mathbf R^{(p+q-1)}$, the Gubser bound holds and the system is gapless, while in the quadrants
\be \label{7.3.2}
	(\ga-\da)(\ga+(p-2)\da)<0\,,
\ee
they descend from near-extremal dilatonic black branes with a point-like charge and horizon $\mathbf R^{(p-1)}\times\mathbf K^{(q)}$, the Gubser bound holds if and only if $q>1$ and the system can either be gapped or not.
There are two limiting cases with infinite Lifshitz exponents:
\begin{itemize}
	\item $\ga=\da$: the metric is conformal to AdS$_2\times \mathbf R^{p-1}$, and can be uplifted to an AdS$_2\times \mathbf R^{p+q-1}$ space-time; the Gubser bound always holds and the system is gapless.
	\item $\ga=-(p-2)\da$: the metric is again conformal to AdS$_2\times \mathbf R^{p-1}$, but this time can be uplifted to the near-horizon limit of a static AdS black $q$-brane with horizon $\mathbf K^q \times \mathbf R^{p-1}$ if $\da^2\leq\d_c^2$; in that case, the Gubser bound always holds and the system is gapless.
\end{itemize}
and a point $\ga=0$, $\da=0$, which admits both the AdS$_2\times \mathbf R^{p+q-1}$ and AdS$_{p+q+1}$ solutions.

There are several possibilities to cover the remaining parts of the $(\ga,\da)$ plane: one may examine what happens in both cases when the higher-dimensional theory contains a $[q+1]$-gauge potential instead of a Maxwell field. Then, in one case, one may obtain some kind of generalised Lifshitz near-extremal, dilatonic black brane\footnote{Note that the existence of Lifshitz black $p$-brane was already noted in \cite{taylor} in a theory with a massive $[p+1]$-form potential.}, while in the other one should recover the well-known near-extremal, dilatonic black $q$-brane wrapped around a torus.

In \cite{Iizuka:2011hg}, the authors study for $p=3$ the response of fermionic probes in the background \eqref{2}-\eqref{555} with action \eqref{action}. In the interior of the quadrant \eqref{LifshitzQuadrants}, their results reproduce a Fermi Liquid-type behaviour, with long-lived quasiparticles but an ${\cal O}(\omega)$ width\footnote{Instead of a width ${\cal O}(\omega^2)$ for Fermi Liquids.}, while they find non-Fermi Liquid behaviour along the lines $\ga=\da$ and $\ga=-\da$. All of these results can now naturally be linked with those of \cite{Iqbal:2011in}, which showed that Lifshitz symmetry allowed for Fermi Liquid behaviour, and earlier ones of \cite{FL}, where the near-horizon AdS$_2\times \mathbf R^2$ emergent geometry was proved to generate non Fermi-Liquid behaviour.

One may surmise that in the other two quadrants, $(\ga-\da)(\ga+(p-2)\da)<0$, the non-Fermi Liquid behaviour found by \cite{Iizuka:2011hg} derives from the generalised Bertotti-Robinson geometry \eqref{7.2.2}-\eqref{7.2.5}.

\section{AdS dilatonic black holes and near-extremal rotating black branes\label{rot}}

In this section, we shall examine black hole solutions which are asymptotically AdS with a non-trivial gauge and scalar field, \cite{Gao:2004tu}-\cite{Hendi:2010gq}\footnote{Different though from those recently reported in \cite{Cadoni:2011nq}.}. The price to pay is to increase the number of exponentials in the potential to three. Then, we shall connect it via its near-extremal limit to the solution \eqref{EMDNEBH1}-\eqref{a29}, showing this constitutes an explicit example of a UV completion of the IR effective action \eqref{action}. Finally, we will argue that they can be given a higher-dimensional interpretation as the Kaluza-Klein sphere reduction of near-extremal rotating black branes, \cite{Cvetic:1999xp}.

\subsection{Generic charged AdS dilatonic black hole}

In this section, we shall focus on a theory with a three-exponential potential, tuned just so that it allows for asymptotically AdS solutions:
\be
	\frac{\mathcal L_{(p+1)}}{\sqrt{-g}} =  R-\half(\partial\phi)^2-\frac14e^{-(p-2)\da\phi}\l(F_{[2]}\r)^2+V(\phi)\,, \label{KKSpAction}
\ee
\bea V(\phi)&=&\frac{2(p-1)^2(p-2)\da^2V_0e^{-\frac{2\phi}{(p-1)\da}}}{p(1+\frac{(p-1)}2(p-2)\da^2)^2}\l[-\frac{(p-2)}{4(p-1)}\l(1-p\frac{(p-1)}2\da^2\r)+\r.\nn\\
&& \l.+e^{\frac{2+(p-2)(p-1)\da^2}{2(p-1)\da}\phi}+\frac{(p-(p-2)^2\frac{(p-1)}2\da^2)}{2(p-1)^2(p-2)\da^2}e^{\frac{2+(p-1)(p-2)\da^2}{(p-1)\da}\phi}\r]. \label{KKSpPotential}
\eea
The black hole solutions of these theories were studied first in \cite{Gao:2004tu,Gao:2004tv}, then generalized  to arbitrary topology \cite{Hendi:2010gq}, and then an arbitrary number of spins on the brane directions were included in the planar case, \cite{Sheykhi:2010pya}. The thermodynamics were studied in \cite{Sheykhi:2009pf,Hendi:2010gq}. Intuitively, we recognize in the gauge coupling $\ga=-(p-2)\da$ the value obtained when reducing a $q$-form field strength down to a two-form field strength, whereas the first exponential of the potential is a signal that a constant curvature space $\mathbf K^q$ has been compactified. We shall see later it is indeed so.

While the potential \eqref{KKSpPotential} looks complicated, it actually displays some simple features:
\begin{itemize}
	\item Such a potential always exhibits a positive local or global minimum at $\phi=0$, with $V(0)=V_0$. Thus, one may expect that asymptotically AdS solutions will exist when the scalar field settles down at such a point.
	\item If $1/p\leq(p-1)\da^2/2\leq p/(p-2)^2$, there is a single global positive minimum at $\phi=0$.
	\item If $1/p>(p-1)\da^2/2$ (respectively $(p-1)\da^2/2> p/(p-2)^2$), then there is also a local maximum for positive (respectively negative) $\phi$.
	\item In all cases, the potential has infinite tails for $\phi\to\pm\infty$.
\end{itemize}
We plot it in each of these ranges in Fig. \ref{Fig:AdSPotential}.

\FIGURE[ht]{
	 \includegraphics[width=0.65\textwidth]{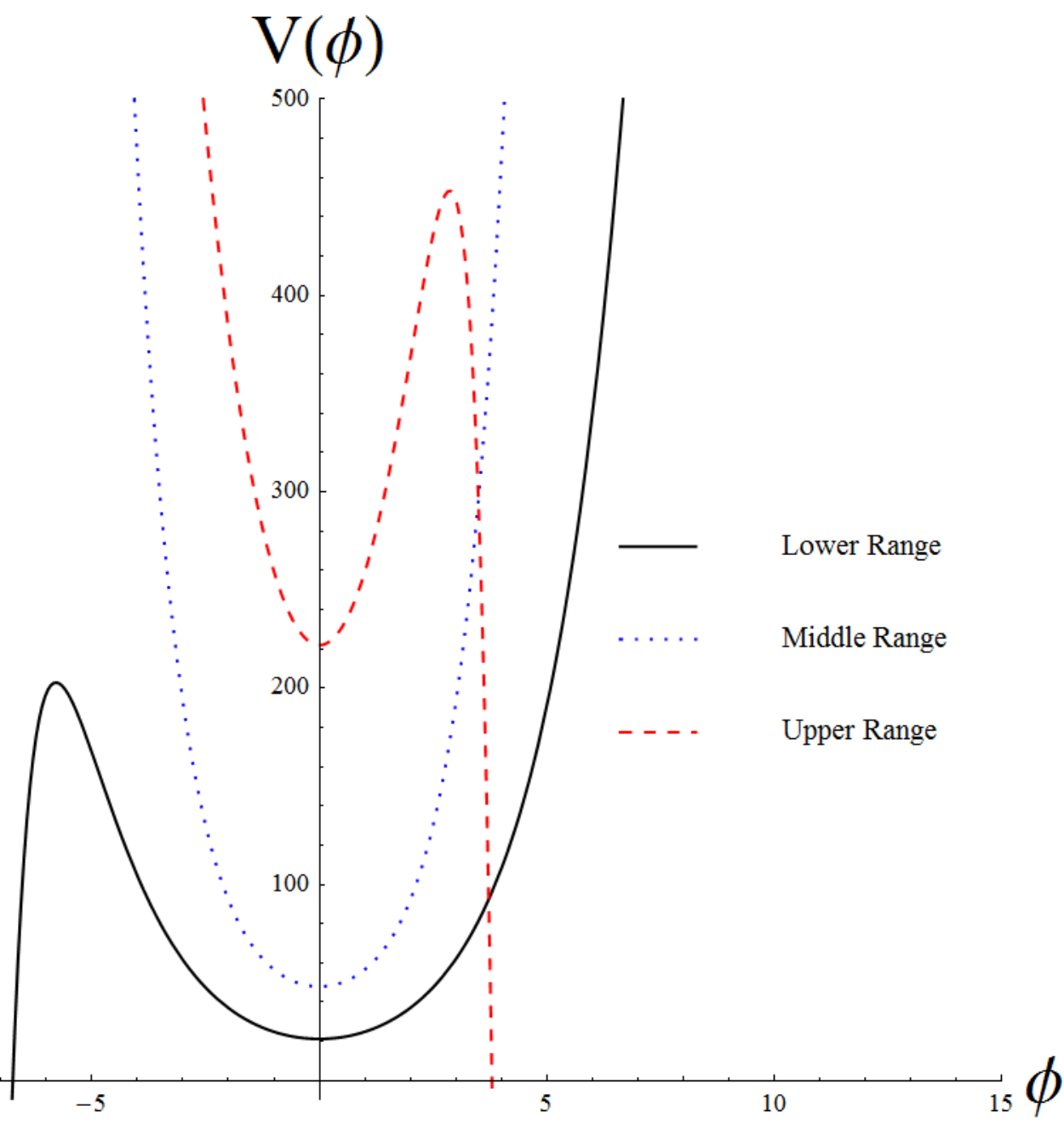}
\caption{The scalar potential \protect \eqref{KKSpPotential} for $p=3$, plotted for representative values of $\da$ in the three ranges $(p-1)\da^2/2<1/p$ (Lower Range, solid black line), $1/p\leq(p-1)\da^2/2\leq p/(p-2)^2$ (Middle range, dotted bue line) and  $ p/(p-2)^2<(p-1)\da^2/2$ (Upper Range, dashed red line). In all three ranges, the potential displays a positive minimum which is an AdS fixed point.}
\label{Fig:AdSPotential}
}

The equations of motion derived from \eqref{KKSpAction} admit asymptotically AdS solutions. We present the non-spinning, arbitrary topology version\footnote{Compared to \cite{Sheykhi:2009pf,Hendi:2010gq}, we have sent $\phi\to\sqrt{(p-1)/8}\phi$, $A\to A/2$, $\al=(p-2)\sqrt{2/(p-1)}/\da$, and changed coordinates to $r^{p-2}=\rho^{p-2}-b^{p-2}$, $b^{p-2}=\sinh^2\ba ~r_0^{p-2}/\kappa$ and $c^{p-2}=\cosh^2\ba ~r_0^{p-2}$.}:
\be
	\ud s^2_{(p+1)}=- h(r)^{-(p-2)\l(\frac2{p-2}-\Ga\r)} f(r)\ud t^2+ h(r)^{\frac2{p-2}-\Ga}\l[\frac{\ud r^2}{f(r)}+r^2 \ud K^2_{(p-1)}\r],\label{AdSBHGeneric}
\ee
\be
	 f(r)=\l(\frac{r}\ell\r)^2 h(r)^{(p-1)\l(\frac2{p-2}-\Ga\r)}+\ka-\l(\frac{r_0}{r}\r)^{p-2}\,,
\label{a41}\ee
\be
	 h(r)=1+\l(\frac{r_0}r\r)^{p-2}\kappa^{-1}\sinh^2\ba\,,
\label{a42}\ee
\be
	e^\phi= h(r)^{-\frac\Ga\da}\,,\qquad \Ga=\frac{2(p-1)\da^2}{2+(p-2)(p-1)\da^2}\,,
\label{a43}\ee
\be
 A=-\sqrt{\frac{(p-1)(p-2)\kappa}{2+(p-2)(p-1)\da^2}}\frac2{(p-2)}\l(1-h(r)^{-1}\r)\coth\ba\,\ud t\,.\label{a44}
\ee
where  $\ell^2$ is the $(p+1)$-dimensional AdS radius $V_0=p(p-1)\ell^{-2}$ and $\kappa$ the normalised curvature of $\mathbf K^{(p-1)}$.

In the $\da=0$ limit, the potential reduces to a pure cosmological constant $V_0$ \eqref{KKSpPotential} and the action \eqref{KKSpAction} to AdS-Einstein-Maxwell. The scalar field \eqref{a43} becomes trivial, and the solution \eqref{AdSBHGeneric}-\eqref{a44} turns into the AdS-Reissner-Nordstr\"om black hole upon making the change of coordinates $\rho^{p-2}=r^{p-2}h(r)$.

The solution above has arbitrary topology, yet it seems like taking the planar limit switches off the gauge field. It is actually not so, by taking the appropriate scaling limit
\be
	\kappa\to0\,,\quad \sinh^2\ba\to\kappa\sinh^2\ba\,,\quad \cosh\ba\to1\,,
\ee
whereupon the solution becomes
\be
	\ud s^2_{(p+1)}=- h(r)^{-(p-2)\l(\frac2{p-2}-\Ga\r)} f(r)\ud t^2+ h(r)^{\frac2{p-2}-\Ga}\l[\frac{\ud r^2}{f(r)}+r^2 \ud R^2_{(p-1)}\r],\label{AdSBHPlanar}
\ee
\be
	 f(r)=\l(\frac{r}\ell\r)^2 h(r)^{(p-1)\l(\frac2{p-2}-\Ga\r)}-\l(\frac{r_0}{r}\r)^{p-2}\,,
\label{a411}\ee
\be
	 h(r)=1+\l(\frac{r_0}r\r)^{p-2}\sinh^2\ba\,,
\label{a421}\ee
\be
	e^\phi= h(r)^{-\frac\Ga\da}\,,\qquad \Ga=\frac{2(p-1)\da^2}{2+(p-2)(p-1)\da^2}\,,
\label{a431}\ee
\be
 A=-\sqrt{\frac{(p-1)(p-2)}{2+(p-2)(p-1)\da^2}}\frac2{(p-2)}\l(1-h(r)^{-1}\r)\,\ud t\,.\label{a441}
\ee

In order to further interpret these solutions, we shall first take a near-extremal limit where the charge parameter $\ba$ is taken to be large. This will have the effect of zooming to the near-horizon region of the extremal black hole.

\subsubsection{The near-horizon region of the extremal black hole}

Taking the limit $\ba\to\infty$ in the solution \eqref{AdSBHGeneric}, and changing to coordinates
\be
	r^{p-2}=\xi^{1+(p-1)(p-2)\da^2/2},
\label{a45}\ee
as well as setting
\be
	\kappa=k_{(p-1)}e^{\frac{-2\phi_0}{(p-1)(p-2)\da}},\quad t=\frac{2(p-2)e^{\frac{(p-1)(p-2)\da^2-2(p-3)}{2(p-1)(p-2)\da}\phi_0}}{2+(p-1)(p-2)\da^2}\bar t,\quad K^{p-1}=e^{\frac{\phi_0}{(p-1)(p-2)\da}}\bar K^{p-1},
\ee
the near-extremal limit of \eqref{AdSBHGeneric} is
\be
	\ud s^2_{(p+1)}=-  f(\xi)\ud \bar t^2+e^{\da\phi} \frac{\ud \xi^2}{f(\xi)}+\xi^{(p-1)\da^2} \ud\bar K^2_{(p-1)}\,,\label{AdSBHGenericNearExtremal}
\ee
\be
	 V(\xi)=\frac{8\tilde\La e^{-\frac{2\phi_0}{(p-2)\da}}\xi^{(p-1)\da^2}}{(p-1)^2\da^2(2-p(p-1)\da^2)}+\frac{(p-2)^2
k_{(p-1)}\xi^2}{(2+(p-2)(p-1)\da^2)^2}-\l(\frac{\xi_0}\xi\r)^{1-(p-1)(p-2)\da^2/2}
\label{a46}\ee
\be
	e^\phi=e^{\phi_0}\xi^{(p-1)\da}\,,\qquad e^{\phi_0}=\l(\frac{k_{(p-1)}}{\sinh^2\ba~r_0^{p-2}}\r)^{\frac{2(p-1)\da}{2+(p-1)(p-2)\da^2}},
\label{a47}\ee
\be
 A=\sqrt{\frac{(p-1)(p-2)k_{(p-1)}}{2+(p-2)(p-1)\da^2}}e^{\frac{p-1}2\da\phi_0}\frac{4\xi^{1+(p-1)(p-2)\da^2/2}}{2+(p-1)(p-2)\da^2}\,\ud t\,,\label{a48}
\ee
which is precisely the scaling solution \eqref{EMDNEBH1} with $\La=0$. Indeed, the effective IR potential derived from \eqref{KKSpPotential} can be reduced to a single exponential,
\be
V_{IR}(\phi)=2\tilde\La e^{-\frac{2\phi}{(p-1)\da}}\,,\qquad 2\tilde\La = -V_0\frac{(p-1)(p-2)^2\da^2\l(2-p(p-1)\da^2\r)}{p(2+(p-1)(p-2)\da^2)^2}.
\ee
Again, taking the planar limit recovers the single exponential, neutral solution \eqref{28} of \eqref{action} once we have made the change $\da\to2/(p-1)\da$. This simply represents the non-commutativity of the near-horizon limit and the planar limit: taking $\ba\to\infty$ in the planar solution \eqref{AdSBHPlanar}-\eqref{a441} turns off the gauge field since there is no extra scale with which to form a fixed ratio. In more pictural terms, the near-horizon limit is a limit where the electric charge blows up at the same time as the horizon radius (\emph{i.e.} its curvature vanishes); if the horizon already has planar topology, this is a singular limit which can only be regularised by switching off the gauge field. On the other hand, when the scalar potential equivalent of a cosmological constant is present, then the two limits may be taken while retaining a non-trivial gauge field, since an extra curvature scale is available, giving rise to the planar, charged, near-extremal scaling solutions \eqref{2}-\eqref{7b}.

Moreover, this provides an explicit UV completion to AdS asymptotics of the single Liouville potential solutions: in order to describe the IR region, \eqref{a45}-\eqref{a48} is all that is needed. Note that to recover the original action \eqref{1}, one should change\footnote{This is consistent with the fact that \eqref{KKSpAction} is derived from \eqref{EFormAction2} by reducing over a curved space, as we shall see next.} $\da\to2/(p-1)\da$, and then the IR effective action derived from \eqref{KKSpAction} corresponds to the specific value $\ga=-2(p-2)\da$.

The next step along the interpretation of the solutions \eqref{AdSBHGeneric}-\eqref{a44} is carried out in the next section, where they are connected with the decoupling limit of asymptotically flat rotating black branes.

\subsection{Sphere reduction of the near-extremal limit of rotating black branes\label{section:KKSphereReduction}}

We will use the results of \cite{Cvetic:1999xp}, section 5 of the paper, to truncate to a single scalar, single gauge field sector the reduction along a sphere $\mathbf S^{q}$, $q=2N-1$, of the decoupling limit of rotating black branes.

The higher-dimensional theory we consider is \eqref{EFormAction} with $n=q$, that is:
\be  \label{EFormAction2}
	S_{(p+q+1)}=\frac{1}{16\pi G_D}\int\ud^{p+q+1}x\,\sqrt{-g}\l[R-\frac{1}{2(p+1)!}G^2_{[p+1]}\r].
\ee
The static black branes were studied in section \ref{section:StaticFlatBranes}, and we shall consider here their rotating version, which is presented in Appendix \ref{section:RotatingBranes}. One may find the necessary details in \cite{Cvetic:1999xp}. By going to the near-extremal limit, the authors of \cite{Cvetic:1999xp} showed that the rotating black brane could be rewritten in just the right way so as to allow for a consistent reduction over the internal sphere, while truncating the massive modes and restricting to a $U(1)$ Cartan subgroup of the generic $\mathbf O({2N})$ Lie group rotating the internal $\mathbf S^{2N-1}$.

In appendices \ref{SUGRAAdS5S5},  \ref{SUGRAAdS4S7} and  \ref{SUGRAAdS7S4}, we take up the explicit Lagrangians worked out by \cite{Cvetic:1999xp} for the cases of $D=10$ SUGRA on $\mathbf S^{5}$ and  $D=11$ SUGRA on $\mathbf S^{4}$, $\mathbf S^{7}$ and show how they may be truncated to a single scalar, single gauge field sector.

Although the authors of \cite{Cvetic:1999xp} write the reduced metric in the generic case, they do not provide the full  reduced Lagrangian. We shall not attempt to work out its expression, which is beyond the scope of this work. We shall simply start from the reduced solution, equations (5.4), (5.5) of \cite{Cvetic:1999xp}, and show how it can be truncated to a single scalar, single gauge field sector, building on the subcases $q=4,5,7$ of appendices \ref{SUGRAAdS5S5}-\ref{SUGRAAdS7S4}. Though this does not prove that the truncation is true for all solutions of the reduced Lagrangian, it is sufficient for our purposes, as we are really interested only in that particular solution:
\be
	\ud s^2_{(p+1)}=-\l(H_1\ldots H_N\r)^{-\frac{p-2}{p-1}}f(r)\ud t^2+\l(H_1\ldots H_N\r)^{\frac{1}{p-1}}\l(\frac{\ud r^2}{f(r)}+r^2\ud R^2_{(p-1)}\r)\,,\label{SqRedRotatingBraneDecoupling}
\ee
\be
	H_i=1+g^2l_i^2\l(\frac{2(N-1)g}{pr}\r)^{p-2},\quad  A^i=\frac1{gl_i\sinh\al}\l(1-H_i^{-1}\r)\ud t\,,
\ee
\be
	f(r)=\l(\frac{2(N-1)gr}{p}\r)^2H_1\ldots H_N-\frac{\mu}{r^{p-2}}\,,
\ee
where $i=1\ldots N$, so that there are $N-1$ independent scalar and $N$ independent gauge fields. The scalars are parametrized in terms of $N$ exponentials $X_i$:
\be
	X_i=e^{-\half \overrightarrow{a}_i\cdot\overrightarrow{\phi}}\,,\qquad X_1\ldots X_N=1, \label{XiDef}
\ee
where the $\overrightarrow{a}_i$ are independent vectors which are fixed by the reduction\footnote{More precisely, they are the root vectors of the Cartan generators of the isometry group of the reduced space.}. These $X_i$ are fixed in terms of the $H_i$ in order to have a solution as:
\be
	X_i=\l(H_1\ldots H_N\r)^{\frac{p}{2(p-1)(N-1)}}H_i^{-1}.
\label{a32}\ee
One may check that the $X_i$ do verify \eqref{XiDef} using $N(p-2)=2(p-1)$.

 We now set all of the charge parameters but two to zero. The two non-zero charge parameters we take to be equal:  $l_1=l_2$, $l_i=0$ for $2<i\leq N$. We also denote $H_1=H_2=h$. It is clear that in this case only two of the gauge fields survive, and moreover they can be combined in a single one. We obtain
\be
	X_1=X_2=h^{\frac{p}{(p-1)(N-1)}-1}\,,\qquad X_{i>2}=h^{\frac{p}{(p-1)(N-1)}}.
\label{a33}\ee
In view of \eqref{XiDef}, this result means that
\be
	(\overrightarrow{a}_2- \overrightarrow{a}_1)\cdot\overrightarrow{\phi}=0\,,\qquad (\overrightarrow{a}_i- \overrightarrow{a}_{i-1})\cdot\overrightarrow{\phi}=0\,\,~~~\forall\,\,3<i\leq N\,.
\label{a34}\ee
Given that the $\overrightarrow{a}_i$ are independent vectors, these conditions can only be accommodated if $\phi_i=0$ $\forall i>1$, and only a single scalar $\phi=\phi_1$ survives. Note that although this does not constitute a rigorous proof of the truncation\footnote{This would entail writing out explicitly the lower-dimensional equations of motion, and show that once all charges but two are turned off, the $N-2$ scalars are not sourced any longer and can also consistently be turned off.}, the procedure describe above reduces correctly to the supergravity cases. Moreover, one can argue that being able to truncate some fields at the level of a particular solution is always consistent \emph{vis-\`a-vis} the equations of motion, although this does not mean that such a procedure will hold generically for \emph{all} of the solutions to this particular set of equations of motion.
Anyway, as our focus is on the near-extremal limit of rotating black branes, the level of detail explained above suffices.

\subsubsection{Dimensional uplift to the near-extremal  limit of the rotating black brane}

Along the lines of \cite{Cvetic:1999xp}, it should be possible to embed the black hole solution \eqref{AdSBHGeneric} in the higher-dimensional theory \eqref{EFormAction2} which, as mentioned before, admits rotating black $(p-1)$-brane solutions, where the $(p+1)$-form field strength is supported by the worldvolume of the brane and rotation is along the space $\mathbf S^{q}$. Indeed, if one sets $(p-2)\da=-\sqrt{2/(p-1)}$, the previous lagrangian \eqref{KKSpAction} reduces to the following:
\be
	\frac{\mathcal L_{(p+1)}}{\sqrt{-g}} =  R-\half(\partial\phi)^2-\frac14e^{\sqrt{\frac2{p-1}}\phi}\l(F_{(2)}\r)^2+V(\phi)\,,
\label{a49}
\ee
	\be V(\phi)=\frac{V_0(p-2)^2}{p(p-1)}\l[e^{-\sqrt{\frac2{(p-1)}}\phi}-\frac{p-4}{(p-2)^2}e^{(p-2)\sqrt{\frac2{(p-1)}}\phi}+\frac{4}{p-2}e^{\frac{(p-3)}{\sqrt{2(p-1)}}\phi}\r], \label{KKSpPotReduced}
\ee
which agrees with \eqref{KKS7Action}, \eqref{KKS5Action} and \eqref{KKS4Action} for $p=3,4,6$. The solution \eqref{AdSBHGeneric} becomes in turn
\bea
	\ud s^2_{(p+1)}&=&- h(r)^{-\frac{2(p-2)}{p-1}} f(r)\ud t^2+ h(r)^{\frac2{p-1}}\l[\frac{\ud r^2}{f(r)}+r^2 \ud K^2_{(p-1)}\r],\label{AdSBHKKSp}\\
	 f(r)&=&\l(\frac{r}\ell\r)^2 h(r)^{2}+k_{(p-1)}-\l(\frac{r_0}{r}\r)^{p-2}\,,\label{a50}\\
	 h(r)&=&1+\l(\frac{r_0}r\r)^{p-2} k_{(p-1)}^{-1}\sinh^2\ba\,,\label{a51}\\
	e^\phi&=& h(r)^{\sqrt{\frac2{p-1}}}\,,\label{a52}\\
	A_{[1]}&=&-\coth\ba\sqrt{2k_{(p-1)}}\l(1-h(r)^{-1}\r)\ud t\,.\label{a53}
\eea
In the planar limit $k_{(p-1)}=0$ and for the values  $p=3$, $p=4$ and $p=6$, the above coincides with the double, equal charge AdS black holes of compactified supergravity theories on $\mathbf S^7$, $\mathbf S^5$ and $\mathbf S^4$, \cite{Cvetic:1999xp}, and for generic $p$ with \eqref{SqRedRotatingBraneDecoupling}. The three exponentials present in the potential can now be explained: one comes from the curvature of the reduced sphere, and there is one surviving per plane with non-zero rotation. As we truncated all charges, \emph{i.e.} all angular momenta but two, this leaves two more scalar exponentials.

Furthermore, this means that the thermodynamic properties of these black holes shall be related to those of the rotating black branes from which they descend. In \cite{Sheykhi:2009pf}, the authors find that in the region of large $\da$ (small $\da$), the spherical black holes are thermally stable (unstable). This seems consistent with the previous remarks that in the former (latter) region, the black holes have AdS-like (Minkowski-like) characteristics\footnote{Taking into account that one should send $\da\to2(p-1)\da$ to recover the conventions of \eqref{action}.}. Similar results are presented in \cite{Hendi:2010gq} for the topological case, and agree with the above and with the results of \cite{cgkkm} at low temperatures.

Finally, this also stresses that where dilaton black holes are concerned, charge may originate from charge in the higher-dimensional theory, or from rotation or boosts.

\section{Scaling in thermodynamics and transport\label{transport}}

The scaling near-extremal solutions have simple power-like thermodynamic functions and conductivity in the IR. This was analyzed in \cite{cgkkm}. We will briefly review it here and we will also
produce the scaling of the AC conductivity for arbitrary frequencies in the near-extremal regime. In all cases, this scaling is manifest in the higher-dimensional incarnation when there is one.

The temperature and entropy of the solutions (\ref{2})-(\ref{a4}) scale as
\bea
T&\sim&  (r_0)^{\frac{w_p}{2(p-1)}-\half(\ga-\d)^2}\sim (r_0)^{u+(2p-3)v-2(p-1)\over 2(p-1)},\\
 S&\sim& (r_0)^{\half(\ga-\d)^2}\sim T^{(p-1)(\ga-\d)^2\over w_p-(p-1)(\ga-\d)^2}\sim T^{(p-1)(u-v)\over u+(2p-3)v-2(p-1)},\label{16}\\
E&\sim&  T^{pu+(p-2)v-2(p-1))\over u+(2p-3)v-2(p-1)},\label{a54}
\eea
so that the near-extremal equation of state is
\be
S\sim E^{pu+(p-2)v-2(p-1)\over (p-1)(u-v)}.
\label{a55}\ee

\subsection{DC conductivities}

There are several regimes for the DC conductivity, depending on the nature of the charge carriers. As discussed in \cite{cgkkm} in the Maxwell case, the DC conductivity, which is the IR limit of the AC conductivity, is ambiguous and further information is needed for its computation.\footnote{In a theory with a quadratic Maxwell term, there is no relation between vev and source for constant electric fields, necessary in order to calculate unambiguously the conductivity, \cite{cgkkm}. This relation is implemented in the probe DBI case, but requiring that the gauge field solution extends in the IR. It can be verified that as we expand the DBI to obtain Maxwell, this source-vev relation disappears. The Maxwell setup need extra information in order for the DC conductivity to be well-defined.}
An unambiguous  computation involved the DBI approach \cite{kob} and provided the two types of asymptotics:

(a) The DC resistivity  of massive charge carriers is dominated by energy loss due to drag and scales as, \cite{cgkkm},
\bea
\rho &\sim& \langle J^t\rangle^{-1}\, T^{m_p}\,,\\
 m_p&\equiv& {2k(p-1)
(\d-\gamma)+2(\d-\gamma)^2\over 2(p-1)(1-\d(\d-\gamma))+(\d-\gamma)^2}
={\pm2k(p-1)\sqrt{u-v}+u-v\over u+(2p-3)v-2(p-1)}\,,
\label{18}\eea
where $\langle J^t\rangle$ is the charge density of carriers on the boudary.

(b) The DC resistivity for light charge carriers is dominated by pair production, is independent of the charge density of carriers and scales as
\be  \label{DCL}
	\rho\sim {e^{(3-p)A(r_h)}\over Z(\phi(r_h))}\sim T^{n_p}
\ee
with
\be
n_p= {\frac{(\ga-\da)[(p+1)\g+(p-3)\da]}{2(p-1)+(\ga-\da)(\ga+(2p-3)\da)}}={(p+1)u+(p-3)v-4(p-1)\over [u+(2p-3)v-2(p-1)]}.
\label{a56}\ee

In the above, $0\leq k\leq\sqrt{2/(p-1)}$ parametrises the dependence of the resistivity on the frame. Since (see \cite{cgkkm})
\be
	g_{\mu\nu}^{(p+q+1)}=e^{k\phi}g_{\mu\nu}^{(p+1)},
\ee
this naturally allows us to find the higher-dimensional expressions for the resistivity by setting $k=-\da$, and $k$ belongs in the correct range for $\da^2\leq\da^2_c$ upliftings. Note that the result for massive carriers depends on the frame, while it does not for massless ones. Then,  for Lifshitz metrics
\be
	m_{p+q+1}= {2(\gamma-\da)(\ga+(p-2)\da)\over 2(p-1)+(\gamma-\da)(\ga+(2p-3)\da))}=\frac{2}z\,,\quad n_{p+q+1}= \frac{p+q+1}{z}\,.\label{9.9}
\ee
which is the expected value in the large mass limit, \cite{Hartnoll:2009ns}. One can also compute the higher-dimensional resistivities in the AdS$_2\times \mathbf R^{p+q-1}$ ($\ga=\da$),  AdS$_{q+2}\times \mathbf R^{p-1}$ ($\ga=-(p-2)\da$) and  boosted AdS cases:
\bea
	\ga=\da \quad &\Rightarrow&\quad m_{p+q+1}=0\,,\;\;  n_{p+q+1}=0\,,\\
	\ga=-(p-2)\da \quad&\Rightarrow&\quad m_{p+q+1}=0\,,\;\;  n_{p+q+1}=q\,,\\
	\ga=\sqrt{\frac{2p}{p-1}}\,,\quad \da=\sqrt{\frac{2}{p(p-1)}}\quad&\Rightarrow&\quad m_{p+2}=\frac4{p+3}\,,\;\; n_{p+2}=1\,.\label{a57}
\eea
One should be careful that this last result hold in $(p+2)$ dimensions. Notice that the $z\to\infty$ and $\ga=\da$ coincide, which is correct since in that limit one recovers the AdS$_2\times \mathbf R^{p+q-1}$ metric from the Lifshitz one. However, the results for the case $\ga=-(p-2)\da$ do not necessarily coincide with that limit, since one would need to work out the uplifting of solution \eqref{2}-\eqref{a4} to a Lifshitz black $(p-1)$-brane. In the limit $z\to 1$, the AdS$_{p+q+1}$ case is recovered.

Let us also revisit what happens in the neutral case, for massless carriers. In that case, our formula corresponds to microscopic fluctuations around the neutral solution and the same formula \eqref{DCL} may be used, in conjunction with equation \eqref{28}. One finds
\be  \label{DCLNeutral}
	\rho\sim T^{\frac{(p-1)\ga\da+(p-3)}{\frac{p-1}2\da^2-1}}.
\ee
Note that one needs to make a choice for the scalar-gauge coupling function $Z(\phi)$ in the action, since this will affect the perturbation. Even though we are working with a neutral background, one needs to select a particular exponent $\ga$, that is  a particular uplifting to $p+q+1$ ($p+2$ in the last case). The same values as previously now yield
\bea
	\ga=\da \quad &\Rightarrow&\quad \rho\sim T^{-(p+q-3)},\\
\ga=-(p-2)\da \quad&\Rightarrow&\quad \rho\sim T^{-(p-q-3)},\\
\ga\da=\frac{2}{p-1}\,, \sqrt{\frac{p-1}2}\da=\sqrt{\frac{1}{p}}\quad&\Rightarrow&\quad \rho\sim T^{-p} \label{a577}
\eea
The first result is valid for fluctuations of a $U(1)$ Maxwell field around a neutral AdS$_{(p+q+1)}$ background, and one may check that it indeed coincides in five dimensions, $p+q=4$, with the linear scaling of the DC conductivity with temperature obtained in various works using either correlator calculations \cite{CaronHuot:2006te}, probe branes \cite{kob} or hydrodynamic gradient expansions \cite{Erdmenger:2008rm,Hur:2008tq}. The third one, in the boosted case, coincides with the one obtained in \cite{skg} by a hydrodynamic expansion. Note that in that case, the DC conductivity scales with the temperature like the shear viscosity for the $p+2$-dimensional AdS black brane, which may be expected since the lower-dimensional gauge field is uplifted to a higher-dimensional metric element of the boundary of the black brane.

Finally, the drag force result for massive charge carriers in the neutral background, which uplifts to AdS$_{p+q+1}$, gives
\be
	\rho\sim  \langle J^t\rangle^{-1}\, T^{\frac{4+2(p-1)k\da}{2-(p-1)\da^2}}= \langle J^t\rangle^{-1}\, T^{\frac{4-2(p-1)\da^2}{2-(p-1)\da^2}}= \langle J^t\rangle^{-1}\, T^2
\ee
which recovers the classic result from \cite{kob} after replacing as prescribed $k=-\da$. It coincides with the $z\to1$ limit in \eqref{9.9}.

\subsection{Scaling of the near-extremal AC conductivity}

We recapitulate here the AC conductivity equation  that was derived in \cite{cgkkm}:
\be
 \partial_r\left(ZC^{p-3\over 2}\sqrt{D\over B} a_i'\right) +{ZC^{p-3\over 2}\left[
 \sqrt{B\over D} \omega^2 - {Q^2\sqrt{DB}\over {Z C^{p-1}}}\right]}a_i=0\,.
 \label{30a}\ee
  with $Q$ defined as
 \be
 A'_t={Q\sqrt{DB}\over ZC^{p-1\over 2}}
   \label{b18}\ee

Restoring the IR scale $\ell$, we substitute the near-extremal charged solution from section \ref{charge} with
\be
 Q^2=e^{(\g-\d)\phi_0}{4 \Lambda v\over u}\sp  T={\ell\Lambda~e^{-\delta\phi_0}\over \pi u}(r_0)^{v-1+{(\g-\d)^2\over 2(p-1)}},
\label{a59}\ee

  \be
  C=e^{2A} = \left({r\over \ell}\right)^{\frac{(\ga-\da)^2}{(p-1)}}\sp
  D=fC=f_0  \left({r\over \ell}  \right)^{v+\frac{(\ga-\da)^2}{(p-1)}} h(r)\sp
  B=1/f={\left({\ell\over r}  \right)^{v}\over f_0 h(r)},
 \label{a60} \ee
  \be
  Z=e^{\g\phi}=e^{\g\phi_0}\left({r\over \ell}\right)^{\g(\da-\ga)},
\label{a61}\ee
  \be
   f_0=\frac{8(p-1)\ell^2\Lambda e^{-\da\phi_0}}{u w_p}\label{a3}
 \sp
 h(r)= \le[1 -\left({r\over r_0\ell}\right)^{-\frac{w_p}{2(p-1)}}\ri],
 \ee
   The entropy is
  \be
  S={1\over 4G}\left({\pi uT\over \ell \Lambda e^{-\delta \phi_0}}\right)^{m}\sp m={(p-1)(u-v)\over u+(2p-3)v-2(p-1)}=\left(1-{\thema\over 2}\right){(p-1)\over z}
  \ee

 We define a new variable $z$ (not to be confused with the Lifshitz exponent), and $z_0$ as
  \be
  z=z_0{\ell\over r}=\left({\pi T u\over \ell\Lambda e^{-\d\phi_0}}\right)^{\hat\kappa} {\ell\over r}\sp \hat\kappa\equiv {1\over v-1+{(\g-\d)^2\over 2(p-1)}}={2\over b+2}\;.
 \label{a63}\ee
  \be
  b={(\g-\d)(\g+(2p-3)\d)\over p-1}={u+(2p-3)v-2(p-1)\over p-1}-2
  \label{b1}  \ee
 In terms of the new variable
 \be
 h(z)=1-z^{c_0}\sp c_0={w_p\over 2(p-1)}\,,
\label{a65} \ee
so that the horizon is at $z=1$ while the boundary at $z=0$.

 Substituting in the equation (\ref{30a}), we obtain
 \be
 z^{-a}\partial_z\left(z^{a}~h(z)\partial_z a_i(z)\right)+
 \left[z^{b}{{\cal W}^2\over h(z)}-{\kappa\over z^2}\right]a(z)=0
 \label{a67}\ee
 with
 \be
 {\cal W}={\omega\over \xi_0}={w_p\over 8\pi(p-1)}{\omega\over T}\sp a={p(\g-\d)^2\over 2(p-1)}={p(u-v)\over 2(p-1)}
  \label{a68}\ee

  \be \kappa={vw_p\over 2(p-1)}={v(p u+(p-2)v-2(p-1))\over 2(p-1)}
  \label{b2}\ee

  Near the boundary the two independent solutions at ${\cal W}=0$ behave as $z^v$ and $z^{-{pu+(p-2)v-2(p-1)\over 2(p-1)}}=z^{-{\kappa\over v}}$.
  The first vanishes at the boundary, as $v>0$ but the second always blows up as $\kappa>0$.

 Therefore, the AC conductivity scales as
 \be
 {\sigma_{AC}\over T}\sim  f\left({\omega\over T}\right).
 \label{a69}\ee

As $\omega\to 0$ we obtain, \cite{cgkkm}, $\sigma \sim \omega^n$ with
\be
 n =\left|\frac{6(1-p)+(\d-\gamma)(p\gamma + (3p-4)\d)}{2(1 -
    p) + (\d - \gamma) (\gamma + (2 p - 3)\d )}\right| - 1={(p-1)(u+v)\over u+(2p-3)v-2(p-1)}>0\,. \label{17}
\ee

In the various cases of interest in this paper, the exponent $n$ reduces to simple expressions
\bea
	\textrm{Lifshitz} \quad&\Rightarrow&\quad n=2+\frac{p+q-3}z\,,\\
	\ga=\da \quad&\Rightarrow&\quad n=2\,,\\
	\ga=-(p-2)\da\quad&\Rightarrow&\quad n=q+2\,,\\
	\ga=\sqrt{\frac{2p}{p-1}}\,,\quad \da=\sqrt{\frac{2}{p(p-1)}}\quad&\Rightarrow&\quad n=\frac{1+3p}{3+p}\,.
\eea

\subsection*{Note added:} As this work was being completed, \cite{Chemissany:2011gr} appeared which contains some overlap with the solutions presented in section \ref{qb}.

 \addcontentsline{toc}{section}{Acknowledgements}
\section*{Acknowledgements}

We would like to thank M. Caldarelli, C. Charmousis, U. Gursoy, B. S. Kim, J. McGreevy, R. Meyer, F. Nitti, A. O'Bannon, C. Panagopoulos, S. Sachdev and K. Skenderis.
We also thank the referees for constructive comments and suggestions.
Preliminary versions of this work were presented in  seminars at Cambridge and Tel Aviv and at the   conference   ``Black Hole Answers for Condensed Matter Questions" held in Leiden from the 14$^{\textrm{th}}$ of June to the 17$^{\textrm{th}}$ of June 2011.

This work was  partially supported by  a European Union grant FP7-REGPOT-2008-1-CreteHEPCosmo-228644 and PERG07-GA-2010-268246.

\newpage
 \addcontentsline{toc}{section}{Appendices}
  \renewcommand{\theequation}{\thesection.\arabic{equation}}
\appendix
\section*{Appendix}

\section{Notations and conventions\label{conv}}
In the following, the most general, higher-dimensional theory we shall consider is $(p+q+1)$-dimensional Einstein gravity plus a cosmological constant and an $(n+2)$-form field strength, with  $n\leq p+q-1$:
\be
	S_{(p+q+1)}=\frac{1}{16\pi G_D}\int\ud^{p+q+1}x\,\sqrt{-g}\l[R-\frac{1}{2(n+2)!}G^2_{[n+2]}+2\La\r].
\ee
We also define the Planck mass in terms of Newton's constant, $M^{p+q-1}=1/16\pi G_D$.

We shall define Einstein spaces $\mathbf K^{p-1}$, with metric $\ud K^2_{(q)}$ and volume element $\ud K^q$, as spaces where the Ricci tensor is proportional to the metric\footnote{Therefore, they are solutions to Einstein's vacuum equations with a cosmological constant. For zero cosmological constant, such spaces will always be Ricci flat.}:
\be
	R^{(p-1)}_{ij} = (p-2)k_{(p-1)}g_{ij}\,, \qquad i,j=1\ldots p-1\,,
	\label{EinsteinSpace}
\ee
as well as constant curvature spaces, for which the Riemann tensor itself is proportional to the metric\footnote{So that the Weyl tensor vanishes.}:
\be
	R^{(q)}_{ijkl}=k_{(q)}\left(g_{ik}g_{jl}-g_{jk}g_{il}\right).
	\label{ConstantCurvatureSpace}
\ee
Both kind of spaces have constant Ricci scalar, which we have defined as
\be
	R_{(q)}=q(q-1)k_{(q)}\,,
	\label{NormalisedCurvature}
\ee
where $k_{(q)}$ is the normalized curvature and is $\pm1,0$ for constant curvature spaces. These can then be classified as the $q$-dimensional round sphere $\mathbf S^{q}$, with metric $\ud\Omega^2_{(q)}$ and volume element  $\ud\Omega^{q}$, the $q$-dimensional plane (or torus) $\mathbf R^{q}$, with metric $\ud R^2_{(q)}$ and volume element $\ud R^q$, and the $q$-dimensional hyperbolic place $\mathbf H^{q}$, with metric $\ud H^2_{(q)}$ and volume element  $\ud H^{q}$. In $(p+q+1>4)$-dimensional Einstein gravity, the geometry of the horizon for black hole solutions gets relaxed to generic Einstein spaces instead of constant curvature spaces. In the following, for simplicity, we will mostly consider the latter, but it should be understood that in many cases the metrics presented can be generalized to comprise generic Einstein spaces on the horizon.

When needed, indices between parenthesis will indicate the dimensionality of spacetime (so that for instance $R_{(p-1)}$ is the $(p-1)$-dimensional Ricci scalar formed with the metric $g_{(p-1)}^{mn}$), while indices in brackets indicate the rank of the form considered (so that $G_{[n+2]}$ is a field strength form of rank $n+2$).

\section{Truncations of supergravity theories to a single scalar field \label{SUGRAAdS}}

In this appendix, we collect results from \cite{Cvetic:1999xp} and show how the various supergravity theories in $D=10,11$ can be truncated to a single scalar and $U(1)$ gauge field upon dimensional reduction along a sphere.

\subsection{$\mathbf S^5$ reduction of type IIB supergravity \label{SUGRAAdS5S5}}

The $\mathbf S^5$ reduction of type IIB supergravity has an $SO(6)$ Yang-Mills gauge group, which can be truncated to an $U(1)^3$ abelian sector. In addition to these three gauge fields, this theory contains the metric and two scalar fields:
\be
	\frac{\mathcal L_5}{\sqrt{-g}} =  R-\half(\partial\varphi_1)^2-\half(\partial\varphi_2)^2+4g^2\sum_i X_i^{-1}-\frac14\sum_i X_i^{-2}\l(F^i_{(2)}\r)^2+\frac14\epsilon^{\mu\nu\rho\sig\la}F_{\mu\nu}^1F_{\rho\sig}^2A^3_\la\,.
\ee
The $X_i$ are defined in terms of the two scalars as
\be
	X_i=e^{-\overrightarrow{a}_i\cdot\overrightarrow\phi}\,, \quad X_1X_2X_3=1\,,
\ee
with dilaton vectors
\be
	\overrightarrow{a}_1=\l(\sqrt{\frac23},\sqrt2\r)\,,\qquad	 \overrightarrow{a}_2=\l(\sqrt{\frac23},-\sqrt2\r)\,,\qquad	 \overrightarrow{a}_3=\l(-2\sqrt{\frac23},0\r)\,.
\ee
Setting $F_{(2)}^1=F_{(2)}^2=F_{(2)}/{\sqrt2}$ allows to consistently set $\varphi_2=0$, implying $X_1=X_2=X_3^{-1/2}$:
\bea
	\frac{\mathcal L_5}{\sqrt{-g}} &=&  R-\half(\partial\varphi_1)^2+4g^2\l(2e^{\frac{\varphi_1}{\sqrt6}}+e^{-\frac{4\varphi_1}{\sqrt6}}\r)-\frac14e^{\frac{2\varphi_1}{\sqrt6}}\l(F_{(2)}\r)^2-\frac14e^{\frac{2\varphi_1}{\sqrt6}}\l(F^3_{(2)}\r)^2+\nn\\
	&&\qquad \qquad\qquad \qquad\qquad \qquad\qquad \qquad\qquad \qquad +\frac18\epsilon^{\mu\nu\rho\sig\la}F_{\mu\nu}F_{\rho\sig}A^3_\la\,.
\eea
If one considers purely electric or purely magnetic solutions of the equations of motion deriving from the above Lagrangian, then the gauge field strength $F^3_{[2]}$ may be set to zero:
\be \label{KKS5Action}
	\frac{\mathcal L_5}{\sqrt{-g}} =  R-\half(\partial\phi)^2+4g^2\l(2e^{\frac{\phi}{\sqrt6}}+e^{-\frac{4\phi}{\sqrt6}}\r)-\frac14e^{\frac{2\phi}{\sqrt6}}\l(F_{(2)}\r)^2\,.
\ee
Then, this describes a two-exponential potential Einstein-Maxwell-Dilaton theory, which admits an AdS$_5$ black hole solution.

\subsection{$\mathbf S^7$ reduction of $D=11$ supergravity \label{SUGRAAdS4S7}}

The $\mathbf S^7$ reduction of  $D=11$ supergravity has an $SO(8)$ Yang-Mills gauge group, which can be truncated to an $U(1)^4$ abelian sector. In addition to these four gauge fields, this theory contains the metric, three scalar fields and three axions. The axions will be sourced by Chern-Simons type terms made up of the various field strengths $F^i_{(2)}$. Though they cannot generically be set to zero in a consistent manner, this can be done if one restricts to purely electric or magnetic solutions, since then all the Chern-Simons type contributions in the equations of motion will vanish. The four-dimensional lagrangian then takes the form
\be
	\frac{\mathcal L_4}{\sqrt{-g}} =  R-\half(\partial\overrightarrow\varphi)^2+8g^2\l(\cosh\varphi_1+\cosh\varphi_2+\cosh\varphi_3\r)-\frac14\sum_{i=1 }^4e^{\overrightarrow{a}_i\cdot\overrightarrow{\varphi}}\l(F^i_{(2)}\r)^2,
\ee
with dilaton vectors
\be
	\overrightarrow{a}_1=\l(1,1,1\r)\,,\quad	 \overrightarrow{a}_2=\l(1,-1,-1\r)\,,\quad	 \overrightarrow{a}_3=\l(-1,1,-1\r)\,,\quad	 \overrightarrow{a}_4=\l(-1,-1,1\r)\,.
\ee
We can now look for a consistent truncation to  a single scalar sector, by inspecting the scalar equations of motion
\be
	\square\varphi_j+8g^2\sinh\varphi_j-\frac14\sum_{i=1 }^4a_i^je^{\overrightarrow{a}_i\cdot\overrightarrow{\varphi}}\l(F^i_{(2)}\r)^2.
\ee
It will only be possible to set, \emph{e.g.} $\varphi_2=\varphi_3=0$ if the Maxwell source term in the above equations vanish. These respectively read for $j=2,3$
\bea
	j=2:&& e^{\varphi_1+\varphi_2+\varphi_3}\l(F^1_{(2)}\r)^2- e^{\varphi_1-\varphi_2-\varphi_3}\l(F^2_{(2)}\r)^2+ e^{-\varphi_1+\varphi_2-\varphi_3}\l(F^3_{(2)}\r)^2- e^{-\varphi_1-\varphi_2+\varphi_3}\l(F^4_{(2)}\r)^2,\nn\\
	j=3:&& e^{\varphi_1+\varphi_2+\varphi_3}\l(F^1_{(2)}\r)^2- e^{\varphi_1-\varphi_2-\varphi_3}\l(F^2_{(2)}\r)^2- e^{-\varphi_1+\varphi_2-\varphi_3}\l(F^3_{(2)}\r)^2+ e^{-\varphi_1-\varphi_2+\varphi_3}\l(F^4_{(2)}\r)^2,\nn
\eea
which upon truncating $\varphi_2,\varphi_3$ become
\bea
	j=2:&& e^{\varphi_1}\l(F^1_{(2)}\r)^2- e^{\varphi_1}\l(F^2_{(2)}\r)^2+ e^{-\varphi_1}\l(F^3_{(2)}\r)^2- e^{-\varphi_1}\l(F^4_{(2)}\r)^2,\nn\\
	j=3:&& e^{\varphi_1}\l(F^1_{(2)}\r)^2- e^{\varphi_1}\l(F^2_{(2)}\r)^2- e^{-\varphi_1}\l(F^3_{(2)}\r)^2+ e^{-\varphi_1}\l(F^4_{(2)}\r)^2.\nn
\eea
Setting further $F_{(2)}^1=F_{(2)}^2=F_{(2)}/\sqrt2$ and  $F_{(2)}^3=F_{(2)}^4=\tilde F_{(2)}/\sqrt2$ cancels theses terms and allows to have a consistent truncation to a single scalar $\phi$. Its equation of motion now is
\be
	\square\phi+8g^2\sinh\phi-\frac14e^\phi\l( F_{(2)}\r)^2+\frac14e^{-\phi}\l(\tilde F_{(2)}\r)^2.
\ee
One can then either set $F_{(2)}=\tilde F_{(2)}$ and get the theory
\be
	\frac{\mathcal L_4}{\sqrt{-g}} =  R-\half(\partial\phi)^2+8g^2\l(\cosh\phi+2\r)-\frac14\l(e^{\phi}+e^{-\phi}\r)\l(F_{(2)}\r)^2,
\ee
or set $\tilde F_{(2)}=0$ to reduce a single-exponential gauge coupling Einstein-Maxwell-Dilaton theory:
\be \label{KKS7Action}
	\frac{\mathcal L_4}{\sqrt{-g}} =  R-\half(\partial\phi)^2+8g^2\l(\cosh\phi+2\r)-\frac14e^{\phi}\l(F_{(2)}\r)^2,
\ee
which admits an AdS$_4$ black hole solution.

\subsection{$\mathbf S^4$ reduction of $D=11$ supergravity \label{SUGRAAdS7S4}}

The $\mathbf S^4$ reduction of $D=11$ supergravity reduces to a theory with $SO(5)$ Yang-Mills gauge group, which can be truncated to an Abelian $U(1)^2$ sector. Provided we restrict ourselves to purely electric or magnetic solutions, the axions of the theory can be set to zero, and the only fields remaining are the metric, two scalars and the two gauge potentials. One can further truncate to a single scalar sector by setting one of the scalar to zero and equating the two gauge fields. The final theory is then:
\be\label{KKS4Action}
	\frac{\mathcal L_7}{\sqrt{-g}} =  R-\half(\partial\phi)^2+g^2\l(4 e^{-\sqrt{\frac{2}{5}}\phi}+4 e^{\frac{3}{\sqrt{10}}\phi}-\half e^{\frac{8}{\sqrt{10}}\phi}\r)-\frac14e^{\sqrt{\frac{2}{5}}\phi}\l(F_{(2)}\r)^2,
\ee
which admits an AdS$_7$ black hole solution.

\section{Rotating black branes\label{section:RotatingBranes}}

The action
\be  \label{EFormAction3}
	S_{p+q+1}=\frac{1}{16\pi G_D}\int\ud^{p+q+1}x\,\sqrt{-g}\l[R-\frac{1}{2(p+1)!}G^2_{[p+1]}\r]
\ee
has rotating black brane solutions, see for instance \cite{Cvetic:1999xp} and references therein, whose worldvolume is supported by a non-trivial $(p+1)$-field strength and whose static limit is the solution \eqref{BlackqBraneNCharge}-\eqref{PhiB1} with a uniform charge along all worldvolume directions, $n=p-1$. For simplicity and to keep with our previous notations, we will only consider the case where the sphere $\mathbf S^q$ in the transverse space is odd, $q=2N-1$. The solution goes as
\be
	\ud s^2=h(r)^{-\frac2p}\l[-\l(1-\frac{2m}{\Delta r^{q-1}}\r)\ud t^2+\ud R^2_{(p-1)}\r]+h(r)^{\frac2{q-1}}\l[\frac{\ud r^2}{H_1\dots H_N-\frac{2m}{r^{q-1}}}+\r.\nn
\ee
\be
	\l.+r^2\sum_{i=1}^NH_i\l(\ud\mu_i^2+\mu_i^2\ud \psi_i^2\r)-\frac{4m\cosh\al}{r^{q-1}h\Delta}\ud t\sum_{i=1}^N\ell_i\mu_i^2\ud\psi_i+\frac{2m}{r^{q-1}h\Delta}\l(\sum_{i=1}^N\ell_i\mu_i^2\ud\psi_i\r)^2\r],\label{D1}
\ee
\be
	H_i=1+\frac{\ell^2}{r^2}\,,\qquad \Delta = H_1\dots H_N\sum_{i=1}^N\frac{\mu_i^2}{H_i},\qquad h(r)=1+\frac{2m\sinh^2\al}{r^{q-1}\Delta},\label{D2}
\ee
\be
	A_{[p]}=\coth\al\l(1-\frac1h\r)\ud t\wedge\ud R_{(p-1)}+\l(1-\frac1h\r)\frac{\sum_{i=1}^N\ell_i\mu_i^2\ud\psi_i}{\sinh\al}\wedge\ud R_{(p-1)},\label{D3}
\ee
where there are $N$ angular momenta $\ell_i$ along the $N$ independent planes of rotation of the sphere, and the charge carried by the form is parametrized by the value of $\al$, with $\al\to 0$ the neutral limit. The sphere is parametrized by $2N$ coordinates $\mu_i,\,\psi_i$ which are not independent as $\sum\mu_i^2=1$. However, this set of coordinates allows to separate neatly the $N$ rotation planes.

One may take the decoupling limit, which zooms on the near-horizon region of the extremal black hole: $\al\to+\infty$, $m\to0$ while keeping $g^{1-q} = 2m\sinh^2\al$ fixed, where $g$ is the radius of the transverse sphere. The metric becomes, after the change of coordinates $g\rho=\frac{p}{q-1}\l(gr\r)^{\frac{q-1}{p}}$
\bea
	\ud s^2&=& \tilde\Delta^{\frac{q-1}{p+q-1}}\l[-\l(H_1\dots H_N\r)^{-\frac{p-2}{p-1}}f(\rho)\ud t^2+\l(H_1\dots H_N\r)^{\frac{1}{p-1}}\l(\frac{\ud\rho^2}{f(\rho)}+\rho^2\ud R^2_{(p-1)}\r)\r]+\nn\\
	&&\qquad\qquad\qquad\qquad\qquad\qquad +g^{-2}\tilde\Delta^{\frac{-p}{p+q-1}}\sum_{i=1}^NX_i^{-1}\l[\ud\mu_i^2+\l(\ud\psi_i+gA_i\r)^2\r],\label{D4}
\eea
\be
	\tilde\Delta=\frac{\sum_{i=1}^NX_i\mu_i^2}{\l(X_i\dots X_N\r)^2}=\Delta\l(H_1\dots H_N\r)^{-\frac{p+q-1}{(p-1)(q-1)}},\quad X_i=\l(H_1\dots H_N\r)^{\frac{p}{(p-1)(q-1)}}H_i^{-1},\label{D5}
\ee
\be
	f(\rho)=\frac{(q-1)^2}{p}g^2\rho^2\l(H_1\dots H_N\r)-\frac{\mu}{\rho^{p-2}}\,,\qquad A_i=\frac{\ell_i\ud t}{gr^2\sinh\al~H_i}\,.\label{D6}\\
\ee
Note that, if all rotations are turned off, the transverse sphere decouples from the internal metric.
As it happens, this is precisely the form needed to Kaluza-Klein reduce along the sphere, with the dilaton vector being proportional to $\tilde\Delta$, \cite{Cvetic:1999xp}.
At some point along the way, in order to be able to reach the above decomposition of the metric in the decoupling limit, we made use of the identity
\be
	2p=(q-1)(p-2)\,,\label{D7}
\ee
which is a necessary condition for truncating the massive Kaluza-Klein modes while retaining the gauge bosons of the symmetry group of the sphere $SO(q+1)$. This condition is very restricitive, since it selects only the supergravity cases described in appendix \ref{SUGRAAdS}. It may be relaxed a little by including a dilaton in the higher-dimensional theory \eqref{EFormAction3}. In section \ref{section:KKSphereReduction}, we shall keep the dimensions arbitrary, as we shall see that the lower-dimensional reduction of \eqref{D4}-\eqref{D6} will be analytically continued to a family of solutions valid in generic dimensions, with a single scalar and gauge field.

\end{document}